\documentclass[a4paper,11pt]{article}
\pdfoutput=1 
\usepackage{amsmath}
\usepackage{jheppub} 
\usepackage{multicol}
\usepackage{scrextend}
\usepackage[dvipsnames]{xcolor}
\usepackage{tikz}
\usepackage[T1]{fontenc} 
\usepackage[vcentermath]{youngtab}
\usepackage{subfig}
\usepackage{xspace}
\usepackage{placeins}
\usepackage{bbm}
\usepackage[utf8]{inputenc} 

\newcommand{\bP}{\mathbbm{P}}
\newcommand{\bZ}{\mathbbm{Z}}

\newcommand{\bE}{\mathbbm{E}}
\newcommand{\bR}{\mathbbm{R}}

\newcommand{\cS}{\mathcal{S}}
\newcommand{\cA}{\mathcal{A}}
\newcommand{\cN}{\mathcal{N}}
\newcommand{\cD}{\mathcal{D}}
\newcommand{\cR}{\mathcal{R}}
\newcommand{\ns}{$n$'s }
\newcommand{\ms}{$m$'s }
\newcommand{\grad}{\nabla}
\newcommand{\TD}[1]{\text{TD}(\ensuremath{#1})\xspace}

\newcommand{\tw}[1]{{\tt #1 }\hspace{-.2cm}}
\newcommand{\tpd}{\ensuremath{\Delta_{\text{TC}}}}
\newcommand{\smd}{\ensuremath{\Delta_{\text{SM}}}}

\newcommand{\SU}[1]{\text{SU}(\ensuremath{#1})}
\newcommand{\SO}[1]{\text{SO}(\ensuremath{#1})}
\newcommand{\USp}[1]{\text{USp}(\ensuremath{#1})}
\newcommand{\Sp}[1]{\text{Sp}(\ensuremath{#1})}

\newcommand{\U}[1]{\text{U}(\ensuremath{#1})}

\newenvironment{definition}
{%
  \setlength{\parskip}{0pt}%
  \par\addvspace{2mm} 
  \renewcommand{\tabcolsep}{0pt}%
  \noindent\begin{tabular}{p{0.15\linewidth}p{0.85\linewidth}}%
  \textit{Definition.~} &%
}%
{%
  \end{tabular}%
  \par\addvspace{2mm}%
}%

\title{\boldmath Branes with Brains: Exploring String Vacua \\ with Deep Reinforcement Learning}

\author[a]{James Halverson,}
\author[a]{Brent Nelson,}
\author[b,c]{Fabian Ruehle}

\affiliation[a]{Department of Physics, Northeastern University,\\Boston, MA 02115, USA}
\affiliation[b]{CERN, CERN, Theoretical Physics Department\\ 1 Esplanade des Particules, Geneva 23, CH-1211, Switzerland}
\affiliation[c]{Rudolf Peierls Centre for Theoretical Physics, Oxford University,\\1 Keble Road, Oxford, OX1 3NP, UK}

\emailAdd{j.halverson@northeastern.edu}
\emailAdd{b.nelson@northeastern.edu}
\emailAdd{fabian.ruehle@cern.ch}

\abstract{
We propose deep reinforcement learning as a model-free method for exploring the landscape of string vacua. As a concrete application, we utilize an artificial intelligence agent
known as an asynchronous advantage actor-critic to explore type IIA compactifications with intersecting D6-branes. As different string background configurations are explored by changing D6-brane configurations, the agent receives rewards and punishments related to string consistency conditions and proximity to Standard Model vacua. These are in turn utilized to update the agent's policy and value neural networks to improve its behavior. By reinforcement learning, the agent's performance in both tasks is significantly improved, and for some tasks it finds a factor of $\mathcal{O}(200)$ more solutions than a random walker. 
In one
case, we demonstrate that the agent learns a human-derived strategy for finding consistent string models.
In another case, where no human-derived strategy
exists, the agent learns a genuinely new strategy that
achieves the same goal twice as efficiently per unit time.
Our results demonstrate that the agent learns to solve various string theory consistency conditions simultaneously, which are phrased in terms of non-linear, coupled Diophantine equations.
}
  
\begin{document} 
\maketitle
\flushbottom

\section{Introduction}

String theory is a theory of quantum gravity that has shed light on numerous aspects of theoretical physics in recent decades, bringing new light to old problems and influencing a diverse array of fields, from condensed matter physics to pure mathematics. As a theory of quantum gravity it is also a natural candidate for  unifying known particle physics and cosmology. The proposition is strengthened by the low energy degrees of freedom that arise in string theory, which resemble the basic building blocks of Nature, but is made difficult by the vast number of solutions of string theory, which arrange the degrees of freedom in diverse ways and give rise to different laws of physics. 

This vast number of solutions is the landscape of string vacua, which, if correct, implies that fundamental physics is itself a complex system. Accordingly, studies of the string landscape are faced with difficulties that arise in other complex systems. These include not only the solutions themselves, which  limit computation by virtue of their number, but also tasks that are necessary to understand the physics of the solutions, which hamper computation by virtue of their complexity. As examples of large numbers of solutions, original estimates of the existence of at least $10^{500}$ flux vacua \cite{Ashok:2003gk} have ballooned in recent years to $10^{272,000}$ flux vacua \cite{Taylor:2015xtz} on a fixed geometry. Furthermore, the number of geometries has also grown, with an exact lower bound \cite{Halverson:2017ffz} of $10^{755}$ on the number of F-theory geometries, which Monte Carlo estimates demonstrate is likely closer to $10^{3000}$ in the toric case \cite{Taylor:2017yqr}.\footnote{The number of weak Fano toric fourfolds that give rise to smooth Calabi-Yau threefold hypersurfaces was recently estimated \cite{Altman:2018zlc} to be $10^{10,000}$, but it is not clear how many of the threefolds are distinct.} In fact, in $1986$ it was already anticipated \cite{LERCHE1987477} that there are over $10^{1500}$ consistent chiral heterotic compactifications. As examples of complexity, finding small cosmological constants in the Bousso-Polchinski model  is NP-complete \cite{Denef:2006ad},  constructing scalar potentials in string theory and finding minima are both computationally hard \cite{Halverson:2018cio}, and the diversity of Diophantine equations that arise in string theory (for instance, in index calculations) raises the issue of undecidability in the landscape \cite{Cvetic:2010ky} by analogy  to the negative solution to Hilbert's $10^{\text{th}}$ problem.  Finally, in addition to difficulties posed by size and complexity, there are also critical formal issues related to the lack of a complete definition of string theory and M-theory. Formal progress is therefore also necessary for fully understanding the landscape. 

For these reasons, in recent years it has been proposed to use techniques from data science, machine learning, and artificial intelligence to understand string theory broadly, and string vacua in particular, beginning with \cite{He:2017aed,Krefl:2017yox,Ruehle:2017mzq,Carifio:2017bov}. Numerous techniques from two
of the three canonical types of machine learning have been applied to a variety
physical problems:
\begin{itemize} 
\item {\bf Supervised learning:} 

Perhaps the best-known type
of machine learning is learning that is \emph{supervised}.
Labelled training data is used to create a model that accurately
predicts outputs given inputs, including tests on unseen data
that is not used in training the model.

Supervised learning makes up the bulk of the work thus far on
machine learning in string theory. In \cite{Ruehle:2017mzq}
it was shown that genetic algorithms can be utilized to optimize
neural network architectures for prediction in physical problems.
In \cite{Carifio:2017bov} it was shown that simpler supervised learning
techniques that do not utilize neural networks can lead to rigorous
theorems by conjecture generation, such as a theorem regarding the
prevalence of $E_6$ gauge sectors in the ensemble \cite{Halverson:2017ffz} of $10^{755}$ F-theory geometries. Supervised learning was also 
utilized \cite{Krefl:2017yox} to predict $a$ central charges in
$4d$ $\cN=1$ SCFTs via volume minimization in gravity duals with
toric descriptions.
In mathematical directions that are also relevant for string vacua,
supervised learning yielded an estimated upper bound on the
number of Calabi-Yau threefolds realized as hypersurfaces in
a large class of toric varieties \cite{Altman:2018zlc}, and has also
led to accurate predictions for line bundle cohomology 
\cite{Ruehle:2017mzq,Klaewer:2018sfl}. See \cite{Liu:2017dzi,Wang:2018rkk,Jinno:2018dek,Bull:2018uow,Rudelius:2018yqi,Jejjala:2019kio}
for additional works in string theory that use supervised learning.

\item {\bf Unsupervised learning:} 

Another type of learning is \emph{unsupervised}. In this case
data is not labelled, but the algorithm attempts to learn features
that describe correlations between data points.

Strikingly, in \cite{Hashimoto:2018bnb} QCD observables were
utilized to learn bulk metrics that give the first predictions of the  $q\overline q$ potential in holographic QCD. The results match lattice data
well, including the existence of the Coulomb, linear confining, and Debye screening phases.\footnote{We describe this work as unsupervised
learning because the learned bulk geometry was encoded in neural network weights, not the neural network outputs that fix boundary conditions for
bulk scalar fields at the black hole horizon.}
In \cite{Cole:2018emh}, topological data analysis (persistent homology)
was utilized to characterize distributions of string vacua represented
by point clouds in low-dimensional moduli spaces. 
In \cite{Mutter:2018sra}, autoencoders were utilized to study the accumulation of minimal supersymmetric
standard models on islands in the two-dimensional latent space
of the autoencoder, suggesting the existence of correlations between
semi-realistic models in the space of heterotic orbifolds.
\end{itemize}
Some techniques in data science do not fit cleanly into these
categories, or the third category we propose to utilize below. 
These include generative adversarial networks \cite{2014arXiv1406.2661G},
 which were utilized to generate effective field theory models
 \cite{Erbin:2018csv}, and network science, which was utilized to
 study vacuum selection in the landscape \cite{Carifio:2017nyb}.

\medskip

In this paper we propose utilizing  deep reinforcement learning (RL) to intelligently  explore string vacua in a model-free manner. Reinforcement learning (RL) is at the heart of many recent breakthroughs in machine learning. What differentiates RL from supervised and unsupervised learning is that, instead of studying a large fixed data set that  serves as training data, RL utilizes an artificial intelligence agent that explores an environment, receiving rewards as it explores states and changing its behavior accordingly. That is, utilizing the basic idea of behavioral reinforcement from psychology, the agent learns how to act properly over time based on received rewards. RL is a mature field that has experienced great progress in recent years as deep neural networks have been utilized in the RL framework, giving rise e.g.\ to AlphaGo \cite{silver2016mastering} and AlphaZero \cite{silver2017mastering}. 

We envision that there are many aspects of RL that could be useful in studies of string vacua.  There are at least three ideas that are central to our proposal: 
\begin{itemize} 
\item  First, the use of neural networks as function approximators for policy and value functions in RL allows for the study of systems with more states than could ever be directly enumerated. The ability to do so seems essential for string landscape studies, based on the numbers quoted above. Success in this direction already exists in the RL literature: for instance, AlphaZero performs at a world-class level, despite the fact that  Go has $\mathcal{O}(10^{170})$ legal board positions.
\item Second, the use of RL allows for the possibility of discovering search strategies that have not been discovered by string theorists. In domains where string theorists have already developed heuristic exploration algorithms, RL could lead to improvements; in new domains, RL may lead to good results while avoiding using time to develop heuristic algorithms.  
\item Third, many RL algorithms have the advantage of being model-free, i.e.\ the same algorithm may lead to good results in a diverse array of environments. That is, RL algorithms can be adapted to new situations simply by telling the agent how to navigate the environment, allowing for fast implementation. 
\end{itemize} 
Finally, given that issues of computational complexity arise in the landscape, one might worry about difficulties it poses for RL. It is hard to address this concern in general, but we note that RL has been successfully utilized \cite{rlnp} to solve instances of NP-complete problems. Similarly, we observe that our agent learns to solve non-linear, coupled systems of Diophantine equations that encode the physical and mathematical consistency conditions we impose on the vacua. Whether RL is able to perform such tasks in general or whether it is due to an underlying structure in these equations which is recognized and learned by the agent is an interesting question, but beyond the scope of this paper. 

For demonstrating the efficacy of RL, we choose a particularly simple string-theoretic setup: our environment is the space of $T^6/(\mathbbm{Z}_2 \times \mathbbm{Z}_2 \times \mathbbm{Z}_{2,O})$ orientifold compactifications of the type IIA superstring with intersecting D6-branes on a toroidal orbifold. An antiholomorphic involution $\mathbbm{Z}_{2,O}$ on the orbifold gives rise to a fixed O6-plane. Cancellation of Ramond-Ramond charge of the O6-plane requires the introduction of D6-branes, which are also subject to K-theory and supersymmetry conditions. If all of these conditions are satisfied, the configuration is a consistent superstring compactification and the relative placements of D6-branes determines a low energy gauge sector that may or may not resemble the Standard Model (SM) of particle physics. From the perspective of RL, different states are defined by different placements of D6-branes, and we define multiple different types of RL agents that differ from one another in how they change the placement of D6-branes. Via appropriate choices of reward function, the agent is incentivized to find consistent configurations that resemble the SM. Though we do not find a SM (which is not guaranteed to exist on this particular space), the RL agent demonstrates clear learning with respect to both consistency and particle physics goals. The RL agents outperform random walkers, in some cases by a factor of $\mathcal{O}(200)$, which serve as our control experiment.

In one case, we demonstrate that
the agent learns a human-derived strategy that utilizes so-called
filler branes. In another case that cannot utilize filler
branes, we find that the strategy utilized by the agent is
about a factor of $2$ more efficient at finding consistent
string models than the filler brane
strategy. This demonstrates the plausibility of utilizing
RL to find strategies in string theoretic environments that are superior to human-derived
heuristics.

This paper is organized as follows. In Section \ref{sec:rl} we provide an introduction to reinforcement learning for the reader, culminating with the asynchronous advantage actor-critic (A3C), which is  used in our study. In Section \ref{sec:IIAEnv} we  describe the IIA environment in detail, including  the orbifold itself, important truncations thereof, and three different implementations of RL agents. Readers familiar
with the physics that are not interested in the
details of the RL algorithm might consider skipping
to Section \ref{sec:SystematicRL}, where we present the  results of our RL experiments in the IIA environment. We
discuss and summarize the results in Section \ref{Sec:conc}.

\section{Basics of Reinforcement Learning}
\label{sec:rl}
Since it is central to our work, we would like to review the basics of RL in this section. We will first review the basic components of an RL system and define a Markov Decision Process (MDP). The MDP describes the interactions of the agent with the environment, and when the MDP is solved the agent has optimal behavior. We will briefly introduce classic techniques in RL that have been utilized for decades. One downside, however, is that these techniques cannot be readily applied in environments with extremely large numbers of states unless only a small subset of the states are sampled. Such situations are helped by the introduction of approximation methods, in particular function approximators. In deep RL, these function approximators are deep neural networks. We will review two types of approximation methods that utilize deep neural networks, value function approximation and policy gradients, and conclude with a discussion of the asynchronous advantage actor-critic (A3C) algorithm that is utilized in our work. For an in-depth introduction to RL, see the canonical text \cite{Sutton98a} or David Silver's lectures \cite{SilverLectures}, which also include recent breakthroughs in deep RL.

\medskip

We present the general ideas before becoming concerned with precise definitions. Reinforcement learning takes place in an \emph{environment}, where an \emph{agent} perceives a subset of the environment data known as a \emph{state} $s$. Based on a \emph{policy} $\pi$, the agent takes an \emph{action} that moves the system to a different state $s'$, and the agent receives a \emph{reward} based on the fitness of $s'$. Rewards may be accumulated as subsequent actions are taken, perhaps weighted by a \emph{discount factor}, and the accumulated discounted reward is called the \emph{return} $G(s)$. The return depends on the state, and there are many possible returns for a given state based on the subsequent trajectory through state space; the expected return is called the \emph{state value function} $v(s)$, and a related function that is more useful for some purposes is the \emph{action value function} $q(s,a)$. There are different classes of RL techniques, but each involves updates to one or more of these functions as the agent explores the environment. These updates improve the agent's behavior, i.e.\ by changing its behavior based on received rewards (or punishments), the agent learns how to act properly in order to carry out its given tasks.

In some cases, an RL agent ends in some final state from which there are no actions. These are \emph{terminal states} and the associated tasks are called \emph{episodic tasks}. In other cases, reinforcement learning tasks are \emph{continuous} or \emph{non-episodic} tasks. For example, an RL agent that learns to play chess may arrive in a terminal state that is a stalemate or a checkmate. Each episode is one game, and the RL agent may learn by studying states, actions, and rewards across many games. There are a number of benchmark RL environments, such as \emph{cart-pole} or \emph{multi-armed bandits}, that are used for testing new RL algorithms. Illustrative codes and videos of these environments and others can be found in the OpenAI gym \cite{DBLP:journals/corr/BrockmanCPSSTZ16} or numerous GitHub repositories.

Finally, one concept central to the success of an RL agent is \emph{exploration vs.\ exploitation}. If an agent usually chooses to exploit its current knowledge about the rewards of the local state space rather than exploring into new regions of state space, it may become trapped at a local reward maximum. Examples abound in the RL literature, but perhaps relevant for physicists is Feynman's restaurant problem, which comes in a few versions. In one, Feynman and his friend hear about an excellent restaurant with $N$ entrees. They have never been to the restaurant, but they are working under the assumption that with perfect knowledge of all entrees there would be an ordered list of entrees according to the reward (flavor) they provide. The first time at the restaurant, they have to explore and try a dish they've never tried. The second time they can try that dish again, exploiting their knowledge of its reward, or they can continue to explore. The problem is, at the $M^\text{th}$ timestep, should they exploit their gained knowledge of the ordered list by ordering their favorite entree thus far, or should they explore? What is the strategy that maximizes the reward? The solution requires a balance of exploration and exploitation that is characteristic of RL problems.

\medskip

We now turn to precise definitions and equations that describe RL systems. Using the notation of Sutton and Barto \cite{Sutton98a}, the central elements of RL are:
\begin{itemize}
\item \textbf{States.} A state represents what the agent measures from the environment. A state is usually written as $s, s',$ or $S_t$, with the convention that $s'$ occurs after $s$, or if there are multiple steps, $t$ denotes the timestep.  The set of states is $\cS$.
\item \textbf{Actions.} The agent acts with an action to move from one state to another. $\cA$ is the abstract set of actions, and $\cA(s)$ is the set of actions possible in the state $s$. A concrete action is denoted by $a$, $a'$ or $A_t$.
\item \textbf{Policy.} A policy is a map from states to actions, $\pi:\cS\to \cA$. A \emph{deterministic policy} $\pi(s)$ picks a unique action $a$ for each state $s$, and a \emph{stochastic policy} $\pi(a|s)$ is the probability of the agent selecting action $a$ given that it is in state $s$.
\item \textbf{Reward.} The reward $R_t\in \bR$ at a given time $t$ depends on the state $S_t$, or alternatively the previous state $S_{t-1}$ and action $A_{t-1}$ that led to the current state and its reward. The goal of an agent is to maximize the total future accumulated reward. The set of rewards is called $\cR$.
\item \textbf{Return.} The return measures accumulated rewards from time $t$,
\begin{align}
\label{eq:Return}
G_t = \sum_{k=0}^\infty \, \gamma^k R_{t+k+1},
\end{align}
  where $\gamma \in [0,1]$ is the \emph{discount factor} and the sum truncates for an episodic task.  The discount factor is used to encode the fact that in some systems receiving a reward now is worth more than receiving the same reward at a later time. For stochastic policies, there may be many trajectories through state space from $s_t$, each with its own associated reward $G_t$.
\item \textbf{Value Functions.} The \emph{state value function} is the expected return given $s$,
\begin{align}
v(s) = \bE[G_t|S_t=s].
\end{align}
It is important to distinguish value from reward, as $v(s)$ captures the long-term value of being in $s$, not the short-term reward. Similarly, the \emph{action value function} is
\begin{align}
q(s,a)=\bE[G_t|S_t=s,A_t=a].
\end{align}
Both may be indexed by a subscript $\pi$ if the trajectories through state space are determined by a policy $\pi$, i.e., $v_\pi(s)$ and $q_\pi(s,a)$. When we refer to the value function, we implicitly mean the state value function $v(s)$.
\item \textbf{State Transition Probabilities.} $p(s'|s,a)$ is the probability of transition to a state $s'$ given $s$ and an action $a$. While in some cases $s'$ is fixed given $s$ and $a$, in other cases it is drawn from a distribution that encodes environmental randomness. 
\end{itemize}

There are two basic types of problems that one encounters in RL, the $\emph{prediction problem}$ and the $\emph{control problem}$. In the prediction problem, the goal is to predict $q_\pi(s,a)$ or $v_\pi(s)$ for a given policy $\pi$. In the control problem, the goal is to find the optimal policy $\pi_*$, i.e.\ the one that optimizes the value functions. We therefore need definitions for these optimizations:
\begin{itemize}
\item An \emph{optimal state-value function $v_*(s)$} is the maximum value function over all policies,
  \begin{equation} v_*(s) := max_{\pi}\,\, v_{\pi}(s).\end{equation}
\item An \emph{optimal action-value function $q_*(s,a)$} is the maximum action-value function over all policies,
  \begin{equation} q_*(s,a) := max_{\pi}\,\, q_{\pi}(s,a).\end{equation}
\item An \emph{optimal policy} $\pi_*(s)$ is a policy for which
  \begin{equation}\pi_*\geq \pi'\qquad \forall \pi',\end{equation}
   where this partial ordering is defined so that \begin{equation}v_\pi(s)\geq v_{\pi'}(s)\,\,\,\forall s \Rightarrow \pi \geq \pi'.\end{equation}   
 \end{itemize}
It is natural to expect that there is a close relationship between optimal policies and optimal value functions. It arises in the context of Markov Decision Processes.

A \emph{Markov Decision Process} (MDP) is a framework by which RL problems may be solved. An MDP is defined by a tuple $(\cS,\cA,\cR,p,\gamma)$. A policy $\pi$ defines the action of an agent in an MDP. 
Important facts about any MDP include:
\begin{itemize}
\item There exists an optimal policy $\pi_*$.
\item All optimal policies achieve the optimal value function $v_{\pi_*}(s)=v_*(s)$.
\item All optimal policies achieve the optimal action-value function $q_{\pi_*}(s,a)=q_*(s,a)$. 
\end{itemize}
There are three types of solutions for the prediction and control problems of MDPs that we will discuss: dynamic programming, Monte Carlo, and temporal difference learning.

To gain some intuition, consider one example of an MDP that is a two-dimensional maze represented by an $N\times N$ grid with $M$ black squares ($N^2-M$ white squares) that the agent cannot (can) travel to. There are therefore $N^2-M$ states, according to which white square the agent occupies.The actions are $\cA=\{U,D,L,R\}$, representing moving up, down, left, and right. For some state $s$, the actions $\cA(s)$ that may be taken may be restricted due to the presence of an adjacent black square. Therefore, a policy labels each square by the probability of executing $U,D,L$ or $R$, and the natural goal for the agent is to solve the maze as quickly as possible. How should the rewards be assigned? One option is to assign $1$ for reaching the terminal state at the end of the maze, and $0$ for all other states. In this case the agent would be incentivized to finish the maze, though not at any particular rate; this is not ideal. On the other hand, if one assigns $-1$ for every square\footnote{It is fine to assign $-1$ to the maze exit because it is a terminal state, so there are no actions that take the agent out of it. The episode ends upon reaching the maze exit.}, then the agent is penalized for each step and it wants to solve the maze quickly. If by ``solving the maze'' we mean doing it quickly, then this is a much better reward structure.

\subsection{Classic Solutions to Markov Decision Processes}
In this section we briefly discuss three classic methods for solving MDPs: dynamic programming, Monte Carlo, and temporal difference learning.

\emph{Dynamic Programming} (DP) is one solution to an MDP that was pioneered by Bellman. We first treat the prediction problem in DP. From the definition of the value function we can derive a recursive expression known as the \emph{Bellman equation for $v_\pi$},
\begin{align}
\label{eqn:BellmanPi}
v_\pi(s) = \sum_a \pi(a|s) \sum_{s'}p(s'|s,a) \,\, [r(s,a,s')+\gamma v_{\pi}(s')],
\end{align}
which allows us to compute the value function recursively. It expresses a relationship between the value of a state and the states that may come after it in an MDP. Note that this is a system of linear equations, and therefore $v_\pi$ can be solved for by matrix inversion. However, via Gauss-Jordan elimination, matrix inversion is an $\mathcal{O}(N^3)$ process for an $N\times N$ matrix, where $N$ is the number of states. Though polynomial time, an $\mathcal{O}(N^3)$ solution is too costly for many environments encountered in high energy theory. In the spirit of RL, it is better to use fast iterative solutions. This can be done via \emph{iterative policy evaluation}, where all states $s$ are looped over and the RHS of \eqref{eqn:BellmanPi} is assigned to the state value until there are no more changes; then the Bellman equation is solved and $v_\pi$ has been found. In practice, convergence to the solution if often fast if $v_\pi$ is updated in real time inside the loop, rather than waiting for the full loop over all states to finish before updating $v_\pi$. A similar Bellman equation exists for $q_\pi(s,a)$, which allows for an iterative policy evaluation that computes the action-value function. 

For solving the control problem, we iterate over two main steps: \emph{policy evaluation} and \emph{policy improvement}. We do this iteration until the policy converges, i.e.\ doesn't change anymore. After evaluating the policy as just discussed, we improve the policy by defining a new policy $\pi'(s)$
\begin{align}
\label{eqn:DPpinewq}
\pi'(s) = \text{argmax}_a \,\, q(s,a),
\end{align}
which is the $\emph{greedy}$ policy. Given a state $s$, the greedy policy greedily chooses the action that maximizes the action-value function. An \emph{$\epsilon$-greedy} policy chooses a random action with probability $\epsilon$ and follows the greedy policy with probability $1-\epsilon$; this has the advantage of encouraging exploration. Though policy improvement is fast, policy evaluation is an iterative algorithm inside the overall iteration for the control problem. This is inefficient. Another solution to the control problem is \emph{value iteration}, which is more efficient. In this algorithm we continue improving the policy via only one loop, over a variable $k$
\begin{align}
v_{k+1}(s) = \max_a \,\, \sum_{s'} p(s'|s,a)\,\,[r(s,a,s')+\gamma v_k(s')].
\end{align}
Note that the policy improvement step is now absent, so we are implicitly doing policy evaluation and improvement at the same time.

Dynamic programming lays the groundwork for the rest of the methods that we will discuss, but it has a number of drawbacks. First, note that for both the prediction problem and control problem we looped over all of the states on every iteration, which is not possible if the state space is very large or infinite. Second, it requires us to know the state transition probabilities $p(s',r|s,a)$, which is difficult to estimate or compute for large systems. Note that in DP there is no agent that is learning from experience while playing one or many episodes of a game; instead the policies are evaluated and improved directly. This is different in spirit from the game-playing central to other techniques.

\medskip

For instance, learning from experience is central in \emph{Monte Carlo} (MC) approaches to estimating the value function. In MC, the agent plays a large number $N$ of episodes and gathers returns from the states of the episode. Then the value function may be approximated by
\begin{align}
\label{eqn:MCv}
v(s)=\bE[G(t)|S(t)=s]\simeq \frac{1}{N}\sum_{i=1}^N G_i(s),
\end{align}
where this value function has been learned from the experience of the agent. MC only gives values for states that were encountered by the agent, so the utility of these methods is limited by the amount of exploration of the agent. The prediction problem is therefore straightforward: given a policy $\pi$, use \eqref{eqn:MCv} to compute $v_\pi(s)$. The control problem again uses policy iteration: as the agent plays episodes policy evaluation is used to calculate $q(s,a)$, from which the policy may be improved via choosing the greedy (or $\epsilon$-greedy) policy~\eqref{eqn:DPpinewq}. Note that since only one episode is played per iteration, the sampled returns are for different policies; nevertheless, MC still converges.

Monte Carlo techniques have important caveats. For instance, many episodes are required to calculate the returns, but if the task is not episodic or the policy does not lead to a terminal state, then the return is not well defined. To avoid this, a cutoff time on episodes can be imposed. MC also leaves many states unexplored. This can be improved by an \emph{exploring starts} method, where different episodes begin from a random initial state, or by improving the policy via $\epsilon$-greedy rather than greedy, which would encourage exploration.

\medskip

Another common method is \emph{Temporal Difference Learning} (TD), which estimates returns based on the current value function estimate. TD utilizes a combination of ideas from MC and DP. Like MC, agents in TD learn directly from raw experience without a model of the environment's dynamics, as required for DP. On the other hand, TD methods update estimates based on learned estimates, as in DP, rather than waiting for the final outcome at the end of an episode, as in MC. This is a major advantage, as TD methods may be applied with each action of the agent, but without requiring a full model of the environment such as the state transition probabilities. The general version of TD is referred to as $\TD\lambda$, where $\lambda\in [0,1]$ interpolates between $\TD0$ and $\TD1$, where the latter is equivalent to MC. Two famous TD algorithms for the control problem are \emph{SARSA} and \emph{Q-learning}. We refer the reader to \cite{Sutton98a} for details but would like to draw an important distinction. An algorithm is said to be \emph{on-policy} if the policy followed by the agent is the policy that is also being optimized; otherwise, it is off-policy. SARSA is on-policy, while Q-learning is off-policy.

\subsection{Deep Reinforcement Learning}
For an infinite or sufficiently large state space it is not practical to solve for optimal policies or value functions across the entire state space. Instead, approximations to policies and value functions are used, which allows for the application of RL to much more complex problems. For example, the game of Go is computationally complex and has $\mathcal{O}(10^{172})$ possible states (legal board positions), but RL yields an agent that is currently the strongest player in the world, AlphaZero \cite{silver2017mastering}.

We will focus on differentiable function approximators, such as those arising from linear combinations of features or from deep neural networks. The use of the latter in RL is commonly referred to as \emph{deep reinforcement learning} (deep RL). All function approximators that we utilize in this paper will be deep neural networks, but the following discussion is more general. We first discuss value function approximation, then policy approximation, and then actor-critic methods, which combine both. Finally, we will review the asynchronous advantage actor-critic (A3C) method~\cite{a3c}, which is the algorithm that we utilize.

\subsubsection{Value Function Approximation} 
Consider \emph{value function approximation}. Here, the approximations associated to the value function and action-value function are
\begin{align}
\hat v(s,w)\simeq v_\pi(s)\,, \qquad \hat q(s,a,w)\simeq q_\pi(s,a)\,, \qquad w\in \bR^n\,,
\end{align}
where $w$ is a parameter vector typically referred to as \emph{weights} for the value function approximation. The advantage is that the weights determine the approximate value function across the entire state space (or action-value function across the entire space of states and actions), which requires much less memory if $n\ll |\cS|$, since one stores the weight vector that determines $\hat v$ rather than an exact value for every state. Another advantage is that it allows for generalization from seen states to unseen states by querying the function approximator.

Suppose first that the value function $v_\pi(s)$ is known exactly. Then one would like to know the mean squared error relative to the approximation $\hat v(s,w)$
\begin{align}
J(w) = \bE_\pi[(v_\pi(s)-\hat v(s,w))^2].
\end{align}
Since the function approximators that we consider are differentiable, we can apply gradient descent with step size $\alpha$ to change the parameter vector in the direction of minimal mean squared error,
\begin{align}
\Delta w = -\frac12 \alpha \grad_w J(w)=\alpha \bE_\pi[(v_\pi(s) - \hat v(s,w)) \grad_w \hat v(s,w)]\,.
\end{align}
The step size $\alpha$ is commonly known as the learning rate. Since we are updating the weights as the agents are exploring, we use stochastic gradient descent,
\begin{align}
\Delta w = \alpha (v_\pi(s) - \hat v(s,w)) \grad_w \hat v(s,w)\,,
\end{align}
which will converge to the minimum mean square error with enough samples.

As an example, consider the case that the function approximator is linear combination of state-dependent features $x(s) \in \bR^n$
\begin{align}
\hat v(s,w) = x(s) \cdot w\,,
\end{align}
where the features are chosen to capture the essential elements of the state.  Then $\grad_w \hat v(s,w)=x(s)$ and
\begin{align}
\Delta w = \alpha (v_\pi(s)-\hat v(s,w))\cdot x(s)\,.
\end{align}
Appropriate feature vectors can be found in many circumstances, and they are very useful when the number of features is far less than the number of states. This seems particularly relevant for string theory studies, where the number of states is extremely large, but the number of features and / or experimental constraints is relatively small.

In reality, we do not know $v_\pi(s)$, or else we wouldn't be bothering to approximate it in the first place. Instead, we will replace the value function with one of the estimators or \emph{targets} associated with MC, $\TD0$, or $\TD\lambda$. Letting $T$ be the target, we have
\begin{align}
\Delta w = \alpha (T - \hat v(s,w)) \grad_w \hat v(s,w)\,,
\end{align}
and then targets associated with MC, $\TD0$, or $\TD\lambda$ are
\begin{align}
\label{eqn:vtargets}
T_\text{MC} = G_t  \qquad T_{\TD0} = R_{t+1} + \gamma \hat v(S_{t+1},w) \qquad T_{\TD\lambda} = G^\lambda_t\,,
\end{align}
where $T_{\TD\lambda}$ is known as the $\lambda$-return. The targets are motivated by incremental value function updates for each of these algorithms, see \cite{Sutton98a} for additional details.

We have discussed methods by which stochastic gradient descent may be used to find the approximate value function $\hat v(s,w)$ and have it converge to having a minimum mean square error, based on a followed policy $\pi$ and associated value $v_\pi(s)$. This is the prediction problem. If we can find the approximate action-value function $\hat q(S,A,w)$ and have it converge to having a minimum mean square error, we will have solved the control problem, as given a converged $\hat q(S,A,w)$ the optimal policy can be chosen greedily (or $\epsilon$-greedily).

We therefore turn to action-value function approximation. If the action value function is precisely known then stochastic gradient descent can be used to minimize the mean squared error. The incremental update to the weights is
\begin{align} 
\Delta w = \alpha (q_\pi(s,a) - \hat q(s,a,w)) \grad_w \hat q(s,a,w)\,,
\end{align} 
which is proportional to the feature vector in the case of linear value function approximation. However, since the value function is not precisely known, the exact action value function in the update is again replaced by a target $T$. For MC, $\TD0$ and $\TD\lambda$, $T$ is the same as the targets in \eqref{eqn:vtargets}, but with the approximate value functions $\hat v$ replaced by the approximate action value-functions $\hat q$.

For both the prediction and control problems, the convergence properties depend on the algorithms used (such as MC, $\TD0$, and $\TD\lambda$), and on whether the function approximator is linear or non-linear. In the case that the function approximator is a deep neural network, the target is chosen to be the loss function of the network.

\subsubsection{Policy Gradients}
We have discussed the use of function approximators to approximate value functions. When doing so, it is possible to converge to an optimal value function, from which an optimal policy is implicit by choosing the greedy policy with respect to the optimal value function.

Another alternative is to use policy based reinforcement learning, where we learn the policy $\pi$ directly rather than learning it implicitly from a learned value function. In particular, a function approximator may be used for a stochastic policy
\begin{align}
\pi_\theta(s,a) = \bP[a|s,\theta]\,,
\end{align}
which gives a probability of an action $a$ given a state $s$ and weight parameters $\theta \in \bR^n$ for the policy approximation.\footnote{They are the analogs of the weights $w$ for the value approximator discussed in the previous section.} We will again assume that our approximator is differentiable, so that policy gradients can point in directions of optimal weight change. Policy gradients maximize the parameters via gradient ascent with respect to an objective function $J(\theta)$ that is related to experienced rewards. The idea is that the objective function provides a measure of how good the policy is, and therefore an optimal policy can be determined by maximizing the objective function. Three common objective functions are 
\begin{align}
\begin{split}
J_1(\theta)&= v_{\pi_\theta}(s_1) = \bE_{\pi_\theta}[G_1]\,,\\
J_{\overline V}(\theta)&=\sum_s d_{\pi_\theta}(s) v_{\pi_\theta}(s)\,,\\
J_{\overline R}(\theta)&=\sum_s d_{\pi_\theta}(s) \sum_a \pi_\theta(s,a) R_s^a\,.
\end{split}
\end{align}
$J_1(\theta)$ is a measure of the expected
return given a fixed start state $s_1$. In
environments where the episode does not end or
there is not a fixed start state,
$J_{\overline V}(\theta)$ computes the average value by
summing over values of given states, weighted
by their probability $d_{\pi_\theta}$ of being visited  while
following
policy $\pi_\theta$; $d_{\pi_\theta(s)}$ is the
\emph{stationary distribution} of the Markov process.
$J_{\overline R}(\theta)$ is the average reward per time
step, where $R_s^a$ is the reward received after
taking action $a$ from state $s$.

To maximize the objective function, the parameters are updated via gradient ascent
\begin{align}
  \Delta\theta = \alpha \grad_\theta J\,,
\end{align}
where $\alpha$ is the learning rate. It is useful
to rewrite policy gradients as
\begin{align}
  \grad_\theta \pi_\theta(s,a)=\pi_\theta \grad_\theta \log  \pi_\theta(s,a),
\end{align}
where $\grad_\theta \log \pi_\theta(s,a)$ is known as the \emph{score function}. Central to optimizing policies via function approximation is the \emph{policy gradient theorem}:
\begin{quote}
  \textbf{Theorem.} For any differentiable policy, for  any of the policy objective functions $J=J_1$, $J=J_{\overline R}$, or $J=J_{\overline V}$,
  \begin{align}
    \grad_\theta J(\theta) = \bE_{\pi_\theta}[\grad_\theta \log \pi_\theta(s,a)\, q_{\pi_\theta}(s,a)].
  \end{align}
\end{quote}
It depends only on the score function and action-value function associated with the policy. In practice $q_{\pi_\theta}(s,a)$ is not known, but can be approximated by MC, $\TD0$, or $\TD\lambda$ as discussed above. An early MC policy gradient algorithm is called REINFORCE~\cite{Williams:1987aaa,Williams:1992aaa}, but it has the downside of being rather slow. To solve this problem, we turn to actor-critic methods.

\subsubsection{Actor-Critic Methods}
The downside of MC policy gradients is that they require waiting until the end of an episode, and are therefore slow. Actor-critic methods solve this problem by updating the policy online, not at the end of an episode. Since online methods are desirable and the action-value function appears in the policy gradient theorem, it is natural to ask whether one could simultaneously use a function approximator for both the action-value function and the policy. Such methods are called \emph{actor-critic} (AC) methods. 

In AC there are two updates to perform: the critic updates the action-value function approximator by adjusting the weights $w$, and the actor updates the policy weights $\theta$ in the direction suggested by the action-value function, that is, by the critic. Letting $\hat \pi_\theta(s,a)$ and $\hat q_w(s,a)$ be the approximated policy and action-value function, the gradient of the objective function and policy parameter update are:
\begin{align}
  \grad_\theta J(\theta)\simeq \bE_{\pi_{\theta}}[\grad_\theta \log \hat \pi_\theta(s,a) \,\,\, \hat q_w(s,a)]\,, \qquad \Delta\theta = \alpha \grad_\theta \log \hat\pi_\theta(s,a)\,\,\, \hat q_w(s,a)\,.
\end{align}
The critic is simply performing policy evaluation using value function approximation, and therefore previously discussed methods are available to AC models.

There is also an important theorem for AC methods. A value function approximator is said to be \emph{compatible} to a policy $\pi_\theta$ if
\begin{align}
\grad_w \hat q_w(s,a) = \grad_\theta \log \pi_\theta(s,a)\,.
\end{align}
The compatible function approximation theorem is
\begin{quote}
  \textbf{Theorem.} If the action-value function is compatible and its parameters minimize the mean squared error, then the policy gradient is exact,
  \begin{align}
    \grad_\theta J(\theta) = \bE_{\pi_\theta}[\grad_\theta \log \pi_\theta(s,a)\,\, \hat q_w(s,a)]\,.
  \end{align}
\end{quote}
In such a case actor-critic methods are particularly accurate.

A \emph{baseline} function $B(s)$ can be utilized to decrease variance and improve performance. Critically, it does not depend on actions and therefore it can be shown that it does not change the expectations in the policy gradient theorem. A particularly useful baseline is the value function itself, $B(s) = v_{\pi_\theta}(s)$. In this case we define the \emph{advantage function}
\begin{align}
A_{\pi_\theta}(s,a) = q_{\pi_\theta}(s,a) - v_{\pi_\theta}(s),
\end{align}
in which case the policy gradient theorem can be rewritten
\begin{align}
 \grad_\theta J(\theta) = \bE_{\pi_\theta}[\grad_\theta\log \pi_\theta(s,a)\, A_{\pi_\theta}(s,a)].
\end{align}
This is an estimate of the advantage of taking the action $a$ in the state $s$ relative to the value of simply being in the state, as measured by $v_{\pi_\theta}(s)$.

\subsubsection{Asynchronous Advantage Actor-Critics (A3C)}  
In this paper we utilize an asynchronous advantage actor-critic (A3C) ~\cite{a3c} to study string vacua. It is a model-free algorithm developed in $2016$ that performs well relative to other algorithms available at the time, such as deep Q-networks \cite{DQN}. As expected based on its name, A3C is an actor-critic method. The central breakthrough of \cite{a3c} was to allow for \emph{asynchronous} reinforcement learning, meaning that many agents are run in parallel and updates are performed on neural networks as the ensemble of agents experience their environments. As an analogy, the idea is that workers (the agents) report back to a global instance (the global policy and/or value functions) in a way that their communal experience leads to optimal behavior. Four different asynchronous methods were studied, and the best performing method was an actor-critic that utilized the advantage function to update the policy, i.e., an A3C. We refer the reader to the original literature for a more detailed account.

For physicists with moderate computational resources, the use of A3C is a significant advantage. This is because many reinforcement learning techniques require specialized hardware such as GPUs or very large systems, whereas A3C may be run on a standard multi-core CPU. Details of our A3C implementation are discussed in the next sections.

We note that we are facing a multi-task reinforcement learning problem, which we tackle with two different methods. In the first method we employ, we check the various goals sequentially, i.e.\ only start checking the $N^\text{th}$ task if the previous $N-1$ tasks are solved. We also only end an episode if all tasks have been solved. However, we do provide increasing rewards for each of the tasks; for example  the $N^\text{th}$ task receives a reward of $10{c\,N}$ with $c$ of order one, in order to incentivize the agent to strive for the larger reward of the next task. In the second method, we learn the $N$ tasks by choosing $N$ different reward functions that are tailored towards one specific task. Since the agents act asynchronously, we simply utilize $N\times M$ workers total, where $M$ workers are learning to solve each of the $N$ tasks~\cite{Birck:2017aaa}.

\section{The Environment for Type IIA String Theory}
\label{sec:IIAEnv}
In this section we formulate the data of a $d=4$, $\cN=1$ compactification of type IIA superstring theory in a form that is amenable for a computer analysis. We begin with a general discussion, and then restrict to the case of orientifolds of toroidal orbifolds.

\subsubsection*{Defining Data} 
A $d=4$, $\cN=1$ orientifold compactification of the type IIA superstring with intersecting D6-branes is specified by: 
\begin{itemize} 
\item A pair $(X,\bar \sigma)$ where $X$ is a compact Calabi-Yau threefold (compact Ricci-flat six-manifold that is also complex and K\" ahler) and $\bar \sigma$ is an antiholomorphic involution which we also call $\bZ_{2,O}$. The fixed point locus is a three-cycle $\pi_{O6}$ that is wrapped by an O6-plane. 
\item A collection $\cD$ of stacks of $N_a$ D6-branes, $a=1,\ldots,|\cD|$, wrapped on three-cycles $\pi_a$ and their orientifold images $\pi_a^\prime$, where $\pi_a$ is a special Lagrangian submanifold, i.e.\  volume minimizing in its homology class. 
\item A Gauss law and a K-theory constraint for D6-brane Ramond-Ramond charge, and a supersymmetry condition; these are necessary in this context for a consistent supersymmetric compactification. 
\end{itemize} 
This data, which partially defines the compactification, is associated with a $d=4$, $\cN=1$ gauge theory sector. 

\subsubsection*{Gauge Group}
The overall gauge group is given by 
\begin{align}
G = \bigotimes_{a=1}^{|\cD|} G_a\,,
\end{align}
where $|\cD|$ is the number of D6 brane stacks and $G_a$ is a non-Abelian Lie group whose type is determined by the intersection of the brane stack with the orientifold plane.
\begin{itemize} 
\item $G_a = \U{N_a}$ if $\pi_a$ and $\pi_{O6}$ are in general position,
\item $G_a = \SO{2N_a}$ if $\pi_a$ is on top of $\pi_{O6}$,
\item $G_a = \USp{N_a}$ if $\pi_a$ is orthogonal to $\pi_{O6}$. 
\end{itemize}

\subsubsection*{Unbroken \U1} 
While each $\U{N_a}$ brane stack contributes a \U1 factor, these can be St\"uckelberg massive and hence not be present as a low energy gauge symmetry\footnote{From the low energy point of view, these symmetries appear as global symmetries that still influence physical observables such as Yukawa couplings.}. For toroidal orbifolds, the generators $T_i$ of the massless \U1s are given by the kernel of the $3\times K$ matrix
\begin{align}
T_i=\text{ker}(N^a m_i^a)\,,\quad i=1,2,3\,,\quad a=1,\ldots,\text{number of \text{U} stacks}\,,
\end{align}
where $K$ is the number of brane stacks with unitary gauge group and the $m_i^a$ are integers characterizing the unitary brane stacks, cf.~Section~\ref{sec:IIAZ2xZ2Orbifold}. Note that for phenomenological reasons, we demand that (at least) one \U1 remains massless, which can serve as the hypercharge of the standard model. Since the rank is $K-3$ generically, this requires in general four $\U{N_a}$ brane stacks.

\subsubsection*{Matter representations} 
Chiral multiplets may arise at brane intersections. The type of matter and its multiplicity depends on the intersection.
\begin{itemize}
\item Bifundamental matter $(\Box_a,\overline{\Box}_b)$ may arise at the intersection of D6-branes on $\pi_{a}$ and $\pi_{b}$, with chiral index $\chi(\Box_a,\overline{\Box}_b)=\pi_a \cdot \pi_b \in \bZ$, where $\Box$ and $\overline \Box$ denote the fundamental and anti-fundamental representation\footnote{For SO and USp groups, these will be the lowest-dimensional irreducible representations.} of the associated stack. Similarly,  $\chi(\Box_a,\Box_b)=\pi_a \cdot \pi_b^\prime \in \bZ$.
\item Matter in the two-fold symmetrized representation ({\tiny$\yng(2)$})$_{a}$ may arise at the intersection of a D6-brane with the orientifold brane, with chiral index $\chi$({\tiny$\yng(2)$})$_{a}=\frac12(\pi_a \cdot \pi_a^\prime - \pi_a \cdot \pi_{O6} ) \in \bZ$.
\item Matter in the two-fold anti-symmetrized representation ({\tiny$\yng(1,1)$})$_{a}$ may arise at the intersection of a D6-brane with the orientifold brane, with chiral index $\chi$({\tiny$\yng(1,1)$})$_{a}=\frac12(\pi_a \cdot \pi_a^\prime + \pi_a \cdot \pi_{O6} ) \in \bZ$. 
\end{itemize}
While this data encodes much of the physics, it is difficult to implement on a computer, as e.g.\ special Lagrangian submanifolds are notoriously difficult to construct explicitly.

\subsection[IIA \texorpdfstring{$\bZ_2 \times \bZ_2$}{Z2xZ2} Orbifold]{IIA  \texorpdfstring{$\boldsymbol{\bZ_2 \times \bZ_2}$}{Z2xZ2} Orbifold}
\label{sec:IIAZ2xZ2Orbifold}
We would like to translate this data into a form that is amenable for a computer analysis. First, we specify to the case that $X=T^6/(\bZ_2\times \bZ_2\times\bZ_{2,O})$, where the $\bZ_2\times \bZ_2$ are the orbifold action and $\bZ_{2,O}$ is the orientifold action. Second, we restrict to the case that the O6-plane and D6-branes wrap factorizable three-cycles, i.e.\ three-cycles that are one-cycles on each of the three $T^2$ factors in $T^6 = T^2 \times T^2 \times T^2$. Each such one-cycle is specified by a vector in $\bZ^2$. We will refer to them as $(n_1,m_1)$, $(n_2,m_2)$, $(n_3,m_3)$, for each of the three $T^2$ factors, respectively. These are the wrapping numbers along the basis of one-cycles $(\pi_{2i-1},\pi_{2i})$. On each $T^2$ we can define a (directed) symplectic intersection product of one-cycles. For a product of three two-tori with wrapping numbers 
\begin{align}
\pi_a = (n^a_1,m^a_1,n^a_2,m^a_2,n^a_3,m^a_3)\,,\quad \pi_b = (n^b_1,m^b_1,n^b_2,m^b_2,n^b_3,m^b_3)\,,
\end{align}
the intersection product is given by
\begin{align}
  I_{ab}=\prod_{i=1}^3 (n_i^a m_i^b-n_i^b m_i^a).
\end{align}

The orientifold action $\bar \sigma$ acts on the basis of one-cycles as
\begin{align}
  \bar \sigma: \pi_{2i-1}\to \pi_{2i-1} - 2b_i \pi_{2i}\,, \qquad
  \bar \sigma: \pi_{2i}\to -\pi_{2i}\,,
\end{align}
where $b_i$ is the tilt parameter.
In addition to the orientifold action we also mod out a non-freely acting $\bZ_2\times\bZ_2$ symmetry with generators $\theta$ and $\omega$ that act on the coordinates $z_i$ of the three tori as
\begin{align}
\begin{split}
\theta:~ &(z_1,z_2,z_3)\mapsto (z_1,-z_2,-z_3)\,,\qquad \omega:~ (z_1,z_2,z_3)\mapsto (-z_1,z_2,-z_3)\,,\\ 
\theta\omega:~ &(z_1,z_2,z_3)\mapsto (-z_1,-z_2,z_3)\,.
\end{split}
\end{align}
There are only two choices for the complex structure of the torus that are compatible with the orbifold and orientifold action: the rectangular torus ($b_i=0$) and the tilted torus ($b_i=\frac12$). The combination
\begin{equation}
\tilde \pi_{2i-1} = \pi_{2i-1}-b_i \pi_{2i}
\end{equation}
is orientifold even, and in the basis $(\pi_{2i},\tilde \pi_{2i-1})$ the wrapping numbers are $(n_i,\tilde m_i)$, where $\tilde m_i = m_i + b_i n_i$. For notational convenience, we also define the real quantities
\begin{align}
\label{eq:Definition UI}
U_0 = R_1^{(1)}R_1^{(2)}R_1^{(3)}\,,\qquad U_i=R_1^{(i)}R_2^{(j)}R_2^{(k)}\,,
\end{align}
with $i,j,k\in \{1,2,3\}$ cyclic and $R_1^{(i)}$ and $R_2^{(i)}$ the radii of the $i^\text{th}$ torus. We furthermore define the combination $\hat b =(\prod_i (1-b_i))^{-1}$, and the products 
\begin{align}
\label{eq:DefinitionXY}
\hat X^0&=\hat b n_1n_2n_3\,, \qquad ~~\hat X^i=-\hat b n_i \tilde m_j \tilde m_k\,,\\
\hat Y^0&=\hat b \tilde m_1 \tilde m_2 \tilde m_3\,, \qquad \hat Y^i=-\hat b \tilde m_i n_j n_k\,,
\end{align}
for $i,j,k\in \{1,2,3\}$ cyclic. The unhatted quantities  are defined in the same way with the factors $b_i$ set to zero. As each stack of D6-branes $a=1,\ldots,|\cD|$ has its own $(n_i,m_i)$ for $i=1,2,3$, the $\hat X$ and $\hat Y$ variables will often carry a subscript $a$ that denotes a particular D6-brane stack. In~\cite{Douglas:2006xy}, the quantities $\hat{X}^I$, $I=0,1,2,3$ are denoted by $P,Q,R,S$, respectively.

Note that if all winding numbers $n_i, m_i$ of a brane stack with $N$ branes have a common multiple $\mu$, the stack can be re-expressed as a stack with winding numbers $n_i/\mu, m_i/\mu$ and $N+\mu$ branes. Therefore, we demand that winding numbers on the torus be coprime, which translates into the condition
\begin{align}
(Y_a^0)^2 = \prod_{i=1}^3 \text{gcd}(Y^0_a,X^i_a)\,.
\end{align}

In terms of these quantities on the orbifold, we can concisely state the various consistency condition we have to impose on the compactification:

\subsubsection*{Tadpole Cancellation} 
The tadpole cancellation condition can be understood as RR charge conservation, i.e.\ we have to balance the positive charge of the D-branes against the negative charge of the Orientifold planes. The conditions read
\begin{align}
\label{eq:TadpoleCondition}
\sum_a N_a \hat X_a^0 = 8\hat b, \qquad
\sum_a N_a \hat X_a^i = \frac{8}{1-b_i}, \qquad i \in \{1,2,3\}\,.
\end{align}

\subsubsection*{K-Theory constraint} 
Another consistency constraint needed to ensure that the string background is well-defined can be derived from K-Theory. It guarantees that the multiplicity of fundamental representations of \USp2 is even and can be written as
\begin{align}
\label{eq:KTheoryCondition}
\sum_{a} N_a \hat Y^0_a\equiv0 \text{ mod }2\,, \qquad (1-b_j)(1-b_k)\sum_a N_a \hat Y_a^i \equiv0 \text{ mod }2\,,
\end{align}
for $i,j,k \in \{1,2,3\}$ cyclic. Violation of this condition will lead to a global gauge anomaly~\cite{Uranga:2000xp} known as Witten anomaly~\cite{Witten:1982fp}. 

\subsubsection*{Supersymmetry} 
The necessary conditions for unbroken supersymmetry (SUSY) read
\begin{align}
\label{eq:SUSYCondition}
\sum_{I=0}^3 \frac{\hat Y^I_a}{U_I}=0, \qquad \sum_{I=0}^3 \hat X^I_a U_I > 0\,.
\end{align}
These conditions are much harder to check than the others, i.e.\ the tadpole, K-theory, spectrum, and gauge group. The latter require linear algebra, while the SUSY conditions require solving a coupled system of equalities and inequalities. We will describe how we implemented the check in Python in Section~\ref{sec:SUSYCheckCQP}.

\subsubsection*{Data Structures} 
We now define concrete data structures that encode the data of one of these type IIA orbifold compactifications.
\begin{definition}
A \emph{plane} is a vector $(n_1,m_1,n_2,m_2,n_3,m_3)\in \bZ^6$ that represents the O6-plane.
\end{definition}
\begin{definition}
A \emph{stack} is a vector $(N,n_1,m_1,n_2,m_2,n_3,m_3)\in \bZ^7$ that represents a D6 stack.
\end{definition}
\begin{definition}
A \emph{state} $s$ is a set $s=(b_1,b_2,b_3,U_0,U_1,U_2,U_3, O,\cD)$, where $b_i\in \{0,\frac12\}$, $U_0,U_i \in \bR_+$, $O$ is a plane, and $\cD$ is a set of stacks. The set of states is denoted $\cS$.
\end{definition}
\noindent These are the data inputs that are central to our analysis.

The particle spectrum is a simple function of a state $s$. The gauge group $G(s)$ is encoded in the brane stacks $\mathcal{D}$ as explained above. The structure of bifundamental matter fields in a state $s$ is encoded in $f(s)\in \bZ^{|\cD|(|\cD|-1)}$. Furthermore there may also be matter fields in $s$ that are in two-index tensor representations. These may be encoded in a vector $t(s)\in \bZ^{2|\cD|}$. The vectors $f(s)$ and $t(s)$ may be combined into a vector encoding all of the matter in $s$, $m(s)\in \bZ^{|\cD|(|\cD|+1)}$. The spectrum $\mathcal{P}(s)$ of a state $s$ is therefore
\begin{align}
\mathcal{P}(s) = (G(s),m(s)).
\end{align}
The computation of $\mathcal{P}(s)$ is fast, as it depends only on simple conditional statements and linear arithmetic. 

Despite the ease with which physical outputs $\mathcal{P}(s)$ can be computed for any state $s\in \cS$, the global structure of $\cS$ is not known, and in fact even its cardinality is not known, though it is finite~\cite{Douglas:2006xy}. In addition to $\mathcal{P}$, we also need to check the K-Theory, tadpole, and SUSY conditions.

\medskip

Let us now put this data into the context of RL. Let $\cS$ be the set of states, $\cA$ the abstract set of possible actions, and $\cA(s)$ be the set of concrete actions on a particular state $s$. We will also use $s_t$ and $a_t$ to denote a state and an action at a discrete time $t$, respectively. 
\begin{definition}
An \emph{action} $a$ is a map $a:\cS \to \cS$ that changes the set of stacks $\cD$.
\end{definition}
Strictly speaking, this action should be called \emph{stacks-action}, since it modifies the brane stacks without changing the compactification space properties such as the tilting parameters $b_i$. Since there are only a few discrete choices, we take the $b_i$ fixed during run time and set up different runs with different $b_i$. From \eqref{eq:TadpoleCondition}, we find that the tadpole cancellation constraints become stronger if we tilt the tori. Thus, one would expect most solutions to appear on three untilted tori. While this is not discussed in the original papers~\cite{Gmeiner:2005vz,Douglas:2006xy} to the best of our knowledge, three untilted tori cannot give rise to an odd number of families. To see this, note that the chiral index can be written as
\begin{align}
\label{eq:EvenChi}
\chi := \sum_{a>b}  I_{a'b} - I_{ab} = 2\sum_{i=1}^3 \hat{X}_i^b \hat{Y}_i^a\,,
\end{align}
with $\hat{X},\hat{Y}$ as defined in \eqref{eq:DefinitionXY}. Since $n_i,m_i$  are integers for untwisted tori, so are $\hat{X},\hat{Y}$, and hence $\chi$ is always even. This was also ``rediscovered'' by the RL agents, which never produced any three generation model (or odd generation model in general) when run on three untwisted tori. This led us to conjecture and prove that this was indeed impossible.

\subsection[Truncated IIA \texorpdfstring{$\bZ_2 \times \bZ_2$}{Z2xZ2} Orbifold]{Truncated IIA \texorpdfstring{$\boldsymbol{\bZ_2 \times \bZ_2}$}{Z2xZ2} Orbifold}
\label{sec:OrbifoldTruncations}
\subsubsection{Truncating state and action space}
\label{sec:SymmetryTruncationb}
To test RL methods in string theory we will study a simplified set of type IIA compactifications where the state space is truncated. Specifically, we take
\begin{equation}
N \in \{0,1,\ldots,N_\text{max}\}, \,\,\, n_i\in \{-n,-n+1,\dots,n-1,n\}, \,\,\, m_i\in \{0,1,\dots,m\},
\end{equation}
with a fixed upper bound $D_\text{max}$ on $|\cD|$. Note that setting $N=0$ in a stack amounts effectively to removing it from $\cD$. Thus, we truncate by restricting to a $D_\text{max}$-stack model where each stack can have at most $N_\text{max}$ branes and the wrapping numbers are restricted according to parameters $n$ and $m$. The values of $N$ have to chosen to allow for standard models, i.e.\ $N_\text{max}\geq3$, and the $m_i$'s are chosen non-negative since the their negatives are automatically included as orientifold images. Since each stack is specified by a vector
\begin{equation}
  d=(N,n_1,m_1,n_2,m_2,n_3,m_3),
\end{equation}
there are $N_\text{max}\times (2n+1)^3\times(m+1)^3$ choices per stack, such that the number of states in the system without taking into account any symmetries is 
\begin{align}
N_\text{states}^\text{all}=\left[N_\text{max}(2n+1)^3(m+1)^3\right]^{D_\text{max}}.
\end{align}
However, this can be reduced by symmetries. We distinguish two inequivalent types of symmetries of a state $s$: 
\begin{itemize}
  \item Symmetries that lead to physically equivalent, indistinguishable models.
  \item Symmetries that connect a state $s$ with a different state $s'$ such that both $s$ and $s'$ are solutions that differ in their properties (e.g.\ in the moduli $U_i$) on a level that is not part of the current analysis but will eventually lead to inequivalent models.
\end{itemize}
Since we are ultimately interested in full solutions, we will only consider symmetries of the first type as true symmetries whose redundancies we want to eliminate. A priori, we can construct an infinite set of states by sending one or more of the parameters $(N_\text{max},D_\text{max},n,m)$ to infinity. While symmetries relate different states, this set will still contain infinitely many inequivalent states. Finiteness of the construction is only guaranteed if one combines symmetries with the physical constraints of tadpole \textit{and} SUSY conditions;
in the current context, this was shown in \cite{Douglas:2006xy}. This interplay\footnote{Note that these discussions focus on a given construction. It is not known, for instance, whether the number of Calabi-Yau threefolds is finite.} has also been observed in other string constructions~\cite{Douglas:2003um,Ashok:2003gk,Acharya:2005ez,Buchbinder:2013dna,Cvetic:2014gia,Nibbelink:2015ixa}. We do not implement this combination of constraints and symmetries to reduce the state space to a finite set, since it is extremely difficult to carry out. Furthermore, the resulting set is most likely still much too large.  Also, we want the machine to learn this connection itself.

The symmetries originate from two sources. First, we can reparameterize the tori. As explained above, due to the orientifold action we need to include $m_i$ as well as $-m_i$. Changing the signs of all three $m_i$ simultaneously corresponds to switching all branes with their orientifold images. Changing signs on two (out of the three) distinct pairs $(n_i,m_i)$ and $(n_j,m_j)$ simultaneously corresponds to an orientation-preserving coordinate transformation on the D6-branes.

Second, we can simply permute and relabel the tori and all their defining properties, which amounts to permuting $\hat{X}$ and $\hat{Y}$. In order to ensure that the physical constraints \eqref{eq:TadpoleCondition} and \eqref{eq:SUSYCondition} remain unchanged we extend the action of the permutation of the $\hat{X}$ and $\hat{Y}$ to the moduli. Note that the symmetry operation that permutes the $\hat{X}$ and $\hat{Y}$ corresponds to a simultaneous 90 degree rotation of two of the three tori,
\begin{align}
 (n_i,m_i)\rightarrow (m_i,-n_i)\qquad \text{and} \qquad (n_i,m_i)\rightarrow (m_j,-n_j)\,.
\end{align}
In order to implement this symmetry, we need to truncate the allowed range of the integers $n_i$ and $m_i$ to the same upper bound, $n=m$. We also need to simultaneously permute the moduli $U_i$ of the tori accordingly. 

We present an upper bound on the number of inequivalent states via the following considerations. First, we look at the three types of symmetries described above:
\begin{align}
\begin{split}
(S1):  &~~(n_i,m_i,n_j,m_j,n_k,m_k) \mapsto (n_i,-m_i,n_j,-m_j,n_k,-m_k)\,,\\
(S2): &~~(n_i,m_i,n_j,m_j,n_k,m_k) \mapsto (-n_i,-m_i,-n_j,-m_j,n_k,m_k)\,,\\
(S3): &~~(n_i,m_i,n_j,m_j,n_k,m_k) \mapsto (m_i,-n_i,m_j,-n_j,n_k,m_k)\,.
\end{split}
\end{align}
Since symmetries $(S2)$ and $(S3)$ leave the winding numbers of one torus invariant, there are three symmetry generators of type $(S2)$ and three symmetry generators of type $(S3)$. By analyzing the group structure, we find that the three generators of $(S3)$ generate a $(\bZ_4)^3$ symmetry. Furthermore, each $\bZ_4$ group of $(S3)$ contains one of the $\bZ_2$ groups generated by $(S2)$ as  a subgroup. Moreover, the three $\bZ_4$ symmetries do not commute with the $(\bZ_2)$ symmetry generated by $(S1)$. Thus, the symmetry operations generate the group $(\bZ_4)^3\rtimes\bZ_2$ of order $128$. Thus, we obtain 
\begin{align}
\label{eq:NStatesUperLimitRough}
N_\text{states}^\text{rough} = \left[N_\text{max}\frac{(2n+1)^6}{128}\right]^{D_\text{max}}\,,
\end{align}
as a first rough estimate for the number of states after symmetry reduction. However, we can further refine this count. First, \eqref{eq:NStatesUperLimitRough} overcounts the number of states since it contains cases in which $(n_i,m_i)=(0,0)$ and cases in which $n_i$ and $m_i$ are not co-prime. On the other hand, it undercounts since e.g.\ $(S1)$ stabilizes cases where all $m_i$ are zero. The first overcounting can be corrected by subtracting
\begin{itemize}
\item $3 (2n+1)^4$ to take into account cases where $(n_i, m_i)$ vanish for one torus,
\item $3 (2n+1)^2$ to take into account cases where $(n_i, m_i)$ vanish for two tori,
\item $1$ to take into account cases where $(n_i, m_i)$ vanish for all three tori.
\end{itemize}
To account for the overcounting, we need to re-instate a factor of 2. Lastly, we are left with the cases in which $n_i$ and $m_i$ are not co-prime. These are  very hard to count, since it requires knowledge of the distribution of primes. However, for  small upper bounds $n$ and $m$,  this doesn't happen very often. Up to this overcounting, we find that the number of states is given by
\begin{align}
N_\text{states}^\text{symm} = \left[N_\text{max} \frac{(2n+1)^6}{128} +(2 n)^3 -\frac{1}{128} [3(2n+1)^4 + 3(2n+1)^2 + 1] \right]^{D_\text{max}}\,.
\end{align}
Even in the most conservative case where we take $N_\text{max}=3$ and $D_\text{max}=4$ (needed to accommodate $\SU3_C$ and $\U1_Y$, respectively), we find that the number of configurations grows very rapidly:
\begin{align}
\label{eq:NStatesSymm}
  \begin{tabular}{|c||ccccc|}
  \hline
   $n=m$													& 1	  	& 2			& 3			& 4			& 5	\\
   \hline
   \text{\phantom{/o}w symm}						& $2.8\times 10^{5}$ & $3.0\times 10^{10}$ & $7.2\times 10^{13}$ & $2.7\times 10^{16}$ & $3.2\times 10^{18}$ \\
   \text{w/o symm}		& $1.8\times 10^{11}$ & $1.1\times 10^{16}$ & $1.9\times 10^{19}$ & $5.6\times 10^{21}$ & $5.5\times 10^{23}$  \\
   \hline
  \end{tabular}
\end{align}
This minimum requirement would in practice exclude many models, such as constructions with more than one hidden sector gauge group and limits the rank of the hidden sector to \SU3. On the other hand, the more hidden sector gauge groups we have the more likely we will find exotically charged particles.

\subsubsection{The Douglas-Taylor Truncation}
\label{sec:DTTruncation}
In this section we perform a different type of truncation where our system is described in the language of A-branes, B-branes, and C-branes\footnote{This is not related to generalized $(p,q)$ 7-branes, which are also referred to as A,B,C-branes.} of Douglas-Taylor~\cite{Douglas:2006xy}. The advantage of this approach is that Douglas-Taylor took into account some necessary conditions for A-branes, B-branes, and C-branes to satisfy the tadpole and supersymmetry conditions, and therefore by using this language of A-B-C-branes we cut down on the number of inconsistent states that are considered.

To carry out the computation of the number of possible states in our truncation, we must define a number of quantities. Let $D_A$, $D_B$, and $D_C$ be the number of A-stacks, B-stacks, and C-stacks that are considered. Let $N_A$, $N_B$, and $N_C$ be the maximum number of branes in any A-stack, B-stack, or C-stack. Let $d_A$ and $d_B$ be the upper bound on the absolute value of any winding number for an A-stack or B-stack. The analogous quantity for C-stacks does not exist because primitivity requires the would-be $d_C=1$, so we do not use it.

\subsubsection*{A-branes}
We first compute an upper bound on the number of possible sets of A-branes~\cite{Douglas:2006xy}. A-branes have four non-vanishing tadpoles $\hat{X}^I$ and there are four possibilities for the signs of the \ns and \ms if one takes into account necessary constraints from tadpole cancellation~\eqref{eq:TadpoleCondition}  and supersymmetry~\eqref{eq:SUSYCondition}. The three \ns may have sign $+++$, in which case the possible signs for the \ms are $+--$, $-+-$, or $--+$. Alternatively, the \ns may have signs $++-$, in which case the \ms must have sign $++-$. So there are four possibilities for sets of signs.  The possible number of sets of A-stacks is less than or equal to
\begin{align}
N_\text{A-stacks}\leq\sum_{i=0}^{D_A}\binom{4N_Ad_A^6}{i},
\end{align}
which follows from the fact that the number of possible A-stacks is $4N_A d_A^6$.

\subsubsection*{B-branes}
We turn to B-branes, which have two non-vanishing tadpoles and two vanishing tadpoles. Direct calculation shows that there are six possible combinations such that there are precisely two vanishing tadpoles, and furthermore tadpole cancellation~\eqref{eq:TadpoleCondition} and supersymmetry~\eqref{eq:SUSYCondition} require that the two non-vanishing tadpoles are positive. These solutions are collected in Table~\ref{tab:B-Brane-Combinations}.
\begin{table}
\centering
\begin{tabular}{|c|c|c|c|}
\hline
$X^0$&$X^1$&$X^2$&$X^3$ \\ 
\hline
$n_1 n_2 n_3$&$-m_2 m_3 n_1$&$0$&$0$\\ 
$n_1 n_2 n_3$&$0$&$-m_1 m_3 n_2$&$0$\\ 
$n_1 n_2 n_3$&$0$&$0$&$-m_1 m_2 n_3$\\ 
$0$&$0$&$-m_1 m_3 n_2$&$-m_1 m_2 n_3$\\ 
$0$&$-m_2 m_3 n_1$&$0$&$-m_1 m_2 n_3$\\
$0$&$-m_2 m_3 n_1$&$-m_1 m_3 n_2$&$0$\\
\hline
\end{tabular}
\caption{Possible winding number combinations for B-branes.}
\label{tab:B-Brane-Combinations}
\end{table}

Next, we have to address the question of how many possible sign choices there are for winding numbers in Table~\ref{tab:B-Brane-Combinations} consistent with the positivity constraint. A brute force calculation verifies the following combinatorics, but we can argue directly using a few useful facts. One is that all of the six solutions have precisely one winding number that appears in both tadpoles, and four that appear in one or the other. So there are five signs to choose. Furthermore, three of the six solutions have one tadpole with a minus sign, and three have minus signs on both tadpoles.

Consider any of the three solutions with only one minus sign. Regardless of whether the repeated quantity is plus or minus, the rest of the variables in one tadpole will have to give a minus, while the rest in the other will have to give a plus, $2$ choices each for a factor of $4$. Then there is the choice associated with the sign of the repeated quantity, for another factor of $2$, bringing us to $8$. This argumentation holds for three of the solutions, bringing us to $24$. They are unique because the different solutions have different entries set to zero. Consider any of the three solutions with two minus signs. Suppose the repeated entry is plus. Then the remaining two variables in each tadpole have to give an overall minus to each tadpole to make the overall tadpole positive. These are $2\times 2$ possibilities since the remaining sets of two variables can each be $+-$ or $-+$. Multiplying by $3$ for the $3$ solutions brings it to $12$. Now suppose the repeated entry is minus. Then the remainder of the variables in the tadpole have to give an overall plus. This gives another $2\times 2 \times 3 = 12$.

All in all, we see that the six solutions allow for a total of $48$ different sign possibilities for the winding numbers. We therefore have that the number of sets of B-stacks is bounded by
\begin{align}
N_\text{B-stacks}\leq\sum_{j=0}^{D_B}\binom{48 N_B d_B^5}{j},
\end{align}
where the number of possible $B$-stacks, $48 N_B d_B^5$ follows from the above combinatorics and the fact that one of the winding numbers must vanish, so it is $d_B^5$ rather than $d_B^6$ as in the case of A-stacks.

\subsubsection*{C-branes}
Now let us consider C-branes, which have one non-vanishing tadpole. This arises from three vanishing winding numbers, and the possibilities are $m_1=m_2=m_3=0$, $m_1=n_2=n_3=0$, $n_1=m_2=n_3=0$, and $n_1=n_2=m_3=0$. By the supersymmetry condition~\eqref{eq:SUSYCondition}, the non-vanishing tadpole must be positive, and in each case there are four choices of signs that render the tadpole positive, so there are four solutions and four sign choices. Thus the possible number of sets of C-stacks is bounded by
\begin{align}
N_\text{C-stacks}\leq\sum_{k=0}^{D_C}\binom{16N_C}{k},
\end{align}
which follows from the fact that there are $16N_C$ possible C-stacks.

In all, the upper bound on the number of orbifold configurations in the truncation is
\begin{align}
N_\text{states}^{DT}\leq\left[\sum_{i=0}^{D_A}\binom{4N_Ad_A^6}{i} \right] \times \left[ \sum_{j=0}^{D_B}\binom{48 N_B d_B^5}{j}\right] \times \left[\sum_{k=0}^{D_C}\binom{16N_C}{k}\right].
\end{align}
Note that $D_A\leq 4N_A d_A^6$, since higher $D_A$ would be adding zero in the sum for any $i>4N_A d_A^6$; similar statements hold for $D_B$ and $D_C$

\subsubsection*{Number of states}
We now study the upper bound as a function of the truncation parameters in order to determine which truncations may be feasible to study. Again, the (very restrictive) minimum requirement is $N_A=N_B=N_C=3$ and $D_A+D_B+D_C=4$, to allow for an $\SU3_C$ gauge group to arise from an A-stack, B-stack, or C-stack and for a massless\footnote{This is based on the argument that a $3\times K$ matrix will have a non-trivial kernel for $K>3$; for very special choices of winding numbers, $K=3$ could be sufficient.} $\U1_Y$, respectively. As in the pure symmetry reduction case~\eqref{eq:NStatesSymm}, even this most conservative upper bound grows quickly with growing $d_A$ and $d_B$. Since the truncation described here takes into account some necessary conditions for  supersymmetry and tadpole cancellation, the numbers are lower than those in the last line of~\eqref{eq:NStatesSymm}, which was an upper bound without any further constraints imposed. However, in this setup, symmetries are only partially accounted for, and hence the number are larger than the first line of~\eqref{eq:NStatesSymm}. In order to quote the numbers, we take all integer partitions of 4 of length three for $D_A+D_B+D_C$ and set $d_A=d_B$. Since the number of states grows with $4d_A^6$ and $48d_B^5$, the size is dictated by $d_A$ for $d_A>d_B$ and by $d_B$ for $d_B\geq d_A$ in the parameter range we consider. The number of states is then given by
\begin{align}
\label{eq:NStatesDT}
  \begin{tabular}{|c||ccccc|}
  \hline
   $d_A=d_B$								& 1								& 2		  						& 3								& 4									& 5\\
   \hline
   												&$7.4\times 10^7$		&$3.6\times 10^{13}$	&$1.5\times 10^{17}$		& $6.2\times 10^{19}$		& $6.9\times 10^{21}$\\
   \hline
  \end{tabular}
\end{align}

The results from~\eqref{eq:NStatesSymm} and~\eqref{eq:NStatesDT} clearly illustrate how large the configuration space is. Even if we allow  winding numbers between $-1$ and $+1$, a complete scan will take considerable time, and a systematic search for winding numbers larger than two is completely unfeasible. This necessitates using other techniques to traverse the string landscape configuration space even for this single choice of compactification manifold. In the following, we will explain the different agents we set up for an analysis with Reinforcement Learning.

\subsection{Different views on the landscape: Environment implementation}
\subsubsection{The Stacking Environment}
\label{sec:StackingEnv}

As explained in the previous sections, we truncate the action and state space available to our agents. The first possibility for traversing the truncated landscape of the $\bZ_2\times\bZ_2$ toroidal orientifold compactifications of Type IIA string theory is based on the Douglas-Taylor truncation outlined in Section~\ref{sec:DTTruncation}. The idea of the stacking environment is to first set an upper bound $D_\text{max}$ of brane stacks we allow to be used. Each of these $D_\text{max}$ stacks can be taken as an A-,B-, or C-brane stack. In addition, we allow the agent to change the number of branes $N_a$ in each stack up to $N_\text{max}$. If the agent sets $N_a$ of any stack to zero, this brane stack is completely removed and an entirely new stack can be added. We thus have the following actions:
\begin{definition}
An \emph{add-brane-action} produces $\cD'$ by selecting a single stack $d_a\in \cD$ and incrementing the number of branes in this stack, $N_a\to N_a+1$.
\end{definition}
\begin{definition}
A \emph{remove-brane-action} produces $\cD'$ by selecting a single stack $d\in \cD$ and reducing the number of branes in this stack, $N_a\to N_a-1$. If $N_a$ reaches zero, the entire stack $d_a$ is removed from $\cD$.
\end{definition}
\begin{definition}
A \emph{new-action} produces a new set of stacks $\cD'$ by adding a new stack $d_a$ to $\cD$ with initially one brane, $N_a=1$. Further branes can be added to this new stack by subsequent add-brane-actions.
\end{definition}

Note that, depending on the state of the environment, some of the actions can be \emph{illegal} actions. Illegal actions are:
\begin{itemize}
  \item Adding a brane to a stack that has already $N_\text{max}$ branes
  \item Creating a new brane stack if there are already $D_\text{max}$ brane stacks
  \item Creating a new brane stack $d_a$ whose winding numbers coincide with those of another stack $d_b\in\cD$ that is already in the model
\end{itemize}
If the agent tries to perform an illegal action, the action is disregarded and the agent is punished as detailed in Section~\ref{sec:RewardFunctions}.

If we denote the number of A,B,C branes by $\mu_A,\mu_B,\mu_C$, the cardinality $N_\text{action}^\text{stack}$ of the action space of the stacking environment is
\begin{align}
\label{eq:NActionStacking}
N_\text{action}^\text{stacking}=D_\text{max}+D_\text{max}+(\mu_A+\mu_B+\mu_C)\,, 
\end{align}
counting the number of add-brane-actions, remove-brane-action, and new-actions, respectively.

\subsubsection{The Flipping Environment}
The flipping environment uses a different strategy to describe the configuration space of D6 brane stacks on the orientifold background. Just like the stacking environment, agents in this environment can increase or decrease the number of branes in any given stack. However, instead of adding/removing entire stacks, the agent in the flipping environment can ``flip'', i.e.\ increment or decrement any of the winding numbers in any of the stacks by one unit. Thus, for this environment, we do not use the distinction of brane types A, B, and C. Instead, we produce any brane stack by increasing/decreasing winding numbers. In order to truncate the state space of this environment, we employ the truncation discussed in Section~\ref{sec:SymmetryTruncationb}. 

The environment has the following four types of actions:
\begin{definition}
An \emph{add-brane-action} produces $\cD'$ by selecting a single stack $d_a\in \cD$ and incrementing the number of branes in this stack, $N_a\to N_a+1$.
\end{definition}
\begin{definition}
A \emph{remove-brane-action} produces $\cD'$ by selecting a single stack $d\in \cD$ and reducing the number of branes in this stack, $N_a\to N_a-1$. If $N_a$ reaches zero, the entire stack $d_a$ is removed from $\cD$.
\end{definition}
\begin{definition}
An \emph{increase-winding-action} produces $\cD'$ by selecting a single stack $d_a\in\cD$ and increasing a single winding number $n_i^a$ or $m_i^a$ by one unit. Depending on the tilting of the torus and the winding number, this increase might be half-integer or integer.
\end{definition}
\begin{definition}
A \emph{decrease-winding-action} produces $\cD'$ by selecting a single stack $d_a\in\cD$ and decreasing a single winding number $n_i^a$ or $m_i^a$ by one unit. Depending on the tilting of the torus and the winding number, this decrease might be half-integer or integer.
\end{definition}
In this case, we allow the agent to ``remove'' a brane stack by setting the number of branes in the stack to zero. Depending on the state the environment is in, there might be the following illegal moves:
\begin{itemize}
  \item Adding/removing a brane from a full/empty brane stack
  \item Flipping a winding number from a state that has zero branes
  \item Increasing/decreasing a winding number beyond its maximum/minimum $n$ or $m$
  \item Changing a winding number of a stack $d_a\in\cD$ such that all winding numbers of the stack $d_a$ match those of another stack $d_b\in\cD$
  \item Changing a winding number such that the resulting winding numbers are not co-prime
\end{itemize}
In the first four cases, we discard the illegal move and punish the agent. The last case is somewhat different. In order to reach some winding configurations, the agent might have to go through a state in which the co-prime condition is violated. Hence, if the agent chooses to perform a winding-action, we increase/decrease the selected winding number by one unit and check the co-prime condition. If this condition is violated, we keep increasing/decreasing the winding number until either the co-prime condition is satisfied or the move becomes illegal since the agent tries to change a winding number beyond the specified cutoff. Also note that, in contrast to the stacking environment, the agent in the flipping environment has to start from a valid brane configuration -- if all winding numbers were set to zero, the agent couldn't reach any valid winding configuration since it can only change one winding number at a time. This is why we start from a random but fixed set of winding configurations for each of the $D_\text{max}$ states, and populate each stack with a random but fixed number of branes $N_a$. 

The number of the actions $N_\text{action}^\text{flipping}$ of the flipping environment is simply
\begin{align}
\label{eq:NActionFlipping}
N_\text{action}^\text{flipping}=D_\text{max}+D_\text{max}+6D_\text{max}+6D_\text{max}\,,
\end{align}
counting the number of add-brane-actions, remove-brane-actions, increase-winding-actions, and decrease-winding-actions, respectively.

\subsubsection{The One-in-a-Billion Search Environments}
Our final environment uses yet another strategy to model the landscape. It is a restriction of the stacking and the flipping environment that ensures the presence of the non-Abelian part of the Standard Model gauge group. In more detail, we set $D_\text{max}$ to four and fix the numbers of branes per stack to $N_a=(3,2,1,1)$. These are the type of brane stacks also considered in~\cite{Gmeiner:2005vz}. The authors identify four possible realizations of the standard model particle content for these brane stacks. Essentially, there is a choice whether the non-Abelian part of the second brane stack realizes an \SU2 or \Sp1 gauge group, which are isomorphic to \SU2 on the level of their Lie algebras. Depending on this choice, the hypercharge generator will be different. Moreover, there are different possibilities to realize some of the particles, for example the right-handed quarks transforming as $(\boldsymbol{\overline{3}},\boldsymbol{1})$ can be realized as $(\bar\Box,1)$ or as $(${\tiny$\yng(1,1)$}$,1)$. For details see~\cite{Gmeiner:2005vz}.

Since the number of stacks as well as the number of branes per stack are fixed, an agent in this environment can just change the winding numbers in the stacks. The one-in-a-billion search agent that is based on the stacking agent will change all 6 winding numbers at once by inserting a brane of type A, B, or C, while the one  based on the flipping agent will just change a single winding number of a single stack at a time. In both cases the number $N_a$ of branes in the stack is kept fixed.
Let us discuss the version based on the stacking agent first. This has the following action:
\begin{definition}
A \emph{change-stack-action} produces $\cD'$ by selecting a single stack $d_a\in\cD$ and exchanging all six winding numbers by new ones from a list of possible A,B,C brane stacks while keeping the number $N_a$ of branes in the stack unchanged.
\end{definition}
\noindent The only illegal move in this environment is to use the same winding numbers in different stacks:
\begin{itemize}
\item Changing all winding numbers of a stack $d_a\in\cD$ such that they match those of another stack $d_b\in\cD$
\end{itemize}
For this version of the one-in-a-billion environment, the number of the actions $N_\text{action}^\text{1:B-stacking}$ is
\begin{align}
\label{eq:NAction1-billion-stacking}
N_\text{action}^\text{1:B-stacking}=4(\mu_A+\mu_B+\mu_C)\,, 
\end{align}
which counts the number of change-stack-actions as $D_\text{max}=4$.

The flipping version of the one-in-a-billion agent has the following actions:
\begin{definition}
An \emph{increase-winding-action} produces $\cD'$ by selecting a single stack $d_a\in\cD$ and increasing a single winding number $n_i^a$ or $m_i^a$ by one unit. Depending on the tilting of the torus and the winding number, this increase might be half-integer or integer.
\end{definition}
\begin{definition}
A \emph{decrease-winding-action} produces $\cD'$ by selecting a single stack $d_a\in\cD$ and decreasing a single winding number $n_i^a$ or $m_i^a$ by one unit. Depending on the tilting of the torus and the winding number, this decrease might be half-integer or integer.
\end{definition}
The illegal moves become:
\begin{itemize}
  \item Increasing/decreasing a winding number beyond its maximum/minimum $n$ or $m$
  \item Changing a winding number of a stack $d_a\in\cD$ such that all winding numbers of the stack $d_a$ match those of another stack $d_b\in\cD$
  \item Changing a winding number such that the resulting winding numbers are not co-prime
\end{itemize}
For this version of the one-in-a-billion environment, the number of the actions $N_\text{action}^\text{1:B-flipping}$ are
\begin{align}
\label{eq:NAction1-billion-flipping}
N_\text{action}^\text{1:B-flipping}=4\times 6\times2=48\,, 
\end{align}
since each of the $D_\text{max}=4$ stacks has 6 winding numbers that can be decreased or increased.

\subsubsection{Comparison of Environments}
The agents in all three environments navigate and ``perceive'' the string landscape differently. There are a number of points we would like to make along these lines:
\begin{itemize}
\item Two states that might be nearby (i.e.\ reachable with a single or very few actions) in one environment might be far away or even unreachable for another environment. Consequently, the way the consistency constraints, the gauge groups, and the spectrum can change with each step is also  different for different environments. For example, while one agent might have to strongly violate tadpole cancellation at an intermediate state in order to move from one consistent state to the next, another might just be able to move along a valley in which the tadpole constraint is kept intact or violated only slightly. Similarly, in one perspective, the majority of states that satisfy the consistency constraints (tadpole, K-Theory, SUSY) might be close to a physically viable state (gauge group, matter content) but not vice versa. That means, some type of states might cluster while others are evenly distributed throughout the landscape.\footnote{While it would be interesting to study whether such clustering occurs, this is beyond the scope of the current paper.}
\item The order or priority in which the agents check the various mathematical and physical constraints can be influenced by the reward function. Since the constraints for all agents are the same, we can use the same reward functions (up to a few differences related to the different illegal actions), which are discussed in Section~\ref{sec:RewardFunctions}. \
\item One perspective on the landscape might be more ``natural'' for an agent to learn than another. Deciding which environment will be best requires a deep understanding of the structure of the landscape, in particular of the way the system of coupled Diophantine equations (arising from our constraints) behaves. Lacking this knowledge, we simply try different approaches. 
\end{itemize}
If one perspective were considerably better than the others, this might tell us about the nature of the landscape (i.e.\ the structure of the underlying mathematical constraints), or which implementation is better suited for Reinforcement Learning. 

Concerning this point, it should be noted that the cardinality of state space is huge in all cases (cf.\ Section~\ref{sec:OrbifoldTruncations}), and the encoding of the data of a state is the same for each agent. Hence, the neural network that predicts the value of a state will get the same input for all environment implementations, and it will have to deal with huge numbers of different states in all cases. However, the cardinality of the action spaces varies considerably between the environments, cf.\eqref{eq:NActionStacking}, \eqref{eq:NActionFlipping}, \eqref{eq:NAction1-billion-stacking}, \eqref{eq:NAction1-billion-flipping}, with the flipping environments having much smaller action spaces. Consequently, training the neural network that predicts the next action is much faster, since the network is smaller.

\medskip

Let us concretely contrast the stacking and flipping environments. The stacking environment has already some necessary conditions built in. However, it needs to take many steps in order to just change the wrapping numbers: For a stack with $N_a$ branes, changing the wrapping numbers requires $N_a$ actions to remove the stack, one action to add a new stack with the new wrapping numbers, and another $N_a-1$ actions to add the branes back onto the stack. 

The agent in the flipping environment, in contrast, can change a single wrapping number with just a single action. However, if the agent wants to change all six wrapping numbers $w_i^a=(n_i^a,m_i^a)$ by a considerable amount to $w_i^{\prime\,a}=(n_i^{\prime\,a},m_i^{\prime\,a})$, it requires at least $\sum_i |w_i^a-w_i^{\prime\,a}|$ actions. If several states in between do not satisfy the co-prime condition, this number will be even higher. 

The way in which the agents in the one-in-a-billion environments can get from a set of winding numbers $w$ to another set $w'$ are the same as for the stacking and flipping agents they are based on. However, they can never reach states with a non-Abelian hidden sector.

\subsection{A3C Implementation via OpenAI Gym and ChainerRL}
For the study of the landscape we use asynchronous advantage actor-critic (A3C) reinforcement learning. The method is based on~\cite{Mnih:2016aa}. It was benchmarked against other RL algorithms such as Deep Q-networks (DQN). Already after 24h of training on a CPU, A3C was found to outperform DQN's that were trained for 8 days on a GPU. The benchmark was carried out using Atari games.

 Since our work is the first application of reinforcement learning to explore the string landscape, there is currently no information on how the performance transfers from their benchmark problems to string theory. It would certainly be interesting to try different RL methods and algorithm implementations and compare their performance against each other. This is, however, beyond the study initiated in this paper.

\begin{figure}
\centering
\includegraphics[width=\textwidth]{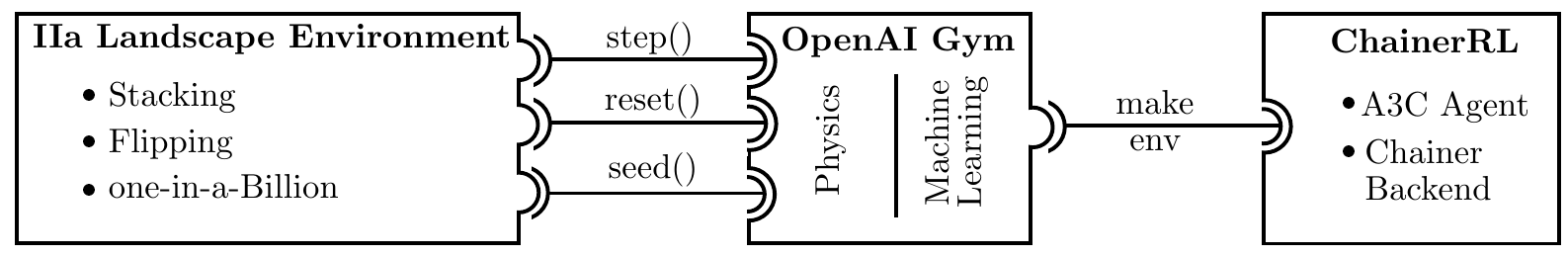}
\caption{Interfacing the physics environments with ChainerRL via OpenAI gym.}
\label{fig:A3CImplementation}
\end{figure}

For the implementation of the algorithm, we use the OpenAI environment~\cite{DBLP:journals/corr/BrockmanCPSSTZ16} in conjunction with the A3C implementation from the ChainerRL library~\cite{chainerLearningsys2015}. The environment class \texttt{Env} in \texttt{gym} is used as an interface between the environment implementation and the A3C agent as implemented in ChainerRL, cf.~Figure~\ref{fig:A3CImplementation}. Inheritance from the \texttt{gym.Env} class requires overriding the following methods\footnote{Since Python does not support interfaces or abstract classes, \texttt{gym.Env} implements these methods to raise a \texttt{NotImplementedError}.} (in order of importance for this project):
\begin{itemize}
\item \texttt{step}: The agent calls this method to traverse the string landscape. The agent calls step with a specific action and expects a new state, a reward, an indicator whether the episode is over, and a dictionary for additional information as its return. 
\item \texttt{reset}: This method is called at the start of each episode and resets the environment to its initial configuration. It returns the start state.
\item \texttt{seed}: This method is used for seeding the pseudo random number generators (RNGs). While the RNGs still produce pseudo-random numbers for all seeds, if an RNG is seeded with the same initial data, it will always produce the same sequence of random numbers. This serves the purpose of reproducibility of runs.
\item \texttt{close}: This allows for final cleanups when the environment is garbage collected or the program is closed; we do not need a special implementation here.
\item \texttt{render}: This allows to render the environment's state and output. We don't use this method to monitor the state of the agent and the environment. Instead, we include outputs directly in the ChainerRL implementation of the A3C agent and in the asynchronous training loop.
\end{itemize}

While the details of our systematic hyperparameter search are given in Section~\ref{sec:SystematicRL}, we discuss here some hyperparameters which we varied initially to find good values but then kept fixed across all experiments (most are default in the ChainerRL implementation). In our implementation, we use $\texttt{processes}=32$ A3C agents that explore the landscape in parallel for $24$ hours or until a combined number of $\texttt{steps}=10^8$ have been performed. Every $\texttt{eval-interval}=10^5$ steps we run the agent for $\texttt{eval-n-runs}=10$ episodes in evaluation mode to monitor its progress. In order to generate the plots in Section~\ref{sec:SystematicRL}, we monitor the states and their properties encountered by the agents while exploring. We use a learning rate of $\texttt{lr}=7\times10^{-4}$ and set $\texttt{weight-decay}=0$. As a cutoff for the sum of the return in~\eqref{eq:Return} we choose $\texttt{t-max}=5$. The policy evaluation network is trained to maximize the log probability (i.e.\ the logarithm of the output of the policy neural network) plus the entropy. We set the relative weight between these training goals to $\texttt{beta}=0.01$, which ensures sufficient exploration at the beginning (since mainly the entropy is maximized) and exploitation towards the end of training (since mainly the policy is optimized). To further ensure exploration, the next actions are not selected greedily but drawn from all action probabilities using the Gumbel distribution.

\section{Systematic Reinforcement Learning and Landscape Exploration}
\label{sec:SystematicRL}
In this section we describe the details of exploring the landscape of type IIA orbifold compactifications with RL. We will perform a series of experiments for the stacking agent, the flipping agent, and the two one-in-a-billion-agents that test the ability of each agent to learn how to satisfy string consistency conditions and find features of the Standard Model. For comparison, we will also implement an agent that picks actions at random, which is implemented by simply returning a zero reward, independent of the actual action taken by the agent. 

For our presentation here, we fix the background geometry to be $T^6/(\mathbbm{Z}_2\times\mathbbm{Z}_2\times\mathbbm{Z}_{2,O})$ with two untwisted and one twisted torus, $b=(0,0,1/2)$ and a fixed orientifold plane. The agent is exploring vacua in this background by changing the winding number of D6 brane stacks as well as the number of branes in the stacks.

We begin by describing the reward functions and associated value assignments we use, including the physical meaning of the rewards. We then describe the details of the experiments performed for the different agents and environments and compare results.

\subsection{Reward Functions}
\label{sec:RewardFunctions}
We will define reward functions according to two natural goals in this system: 1) finding fully consistent string models, by which we mean states that satisfy the tadpole cancellation \eqref{eq:TadpoleCondition}, K-theory \eqref{eq:KTheoryCondition}, and supersymmetry \eqref{eq:SUSYCondition} conditions,  and 2) finding models that are as close to the Standard Model as possible (SM-like). The reward functions are organized according to whether consistency is prioritized over being SM-like, vice versa, both are equally prioritized, or only one or the other is checked. For brevity we will refer to these as CONSISTENCY-SM, SM-CONSISTENCY, SIMULT, CONSISTENCY, and SM respectively. We will describe each in more detail below.

Before doing so, we first introduce various features that play a role in the reward function. There are two binary features, according to whether or not the supersymmetry conditions \eqref{eq:SUSYCondition} are satisfied, and also whether or not the K-theory conditions \eqref{eq:KTheoryCondition} are satisfied. Two other features are what we refer to as the tadpole distance (\tpd) and the SM distance. The tadpole distance is a measure of the distance from tadpole cancellation,
\begin{align}
\label{eq:TadpoleDistance}
\tpd := |8-P| + |4-Q| + |4-R| + |8-S|\,,
\end{align}
which is zero iff the tadpole cancellation conditions \eqref{eq:TadpoleCondition} for one twisted torus are satisfied. Conceptually, this is a measure of the amount of brane charge that must be added to the system. Note that a more refined measure, such as the separate distances for $P$, $Q$, $R$, and $S$ would also have been conceivable.
  
The Standard Model distance (\smd) is more involved. It is a measure of the distance from the Standard Model gauge symmetry and spectrum. Since SM gauge factors $\SU3_{C}$ ($\SU2_L$) arise from D6-brane stacks carrying $\U3$ ($\U2$ or $\Sp1$), respectively, we can count the number of missing gauge group factors if the model has no $\U3$ stack, or no $\U2$ or $\Sp1$ stack. Furthermore, the model has to have a massless $\U1$ symmetry that realizes the weak hypercharge $\U1_Y$. If there is no missing group factor, we can compute all possible ways of labeling the matter spectrum of the D6-brane theory with the three families of $\SU3\times \SU2$ representations of the SM. Each labeling has an associated number of exotic particles charged under $\SU3\times \SU2$. The possibility of realizing the standard model is non-unique for the following reasons:
\begin{itemize}
  \item There might be more than one $\SU3$ brane stack that can serve as the strong sector of the Standard Model.
  \item There might be more than one $\SU2$ or $\Sp1$ brane stack that can serve as the weak sector of the Standard Model.
  \item Standard model particles might be realized in different ways (e.g.\ the $\boldsymbol{\bar3}$ of $\SU3$ can be realized as a complex conjugated representation or a two-fold anti-symmetrized representation).
  \item If there are additional (vector-like) states, there is a choice which ones are considered as SM particles and which ones as exotics (this will influence the choice for the $\U1$ hypercharge generator).
\end{itemize}
We define the number of exotics to be the number of non-Standard Model particles for the best possible assignment of the Standard Model sector. Written as left-chiral fermions in representations of $\SU3 \times \SU2 \times \U1$, the minimal supersymmetric Standard Model spectrum is
\begin{align}
\label{eq:SMIrreps}
  3 \times [ (3,2)_{\frac16} +  (\overline 3, 1)_\frac13 + (\overline 3, 1)_{-\frac23} + (1,2)_{-\frac12} + (1,1)_1] + 1 \times [(1,2)_{-\frac12} + (1,2)_{\frac12}].
\end{align}
and if the state exhibits particles not listed here, we call them exotics.\footnote{If right-handed neutrinos $(1,1)_0$ are present, these are not counted as exotics.}

In order to denote the consistency and particle physics properties of the states encountered by the agents we use the following shorthand notations:
\begin{itemize}
\item TC: the D6-brane tadpole cancellation condition \eqref{eq:TadpoleCondition} is satisfied.
\item TCK: the tadpole cancellation conditions and also the K-theory constraints \eqref{eq:KTheoryCondition} are satisfied.
\item TCKS: the tadpole cancellation, K-theory constraints, as well as the SUSY conditions \eqref{eq:SUSYCondition} are satisfied. Note: states with this label are fully consistent supersymmetric string compactifications.
\item SM GG: the state contains stacks of branes that realize $\SU3\times \SU2$, as well as at least one massless $\U1$.
\item TCKS+SM: the state is a fully consistent string model with $\SU3\times \SU2$, a massless $U(1)$,
and at least one of the following: three families of left-handed quarks, three families of right-handed quarks,
or three families of leptons plus Higgs doublets. Other Standard
Model and / or exotic particles may or may not
be present; however, arbitrary combinations are not
possible, since the anomaly constraints are satisfied. 
\end{itemize}

Let us now explain the different reward functions in more detail. Each reward is initialized to zero and then changes according to the following ordered sets of SM and/or consistency checks.
\subsubsection*{CONSISTENCY}
CONSISTENCY comprises checking for a vanishing tadpole, as well as the K-theory and SUSY constraints. We  check the constraints in this order, since especially the SUSY constraint is very expensive to check and we want to avoid checking it if the configuration is already inconsistent for other reasons. The consistency reward $R^\text{C}$ is thus
\begin{align}
\label{eq:ConsistencyReward}
R^\text{C}=\left\{
\begin{array}{ll}
0&\text{if~}0<\tpd\leq8\\
-\tpd\times\tw{tadpoleDistanceMultipler}~~& \text{if~}\tpd>8\\
\tw{TC\_Reward}&\text{if~TC,~i.e.~}\tpd=0\\
\tw{TCK\_Reward}&\text{if~TCK}\\
\tw{TCKS\_Reward}&\text{if~TCKS}
\end{array}
\right.\,.
\end{align}
Note that for small enough tadpole mismatches, $0<\tpd\leq8$, the agent is neither rewarded nor punished. If the mismatch is too large, the punishment (note the minus sign in the reward) is proportional to the distance from a tadpole-cancelling state. If the tadpole is cancelled, a large reward is awarded. If in addition, the K-Theory constrained is satisfied, an even larger reward is given, and if on top of that the SUSY conditions are met a yet larger reward is returned. Note that we check these constraints sequentially, so a state that satisfies SUSY but not the tadpole might actually receive no reward or even a punishment.

\subsubsection*{SM}
SM comprises sequential checks for the particle physics properties of the model. First, it is checked whether the gauge group $\SU3\times\SU2\times\U1$ of the Standard Model is realized. If not, the punishment is proportional to the number $\Delta_\text{GG}$ of missing gauge groups. If all three gauge group factors are present, the agent is awarded a reward and the irreducible representations and the multiplicity of all massless particles are determined and compared against the Standard Model content~\eqref{eq:SMIrreps}. If several assignments of potential Standard Model matter is possible, the one that closest resembles the actual spectrum is chosen, and the punishment is proportional to the number $\Delta_\text{EX}$ of extra (or missing) particles in the spectrum (which we call exotics by a slight abuse of terminology) as compared to the Standard Model. If the model has the exact number of Standard Model Particles, the agent receives a big reward. The Standard Model reward $R^\text{SM}$ is thus
\begin{align*}
R^\text{SM}\!=\!\left\{
\begin{array}{ll}
-\Delta_\text{GG}\times\tw{missingGroupFactorDistance} &\text{if~}\Delta_\text{GG}\neq0\\
\tw{SMlike\_Reward}\;-\!\Delta_\text{EX}\times\tw{missingParticleDistance}~~& \text{if~}\Delta_\text{GG}=0,\Delta_\text{EX}\neq0\\
\tw{SM\_Reward}& \text{if~}\Delta_\text{GG}=\Delta_\text{EX}=0
\end{array}
\right.\,.
\end{align*}
Note the minus sign in the first and second line that lead to a punishment rather than a reward.

\subsubsection*{SIMULT}
In SIMULT, we check --- as the name suggests --- the consistency and particle physics properties of the model simultaneously and reward or punish the agent according to both aspects. This is computationally expensive since it requires finding all possible Standard Model particle realizations and checking the SUSY constraints. The reward is simply
\begin{align*}
R^\text{SIMULT}=R^\text{C}+R^\text{SM}\,.
\end{align*}

\subsubsection*{CONSISTENCY-SM}
CONSISTENCY-SM first performs all consistency checks and only proceeds to check SM properties if all consistency constraints are satisfied, i.e.\ if the model is TCKS. This is more efficient than SIMULT, since checking all possible Standard Model realizations is computationally rather expensive. The reward structure is
\begin{align*}
R^\text{C-SM}\!=\!\left\{
\begin{array}{ll}
R^\text{C} &\text{if~not TCKS}\\
R^\text{C}+R^\text{SM} \quad&\text{if TCKS}
\end{array}
\right.\,.
\end{align*}

\subsubsection*{SM-CONSISTENCY}
SM-CONSISTENCY is similar to CONSISTENCY-SM, but with the order of the sequential checks inverted. This means we first check all particle physics properties and only once these are satisfied we proceed to checking consistency of the model. This seems less intuitive for a physicist, since usually we want to ensure that the model is consistent in order to be able to trust our matter computations. If e.g.\ SUSY is broken, computation of the massless particle spectrum changes. Also, if the tadpole is not cancelled, the theory will have anomalies. Nevertheless, we try this order to see whether or not it is beneficial for the learning process of the agent. Also, this is much more efficient than checking consistency as for a model that e.g.\ does not even have the Standard Model gauge group, since especially the SUSY consistency check is computationally very expensive. The reward structure is
\begin{align*}
R^\text{SM-C}\!=\!\left\{
\begin{array}{ll}
R^\text{SM} &\text{if~}\smd\neq0\\
R^\text{SM}+R^\text{C} \quad&\text{if~}\smd=0
\end{array}
\right.\,.
\end{align*}

\subsubsection*{STC}
STC is another reward function
that checks for consistent
compactifications, specifically for a vanishing tadpole
and also the SUSY constraints. Both constraints are
in each case, unlike CONSISTENCY which only checked
the SUSY constraints if the tadpole conditions
are satisfied. This will be relevant in Section
\ref{sec:filler} when we study whether the agent can
learn a human-derived strategy using so-called filler branes.
The consistency reward $R^\text{STC}$
is initialized to $0$ and is incremented as
\begin{align}
\label{eq:STCReward}
R^\text{STC}+=\left\{
\begin{array}{ll}
0&\text{if~}0<\tpd\leq8\\
-\tpd\times\tw{tadpoleDistanceMultipler}~~& \text{if~}\tpd>8\\
\tw{TC\_Reward}&\text{if~TC,~i.e.~}\tpd=0\\
\tw{S\_Reward}&\text{if~S}\\
\tw{STC\_Reward}&\text{if~STC}
\end{array}
\right.\,.
\end{align}

The output module saves state information at regular time intervals in our experiments, according to information that is stored about states as they are encountered. For reasons of timing, the information that is saved depends on the reward function. For instance, checking the supersymmetry conditions involves solving a constrained quadratic programming problem (see Section \ref{sec:SUSYCheckCQP}), and therefore the SUSY condition should be checked judiciously. Consider as an example experiments that utilize SM-CONSISTENCY. There, the SUSY conditions are not checked unless the SM GG part of the reward function passes. Therefore states in such experiments could satisfy the SUSY condition but not be labelled as such, because the state has not passed the other sequential checks in the reward function.

\subsection{SUSY Conditions and Constrained Quadratic Programming}
\label{sec:SUSYCheckCQP}
While most physical and consistency constraints can be checked using linear algebra, the SUSY conditions~\eqref{eq:SUSYCondition} require solving a set of coupled equations and inequalities. 

In most cases, this can still be reduced to a simple linear algebra problem via an algorithm that we refer to as the BCD algorithm. The supersymmetry conditions \eqref{eq:SUSYCondition} include an equality
that may be rewritten as
\begin{align}
\frac{1}{U_0}\left(\hat Y^0_a + \hat Y^1_a \frac{U_0}{U_1}+ \hat Y^2_a \frac{U_0}{U_2}+ \hat Y^3_a \frac{U_0}{U_3} \right) = 0,
\end{align}
where the ratios of moduli $U_I$ are independent of the brane-stack index $a$. Defining the ratios of moduli
\begin{align}
j=\frac{U_0}{U_1}\,,\qquad k=\frac{U_0}{U_2}\,,\qquad l=\frac{U_0}{U_3}\,,
\end{align}
we may rewrite this as 
\begin{equation}
A_a + B_a \, j + C_a\,  k + D_a \, l=0, \qquad \forall a.
\end{equation}
This equation defines a hyperplane in moduli space, and as the number of brane stacks goes up, it is increasingly likely that there are no solutions to the system of hyperplane constraints. More specifically, in certain circumstances we may check the SUSY conditions quickly, as follows. First, find a triple of brane stacks such that the three hypersurface equations define a matrix equation with full rank. Next, invert that matrix to solve uniquely for the $(j,k,l)$ consistent with supersymmetry. Finally, check for those specific $(j,k,l)$ whether the rest of the supersymmetry conditions are satisfied. This BCD-algorithm can be applied whenever such a full-rank matrix exists. 

If the BCD algorithm cannot be applied we need to check existence of a common solution to the system of equations and inequalities~\eqref{eq:SUSYCondition} by other means. To do so, we need to find a solution subject to a positivity constraint on the variables of the problem. We phrase the problem as a constraint minimization problem that can be solved using the optimization implementations from the \texttt{scipy} library of Python. By benchmarking different approaches we find that \textit{Sequential Least Squares Programming} (SLSQP) works best for our purposes. In order to rewrite~\eqref{eq:SUSYCondition} as an SLSQP problem, we minimize
\begin{align}
\label{eq:SUSY_SLSQP}
f(t^I)=(t^0\hat{Y}^0+t^1\hat{Y}^1+t^2\hat{Y}^2+t^3\hat{Y}^3)^2\,,
\end{align}
where $t^I:=1/U^I$ with $U^I$ defined in~\eqref{eq:Definition UI}. Note that instead of solving an equation, we are minimizing a scalar function. This is equivalent to solving the equation since we minimize the square of a real equation. This is bounded from below by zero, which is a solution to the original problem.

But we are not done yet: we have to solve this minimization problem subject to two constraints. First, since the $U^I$ are a product of three radii, they have to be positive. Furthermore, in order to trust the supergravity approximation, they should not become too small and hence the $t^I$ cannot become too large. Hence we have upper and lower bounds on the $t^I$. Second, we need to solve the inequalities of the SUSY constraints, i.e.\ the second set of equations in~\eqref{eq:SUSYCondition}. We do this by specifying the constraints on the $t^I$ as bounds and the SUSY inequalities as constraints in SLSQP. We then minimize \eqref{eq:SUSY_SLSQP} numerically starting from a random initial guess and check that there exists a minimum sufficiently close to zero. Since the algorithm might fail to converge at all or might converge to a local minimum rather than to zero, we repeat this step 10 times. If no solution sufficiently close to zero is found it is assumed that no solution to the SUSY conditions exist. This might disregard perfectly fine models, but since the minimization procedure is significantly impacting the runtime, we find this to be a good compromise. During our benchmarks, we have analyzed thousands of configurations and compared them against methods that are much slower but guaranteed to converge to the global minimum, and we have never observed any discrepancy.

\subsection{Neural Network Architecture}

\begin{figure}
\centering
\subfloat[Policy evaluation network]{\label{fig:piNN}\includegraphics[width=0.45\textwidth]{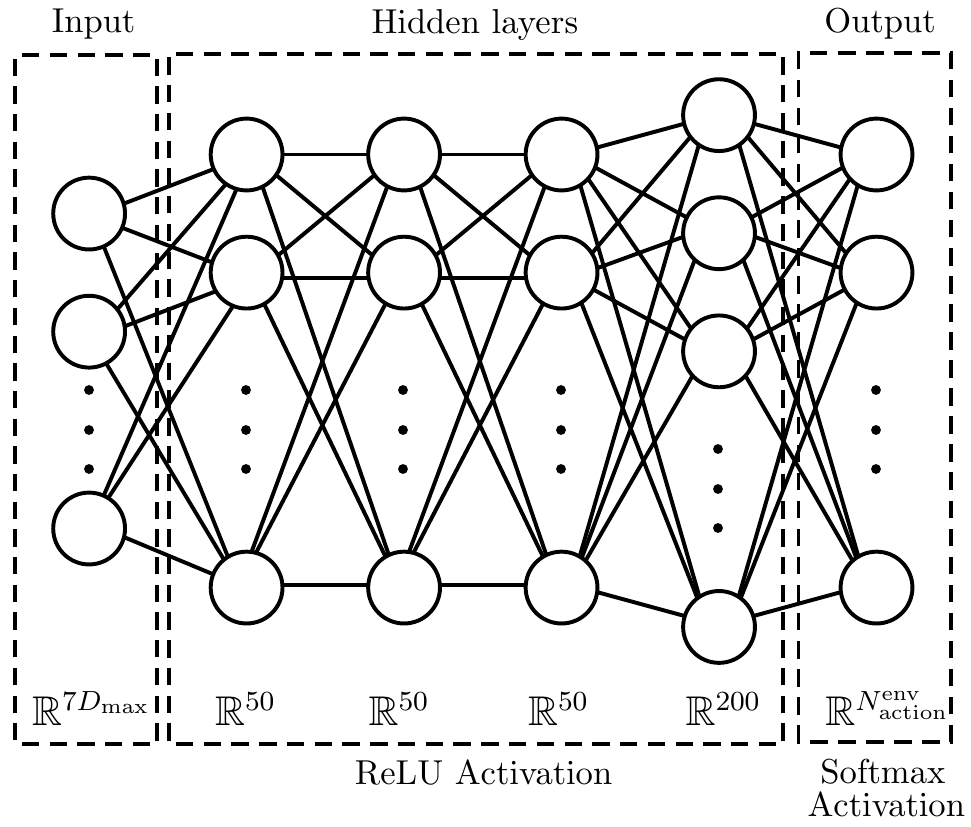}}\qquad
\subfloat[Value evaluation network]{\label{fig:valueNN}\includegraphics[width=0.45\textwidth]{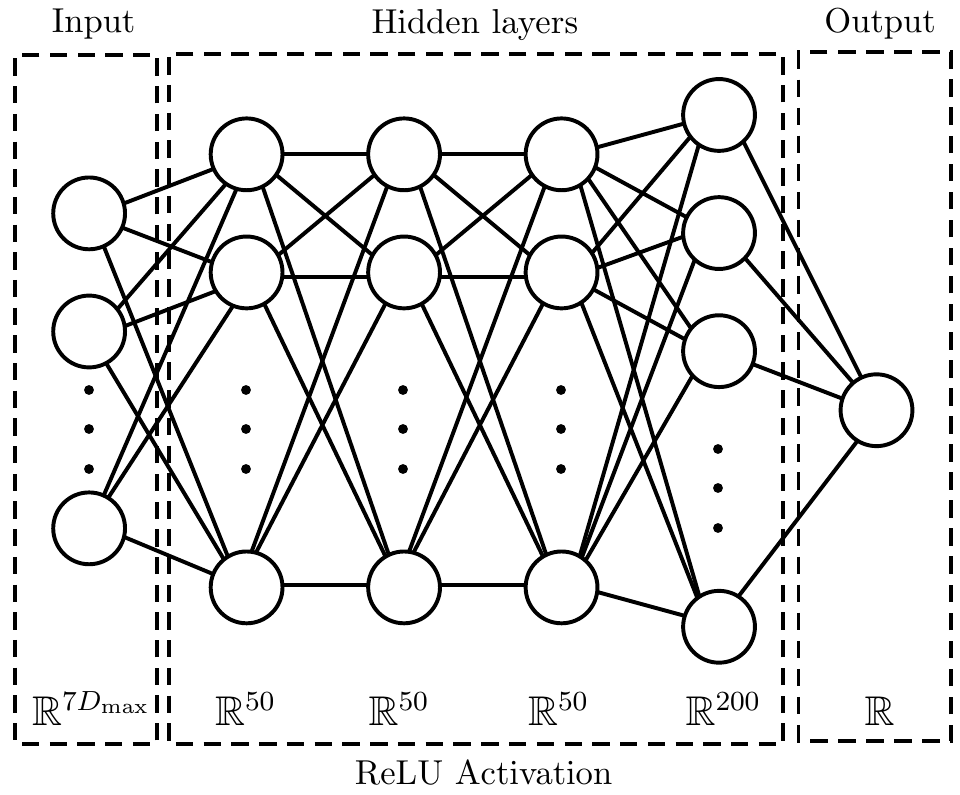}}
\caption{The neural network architectures used for policy and value evaluation.}
\label{fig:NNOverview}
\end{figure}

We tried different neural network architectures for the value and policy evaluation networks. We tried deep feed-forward neural networks with different numbers of hidden layers, different numbers of nodes per layer, and different activation functions. We furthermore tried recurrent neural networks with a Long Short Term Memory (LSTM) layer. We found that the performance is not too sensitive to the hyperparameters of the network. The networks we used for value and policy evaluation are given in Figure~\ref{fig:NNOverview}.

The policy network gets the current state the agent is in as its input, i.e.\ a vector\footnote{Remember that each brane stack is specified by the number of branes $N_a$ and six (half-)integer winding numbers.} in $\mathbbm{R}^{7 D_\text{max}}$. This is fed into four hidden layers, the first three have 50 nodes and a ReLU activation function, while the fourth layer has 200 nodes and a ReLU activation. The output layer is a vector in  $\mathbbm{R}^{N_\text{actions}^\text{env}}$, where the number of actions differs for different environments, cf.\ \eqref{eq:NActionStacking}-\eqref{eq:NAction1-billion-flipping}. The softmax layer is used to assign a probability to each action, see Figure~\ref{fig:piNN}.

The value network also receives the current state as a vector in $\mathbbm{R}^{7 D_\text{max}}$, followed by four hidden layers, again three with 50 nodes, one with 200 and all with ReLU activation. The output layer is just a single node, which corresponds to the value of the given state, see Figure~\ref{fig:valueNN}.

For the LSTM layer, we fed the entire state sequentially to the LSTM and selected the last output. The rationale behind this was to get a prediction once the network has seen the entire state. Since the physics is invariant under permutations of the stacks in a state, and the LSTM ``averages'' over the input, the hope was that the result would be permutation invariant and could thus help the network learn faster. While the results improved a bit, the training time increased as well considerably, such that we abandoned this approach.

\subsection{Learning to solve string consistency conditions \label{sec:consistency}}
Before we tackle the full-fledged analysis we benchmark the agents with regard to whether they can learn to solve string consistency conditions. To this end, we run 32 stacking agents with reward function CONSISTENCY~\eqref{eq:ConsistencyReward}, and end the game once a consistent model is found. In order to get better statistics, we run the agents for a background with three untwisted tori; these allow for more possibilities to satisfy the tadpole, but do not allow for an odd number of families.

Note that tadpole cancellation and the K-theory constraint lead to a coupled system of Diophantine equations in the winding numbers $(n_i^a,m_i^a)$. The SUSY constraints contain in addition qualities and inequalities for the three real parameters $(i,j,k)$, which are subject to positivity constraints. However, we are not interested in moduli stabilization at this point, which means that the SUSY conditions will leave flat directions. Hence, we can find rational solutions (or integer solutions, after clearing the denominator) for $(i,j,k)$as well. This way, we obtain a coupled, nonlinear system of Diophantine equations and inequalities in the winding numbers and K\"ahler parameters.

To analyze how well the agents perform in solving this system, we perform 10 test runs every $10^5$ steps and compute the average score as well as the entropy of states encountered in these runs (this data is recorded automatically by ChainerRL). We run the stacking agent with $(D_\text{max},D_A,D_B,\gamma)=(7,2,1,0.99)$ and 32 workers for roughly $3\times10^7$ steps. For the sake of illustration, we start with the single objective for the agent to solve the Diophantine equations associated with the tadpole cancellation condition~\eqref{eq:TadpoleCondition}. We use a reward of $\texttt{TC\_Reward}=10^6$ and end the episode once a tadpole cancelling model is found or after $10^4$ steps. The analog of~\eqref{eq:TadpoleCondition} for $\tpd$ on three untwisted tori is
\begin{align}
\label{eq:TadpoleDistanceUntwisted}
\tpd := |8-P| + |8-Q| + |8-R| + |8-S|\,.
\end{align}

\begin{figure}
\centering
\subfloat[Mean score.]{\label{fig:TCscore}\includegraphics[height=0.35\textwidth]{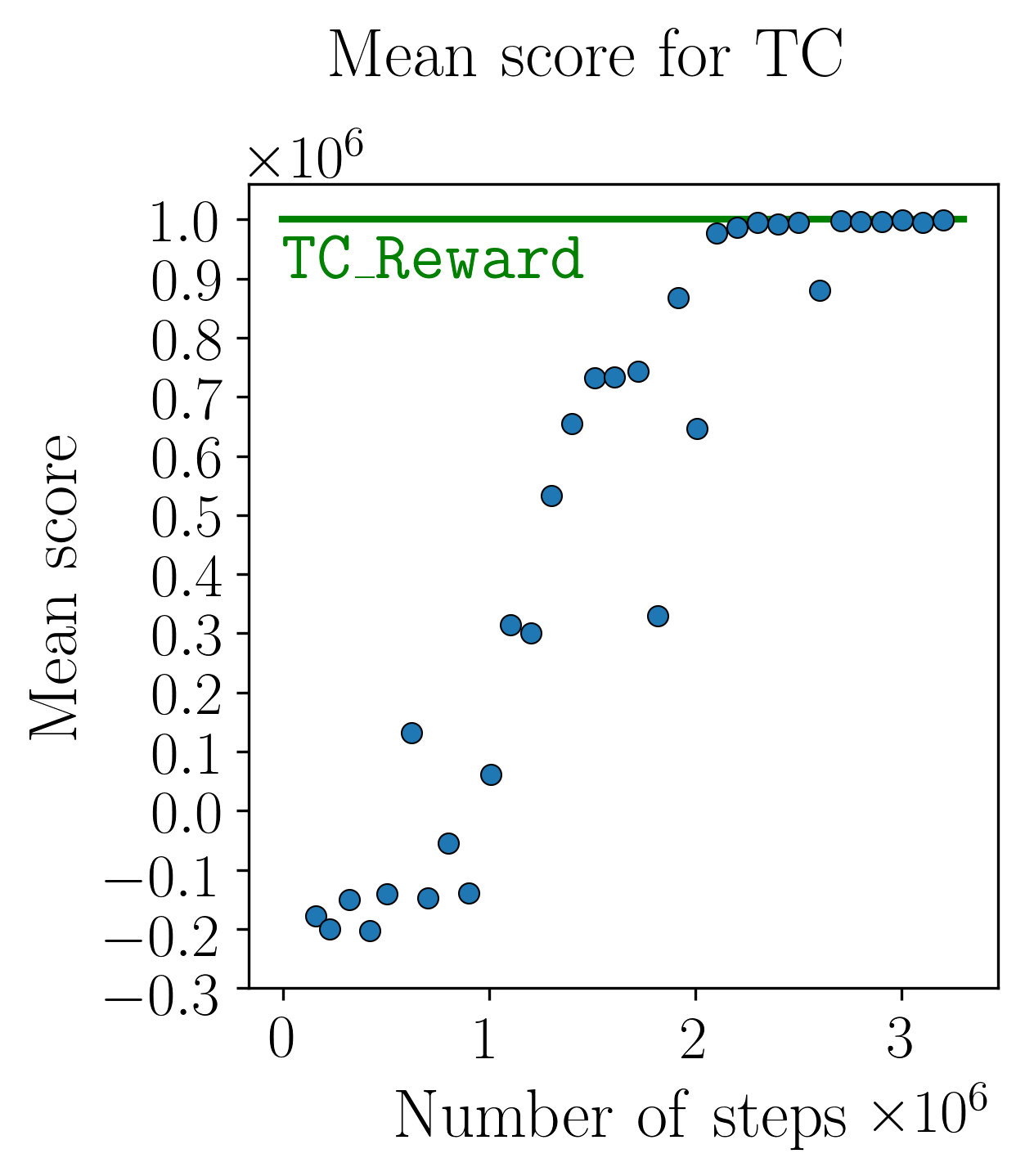}}~~
\subfloat[Average number of steps.]{\label{fig:TCsteps}\includegraphics[height=0.35\textwidth]{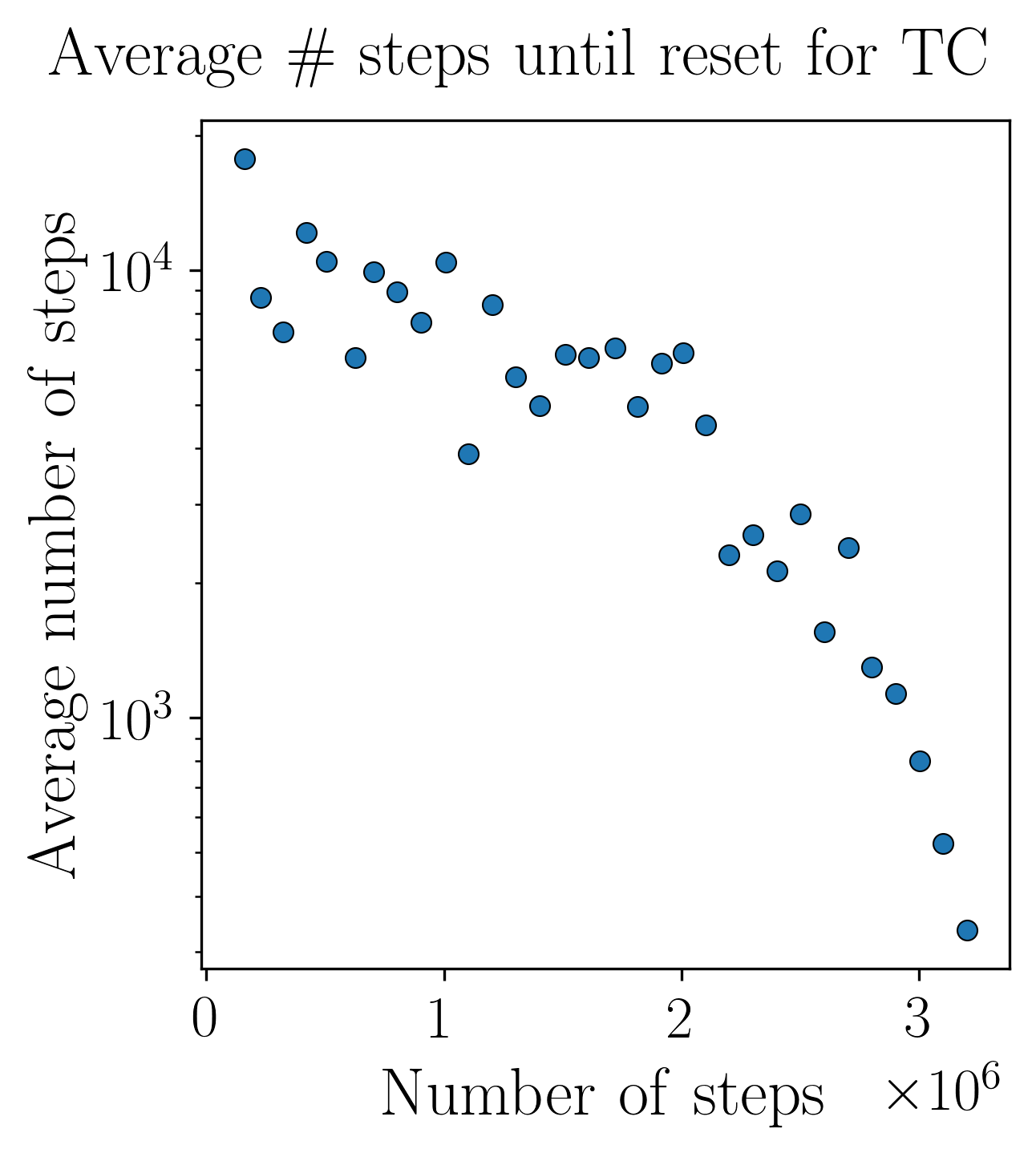}}~~
\subfloat[Entropy.]{\label{fig:TCentropy}\includegraphics[height=0.35\textwidth]{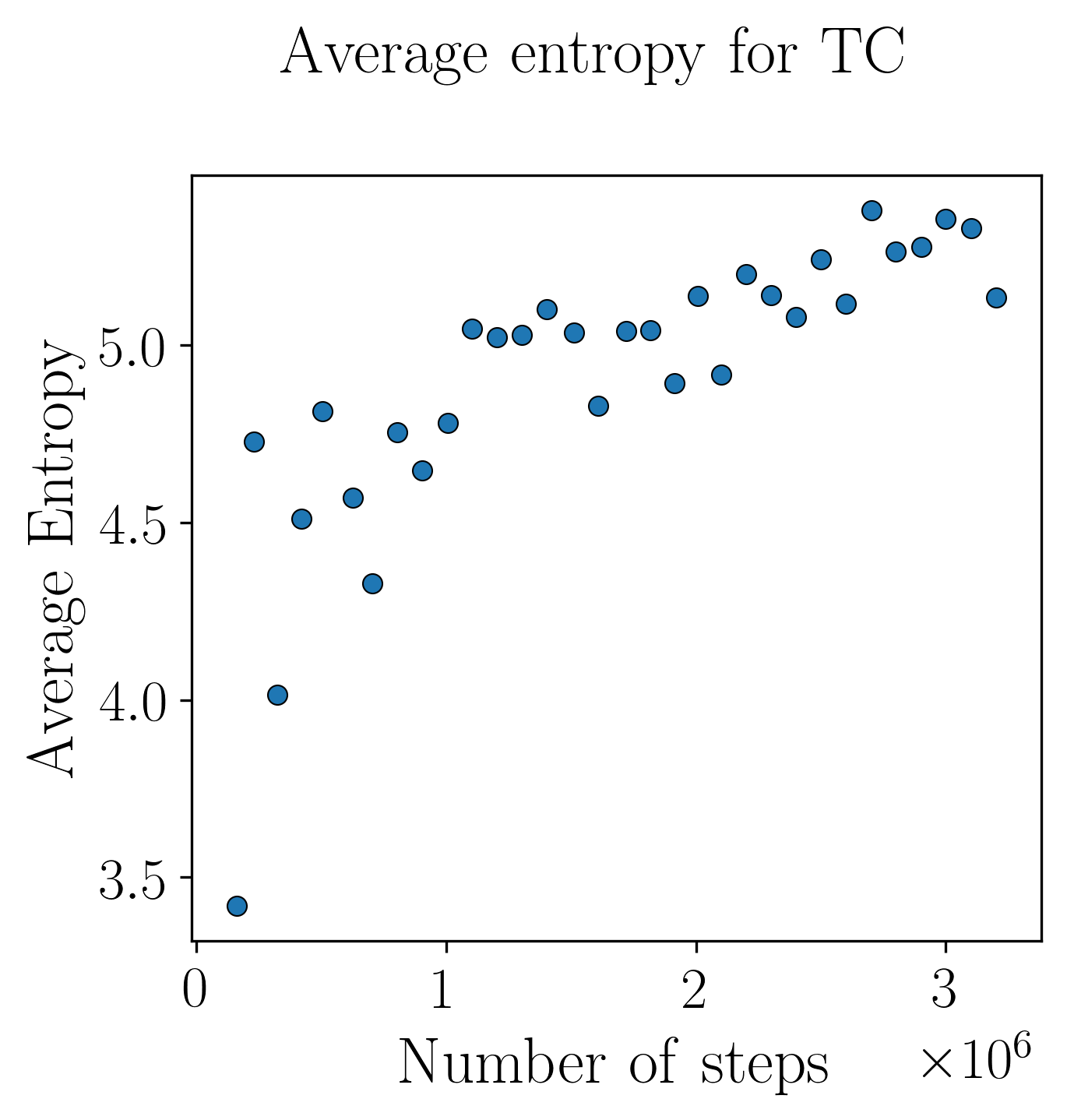}}
\caption{Plots illustrating how the agent learns to solve the IIA tadpole constraint.}
\label{fig:LearningTC}
\end{figure}

The results of the run are shown in Figure~\ref{fig:LearningTC}. We plot on the $x$-axis the total number of steps all agents have taken together. All data in the plots is recorded for the 10 evaluation runs, which occur every $10^5$ steps. Let us explain the plots in more detail. 

Figure~\ref{fig:TCscore} illustrates that the agent learns to solve TC around roughly $\mathcal{O}(10^6)$ steps.  Note that the punishment~\eqref{eq:TadpoleDistanceUntwisted} is an order $\mathcal{O}(10)$ number  (unless the distance is smaller than 8, in which case the agent is not punished at all) while the reward is $10^6$. Initially, the agent randomly performs actions leading to states that do not satisfy the tadpole. Each such action is punished by an order one number. After $10^4$ steps, the episode ends and the agent is reset. At that point the agent will have received a total negative reward of order $\mathcal{O}(10^5)$. This explains the average reward score in the first $10^6$ steps. After that,the agent starts to learn how to solve the tadpole constraint. If it solves the constraint in $k$ of the $10$ test runs, it will receive an average reward of $0.1(k\times 10^6-(10-k)\times 10^5)$, which explains why the points for the average score occur around $k\times10^5$ for $k\in[-1,10]$, depending on how often the agent manages to solve the constraint within the 10 test runs.

Figure~\ref{fig:TCsteps} illustrates that the agent is not only learning to solve the tadpole constraint, but that it is getting more and more efficient in doing so. We show on the $y$-axis the average number of steps per episode. An agent that never finds a TC state is reset after $10^4$ steps, so this is the upper bound in this plot. Note that we record only finished episodes, so the actual number can be somewhat larger if the agent is just about to finish an episode when the data is written to disk; in the worst case, this can lead to a factor of 2 in the averaged number of steps. In the beginning, the average number of steps is around $\mathcal{O}(10^4)$, indicating that most agents are reset without finding TC states. After about $10^6$ steps, as  the agent learns how to solve TC, the average number of steps per episode drops to $\text{few}\times10^3$.  At around $2\times10^6$ steps, the agent solve TC in all test runs (cf.\ Figure~\ref{fig:TCscore}). Nevertheless, the agent is still improving its efficiency, i.e.\ the number of steps it needs to take in order to find tadpole cancelling solutions. At the end, the average number of steps has dropped to $\mathcal{O}(100)$. Note that, due to the reward structure, the agent is not incentivized to solve the TC constraints in as little steps as possible (which is $\mathcal{O}(10)$) as long as it stays close enough (within a total distance of $\tpd\leq8$) to a tadpole cancelling solution.

We also want to know whether the agent is actually exploring the landscape and using its learned heuristics to solve the Diophantine equations or whether it is just randomly stumbling upon a solution and keeps reproducing that (exploration vs exploitation). As a measure for how diverse the solutions found by the agents are we look at the entropy of the agents in Figure~\ref{fig:TCentropy}. As we can see, the entropy is roughly constant (if anything, it is increasing over time), which indicates that the agent takes different actions and thus arrives at different states. We also confirm this by explicitly looking at the solutions the agents finds. Since we are using the stacking agent, which is based on the A,B,C brane construction, we know that the solutions are genuinely different and not related by a symmetry action to one another.

\begin{figure}
\centering
\includegraphics[height=0.45\textwidth]{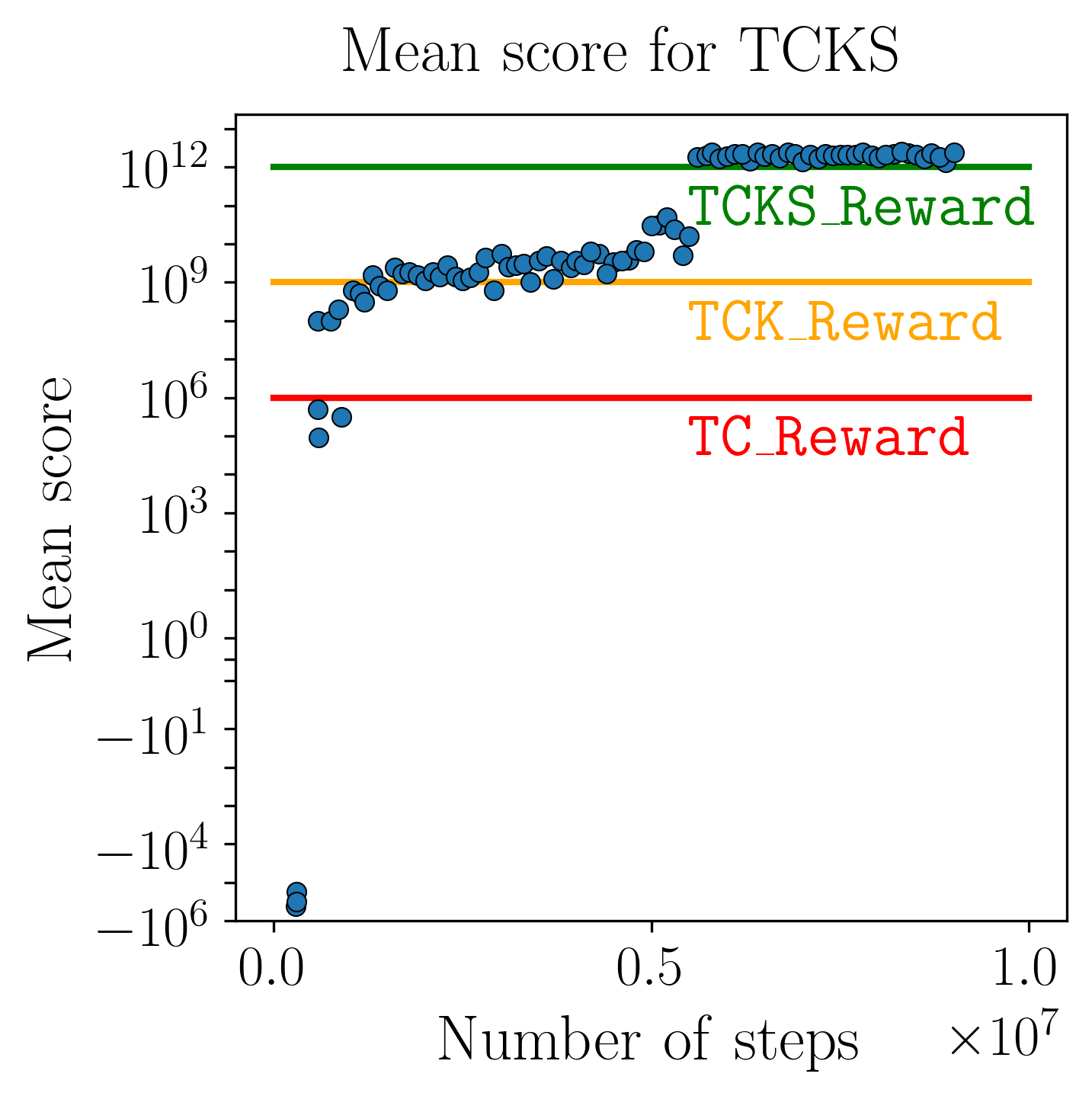}
\caption{Plots illustrating how the agent learns to solve multiple RL-tasks: First to solve the IIA tadpole constraint, then the K-theory constraint, and finally the SUSY constraints.}
\label{fig:LearningTCKS}
\end{figure}

Finally, we show the average score for a multi-tasking agent that successively learns to solves tadpole cancellation, K-Theory, and SUSY in Figure~\ref{fig:LearningTCKS}. In the beginning, the agent does not solve any of the consistency requirements and is receives a punishment proportional to the tadpole distance as in the TC case, thus ending up at $-10^5$. Again, after having taken around $10^6$ steps, the agent has learned how to solve TC, for which it receives the $\texttt{TC\_Reward}=10^6$ and is now also testing for the K-theory constraint. Once it receives feedback on its performance with regard to K, it learns to solve TC and K simultaneously between $10^6$ and $5\times10^6$ steps, which is rewarded with $\texttt{TCK\_Reward}=10^9$. Once TCK is solved, the SUSY constraints start to be checked. After $6\times10^6$, the agent learns to incorporate these as well, leading to fully consistent TCKS models and a reward of $\texttt{TCKS\_Reward}=10^{12}$.

We can also demonstrate learning of the different constraints by studying the relative frequency with which the agent finds models that satisfy the various constraints. We find that in the beginning for less than $3\times10^6$ steps, when the agent has not yet learned to produce models that satisfy the TC or K constraint, the ratio between models with TC and TCK is $1:5$. This is consistent with the statistics of ~\cite{Gmeiner:2005vz}, where the authors also find a reduction factor of 5 from imposing the K theory constraint on the untwisted torus (however, they impose K theory last, i.e.\ their models satisfy already the SUSY constraint). At the end of the run, the reduction factor has dropped to 3, indicating that the agent is doing better in finding models that satisfy the K-theory constraint as compared to randomly sampling the landscape. Of course, our numbers are too small for reliable statistics, but since we already reproduce the factor of 5, we are optimistic that our sampling size is sufficient. Likewise, we see a drop in the ratio of TCK to TCKS from initially around 5 down to 3 as soon as the agent learns to take SUSY into account.

\subsection{Learning a Human-Derived Strategy: Filler Branes \label{sec:filler}}

\begin{figure}[t]
\centering
\subfloat[Finding STC Models.]{\label{fig:STCgrowth}\includegraphics[height=0.45\textwidth]{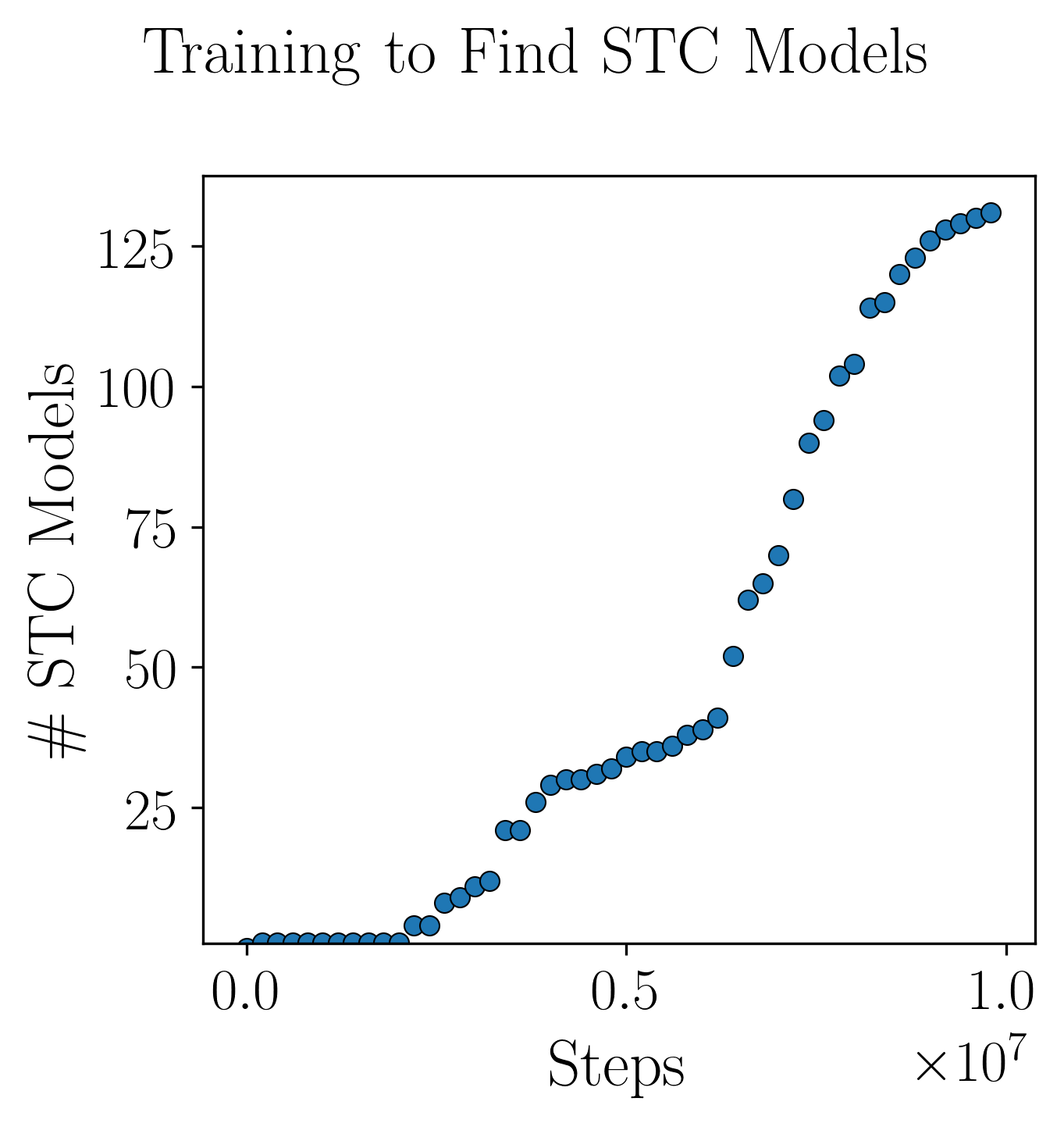}}~~
\subfloat[Using filler branes.]{\label{fig:BraneTypegrowth}\includegraphics[height=0.45\textwidth]{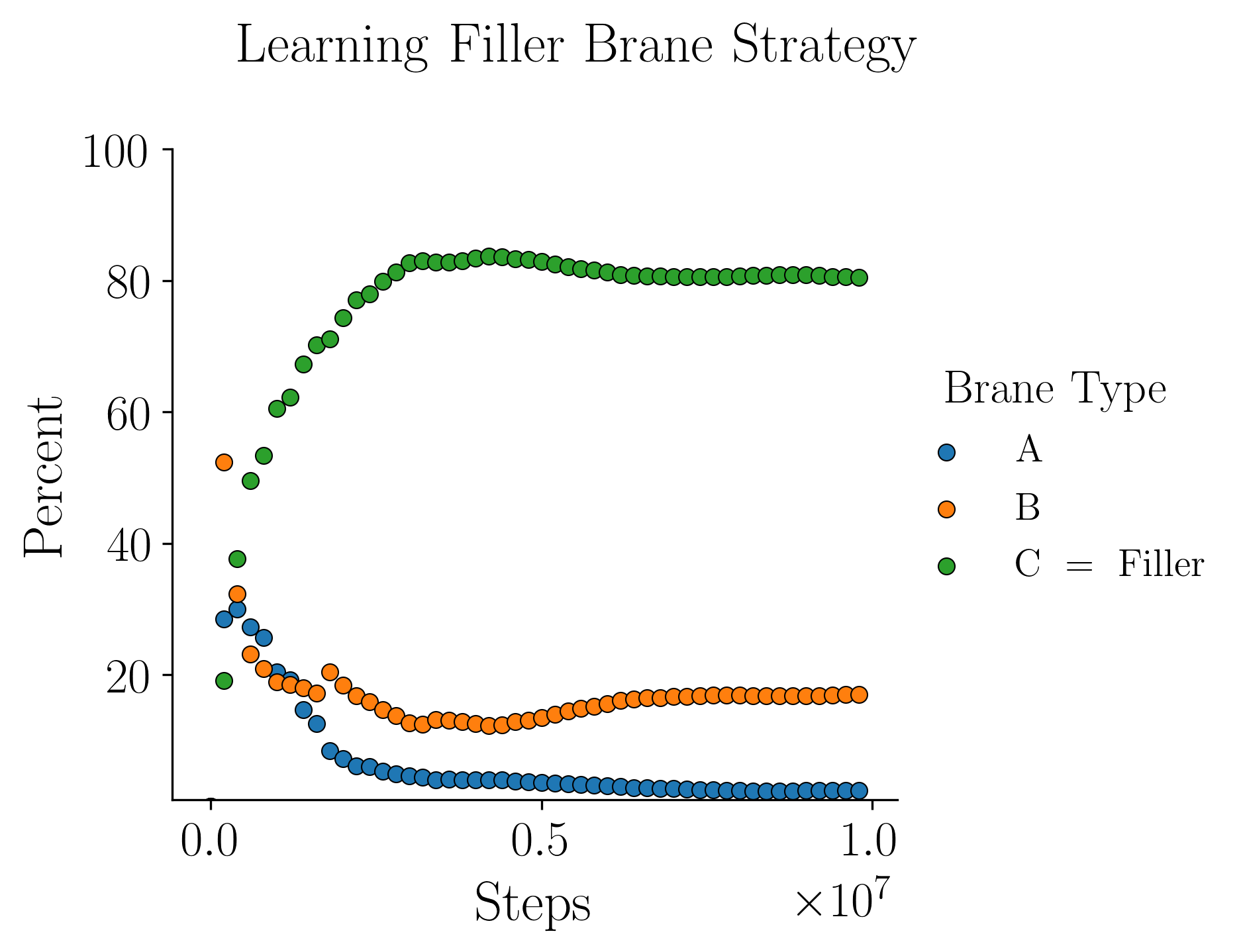}}~~
\caption{ RL learning a human-derived
strategy to solve SUSY and tadpole conditions.}
\label{fig:LearningFiller}
\end{figure}

The last section demonstrates that the RL agent learns a strategy to solve the coupled Diophantine equations in the TCKS setup. There is no human-derived strategy for doing this, and we are not attempting to find out the strategy employed by the agent, which in general falls
in the realm of intelligible AI, an area of active research. 

Instead, we look at a slightly modified setup in which humans have derived a strategy to partially decouple the system of equations. The strategy is to use so-called ``filler" branes (see, e.g., \cite{Cvetic:2004ui}). These are D6-branes that do not contribute to the supersymmetry conditions, but do contribute to the tadpole cancellation conditions. Therefore, one may add filler branes to supersymmetric D6-brane configurations in order to try to satisfy the tadpole cancellation conditions, but without spoiling the supersymmetry conditions. In the language of \cite{Douglas:2006xy}, it is C-branes that do not contribute to the SUSY conditions, and therefore should be identified as filler branes. The filler brane strategy cannot be utilized in the setup of Section \ref{sec:consistency} (which sought to solve the tadpole cancellation conditions first since it used CONSISTENCY), since the strategy is only useful in helping to solve the tadpole conditions when the SUSY conditions are already satisfied, not the other way around.

Our goal is therefore to utilize a different reward function in order to investigate whether an RL agent can learn the filler brane strategy as (part of) its solution approach, i.e.\ to use C-branes to solve the supersymmetry and tadpole cancellation conditions. We will utilize the STC reward function, which allows filler branes to potentially be a useful strategy because it always checks both the SUSY and tadpole conditions. More specifically, our experiment utilizes the STC reward function with $\tw{tadpoleDistanceMultipler}\,\,=10$, $\tw{TC\_Reward}\,\, =10^6$, $\tw{S\_Reward}\,\, =10^1$, and $\tw{STC\_Reward}\,\, =10^{14}$. We run an A3C on the untilted torus with $32$ workers for a maximum of $10^8$ steps or $24$ hours, whichever comes first, with $(d_A,d_B)=(2,1)$ and a maximum number of stacks $|\mathcal{D}|=10$. As training progresses, we keep track of configurations that are S, TC, or STC (corresponding to solving SUSY, tadpole, or both) and the percentage of A-branes, B-branes, and C-branes that are utilized in those solutions up to that point in the training. For this truncation with$(d_A,d_B)=(2,1)$ there are $108$ A-brane cycles, $48$ B-brane cycles, and $16$ C-brane cycles, respectively. A random walker that adds a new brane stack would therefore utilize these types $62.8\%$, $27.9\%$, and $9.3\%$ of the time, respectively.

The results of the experiment are presented in Figure \ref{fig:LearningFiller}. We first notice that the agent only takes about $10^7$ steps and its run is stopped due to $24$ hours time constraint. As we shall see below, agents that utilize a different reward function reach $10^8$ steps before the 24 hours expire. 
This is due to the fact that for the STC reward function the SUSY condition is checked at every step, and it is computationally expensive. 
We see from Figure \ref{fig:STCgrowth} that the agent does not begin finding STC models until around $2.5$ million steps, and does so at a reasonable rate thereafter, finishing with over $125$ STC models after $10^7$ steps. On the other hand, from Figure \ref{fig:BraneTypegrowth} we see that the agent is using about $50\%$ A-branes, $30\%$ B-branes, and $20\%$ C-branes after $200,000$ steps; even by this time, it has already deviated from the percentages that a random walker would utilize. However, the percentage of C-branes utilized goes up dramatically for the next few million steps, until the agent has utilized over $80\%$ C-branes by the time it has taken $2.5$ million steps, which is about when it begins to find STC models. From that point, the agent finds STC models consistently, continuing to utilize C-branes $80\%$ of the time. These percentages of A-branes, B-branes and C-branes utilized are across all models recorded by the agent, including S, TC, and STC. However, if one restricts to studying the percentages only for STC models, it is still the case that over $80\%$ of the branes are C-branes.

In obtaining this result, we note that it depends critically on the choice of $\tw{S\_Reward}\,\,\, $ relative to $\tw{TC\_Reward}\,\,\,$ and $\tw{STC\_Reward}\,\,\,$. Specifically, if instead of our parameters one instead changes $\tw{S\_Reward}\,\,\,$ to $10^6$, matching $\tw{TC\_Reward}\,\,\,$, then no STC models are found. This is simply because the agent can receive a large reward move after move by adding a C-brane to a SUSY solution, which maintains the SUSY property and adds the large $\tw{S\_Reward}\,\,\,$. This takes the system far away from tadpole cancellation, but the agent does not mind because maintaining SUSY is very rewarding. This explains the absence of STC models for this choice of parameters, and was what led us to instead choose $\tw{S\_Reward}\,\,\, = 10^1 $.  

Summarizing, the agent has clearly learned the human-derived strategy that utilizes filler branes. Despite its successes, our results below will not use this strategy because of the large amount of time it takes to check the SUSY condition.

\subsection{Systematic RL Stacking Agent vs.\ Random Agent Experiments}

In order to find promising hyperparameter settings we perform a box search over (part of the) hyperparameter space. We do this for the stacking agent and then apply the best set to the flipping and the one-in-a-billion agents. In total, we perform $108$ experiments for the stacking agent, each with $32$ workers. Each experiment requires making the following choices:
\begin{itemize}
\item Discount factor $\gamma \in \{.99,.9999,.999999\}$.
\item Brane bounds $(d_A,d_B) \in \{(2,1),(2,2),(3,1)\}$.
\item Reward type given by SIMULT, CONSISTENCY-SM, or SM-CONSISTENCY,
  or a multi-task A3C agent. In the multi-task agent, half of the workers
  use CONSISTENCY as their reward function, and half use SM.
\item One of three possible reward value assignments as detailed in Tables 
\ref{tab:valssimult}, \ref{tab:valsconsec1}, or \ref{tab:valsconsec2}
in the appendix.
\end{itemize}
Each experiment trains the agent for $10^8$ steps, or $24$ hours, whichever comes first.\footnote{The latter is due to computer cluster limitations.} Output is saved according to the discussion at the end of Section \ref{sec:RewardFunctions}, and we remind the reader that the checks that are performed, and hence the properties of the models specified in the output, depends on the choice of reward function.

In order to determine how well the stacking agent is performing on this problem,
 we must compare to a control study. 
For that reason, we perform $9$ experiments for the random agent. In each experiment, we choose:
\begin{itemize}
\item Brane bounds $(d_A,d_B) \in \{(2,1),(2,2),(3,1)\}$.
\item Reward type given by SIMULT, CONSISTENCY-SM, or SM-CONSISTENCY,
  or a multi-task A3C agent.
\end{itemize}
Note that the reward value returned by the reward function does not affect the random
agent in any way. Instead, we must choose a reward function so that the checks that are performed (and hence the output information) can be compared to the results of the RL agents in a meaningful way.

\begin{figure}[t]
\centering
\includegraphics[width=0.81\textwidth]{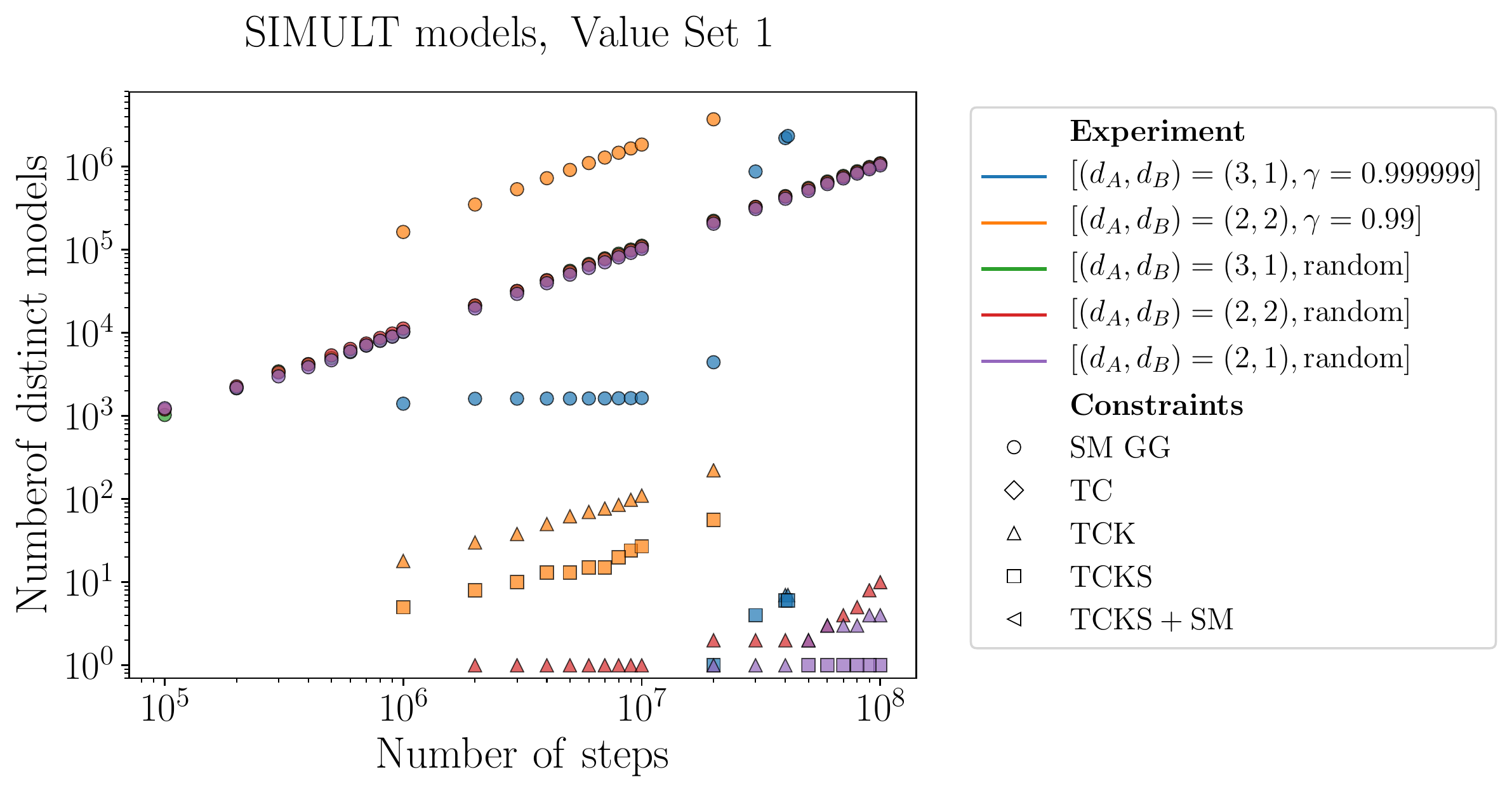} \\
\includegraphics[width=0.81\textwidth]{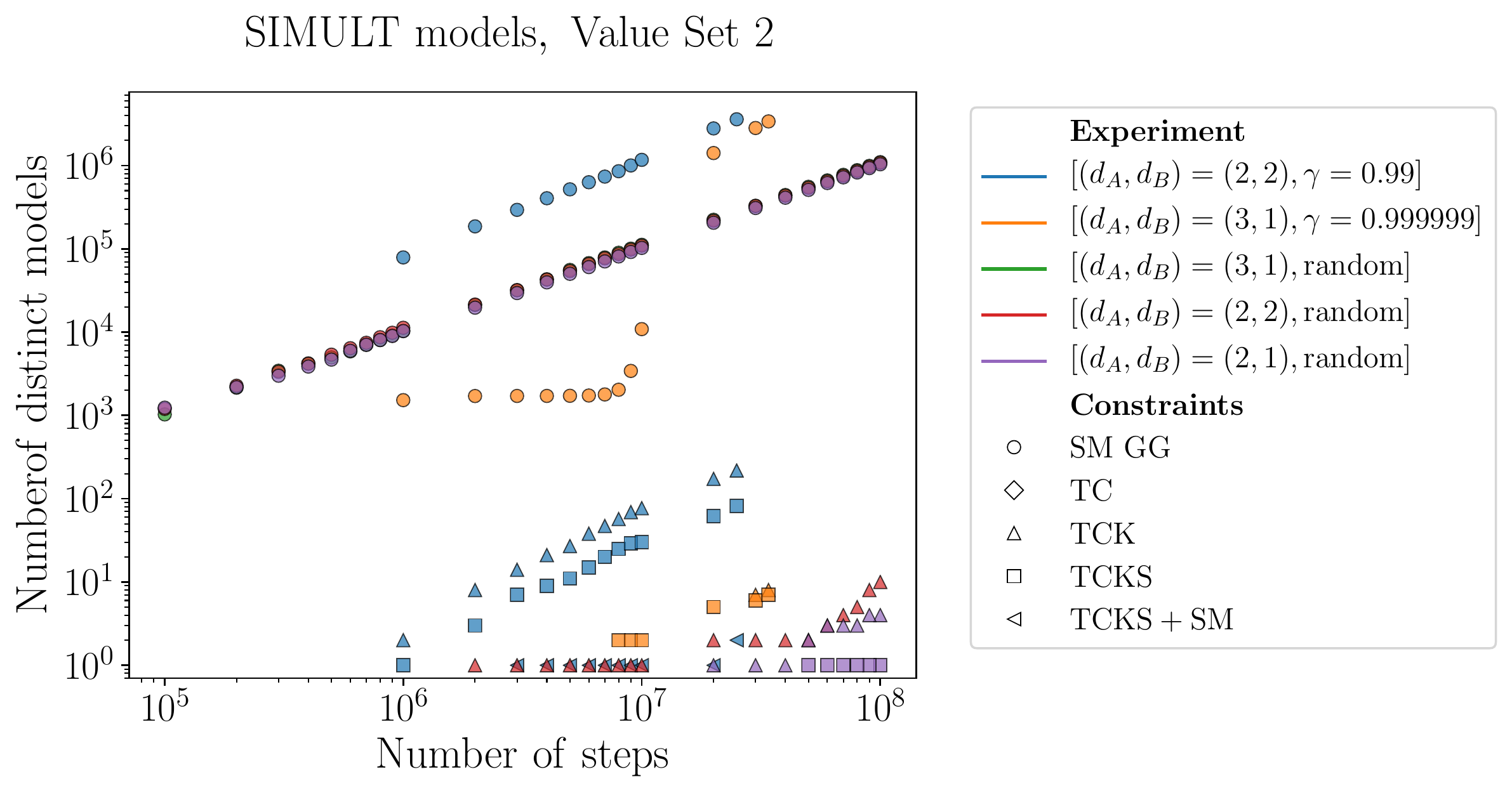} \\
\includegraphics[width=0.81\textwidth]{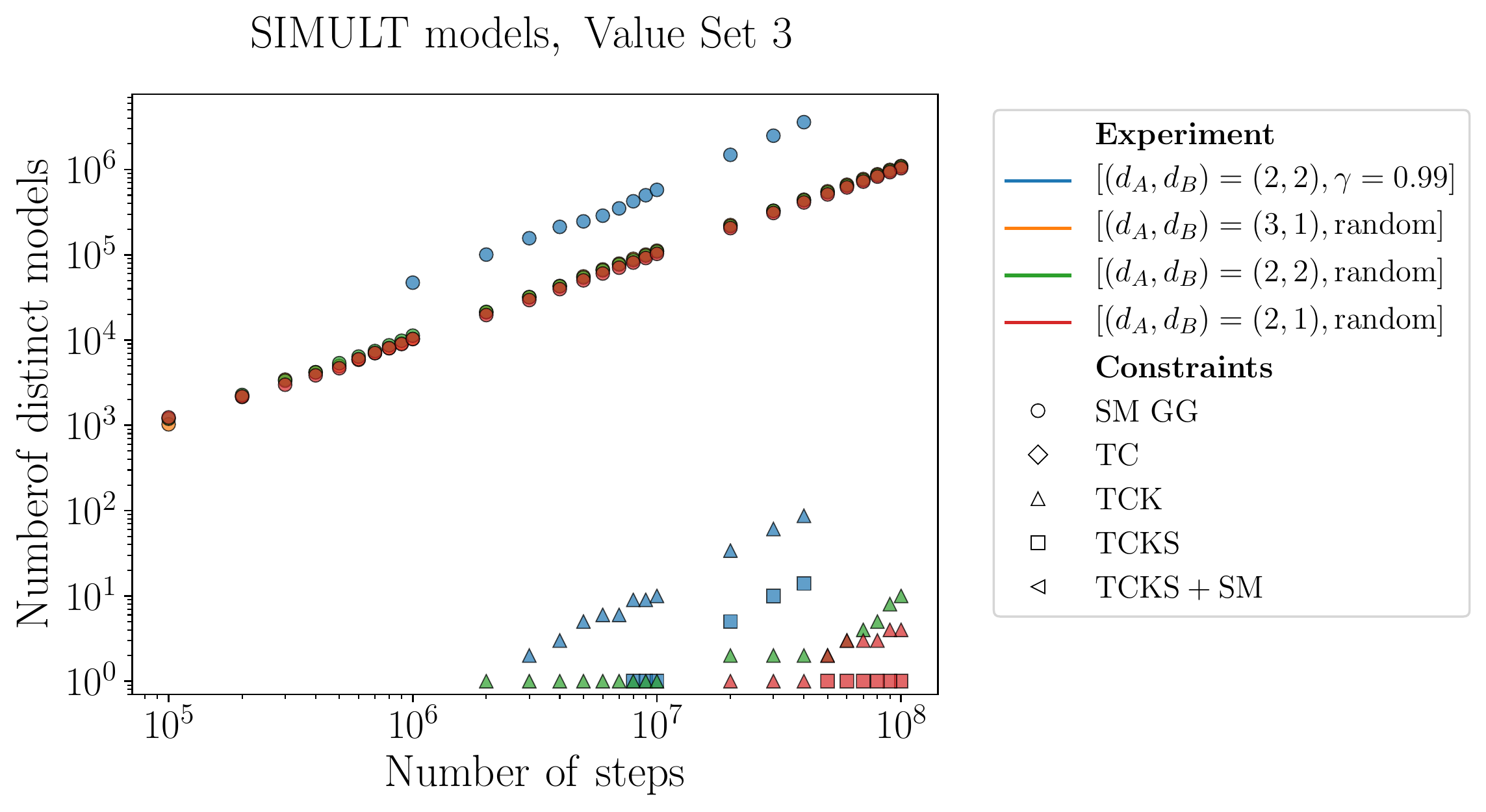}
\caption{Experimental results for reward function SIMULT. Results depend on choices of
value sets as in Table \ref{tab:valssimult}. Results for value sets $1$, $2$, and $3$ are
at top, middle, and bottom, respectively.}
\label{fig:stackvsrandomsimult}
\end{figure}

\begin{figure}[t]
\centering
\includegraphics[width=0.8\textwidth]{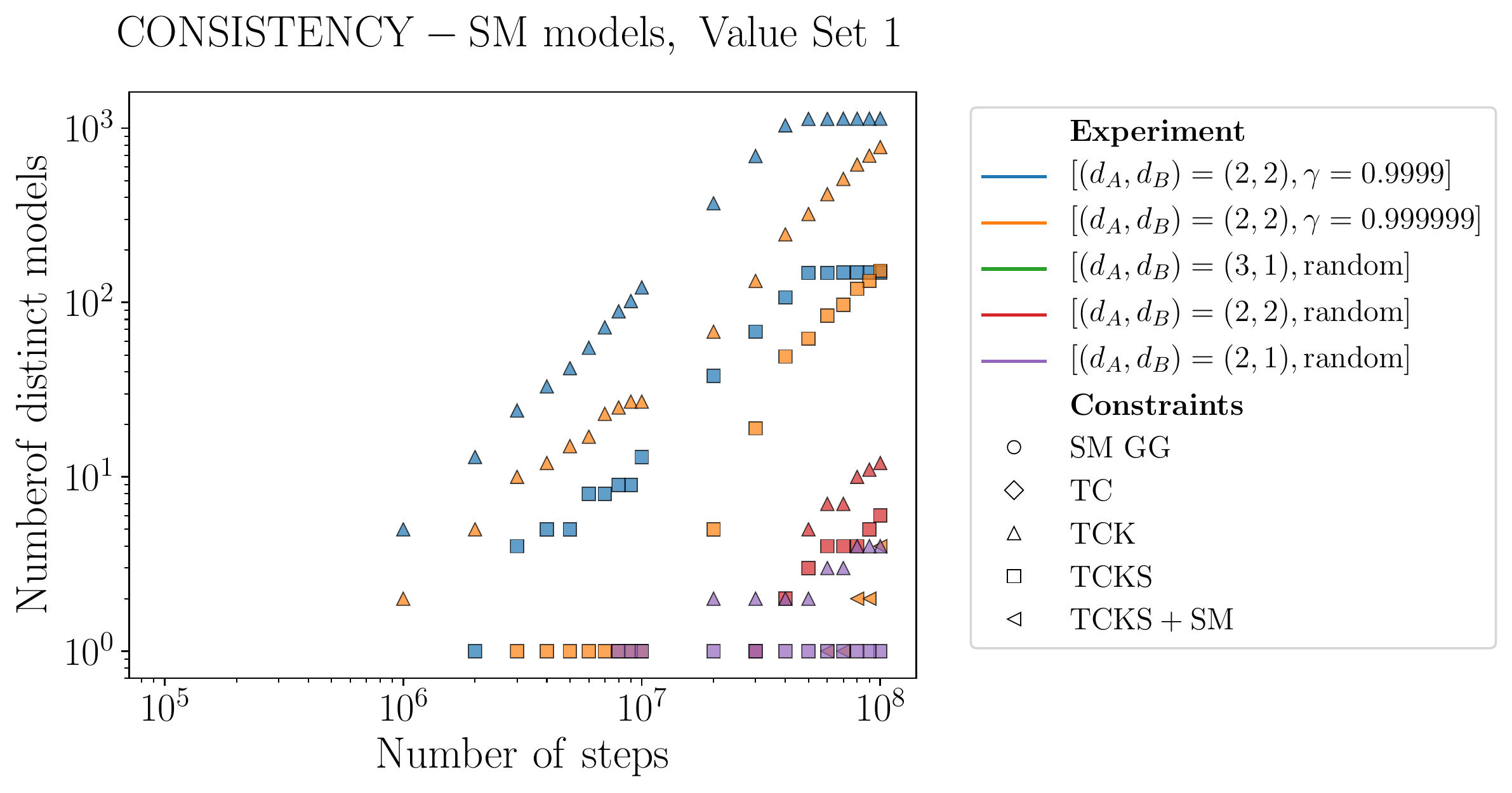} \\
\includegraphics[width=0.8\textwidth]{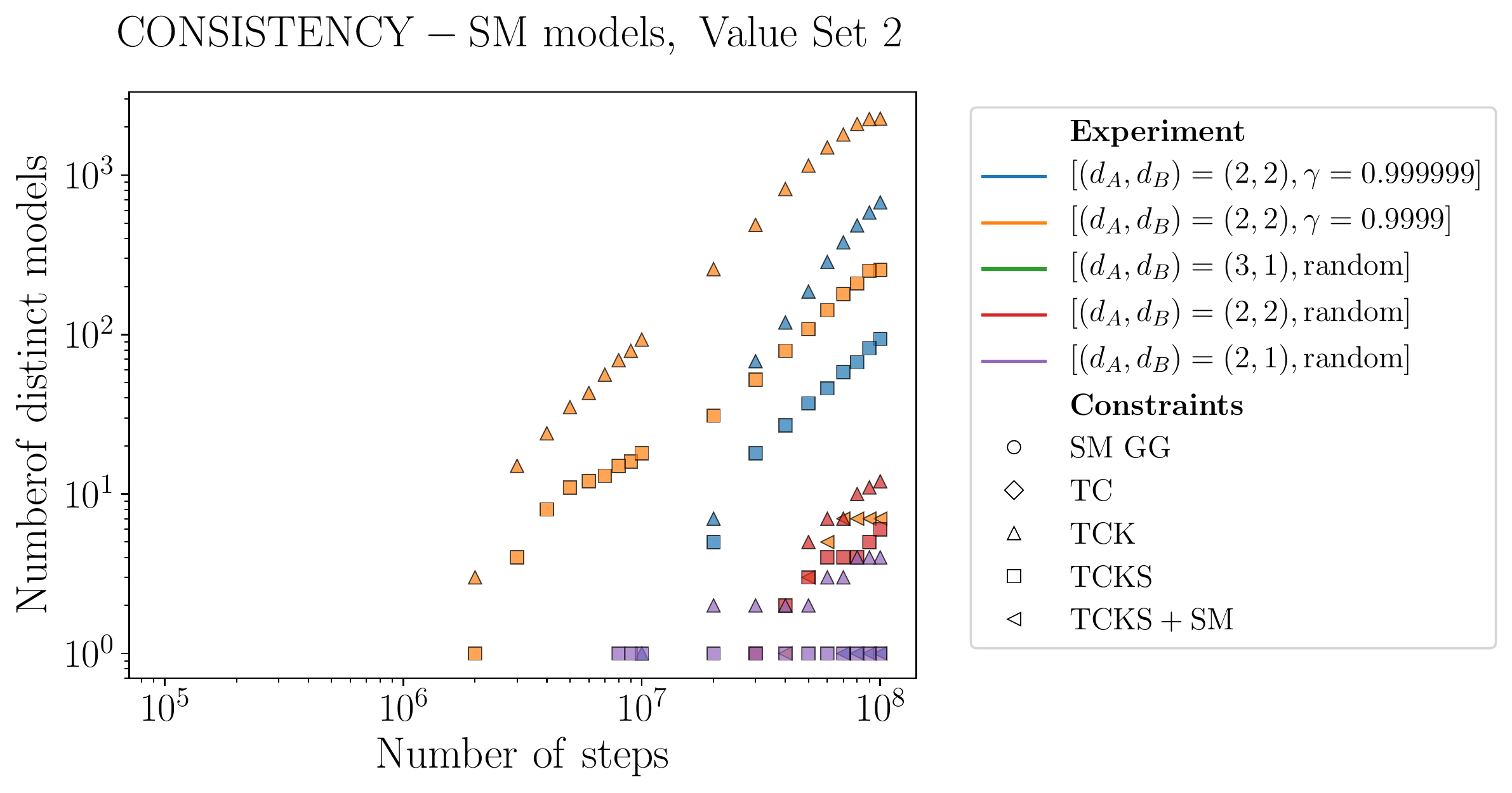} \\
\includegraphics[width=0.8\textwidth]{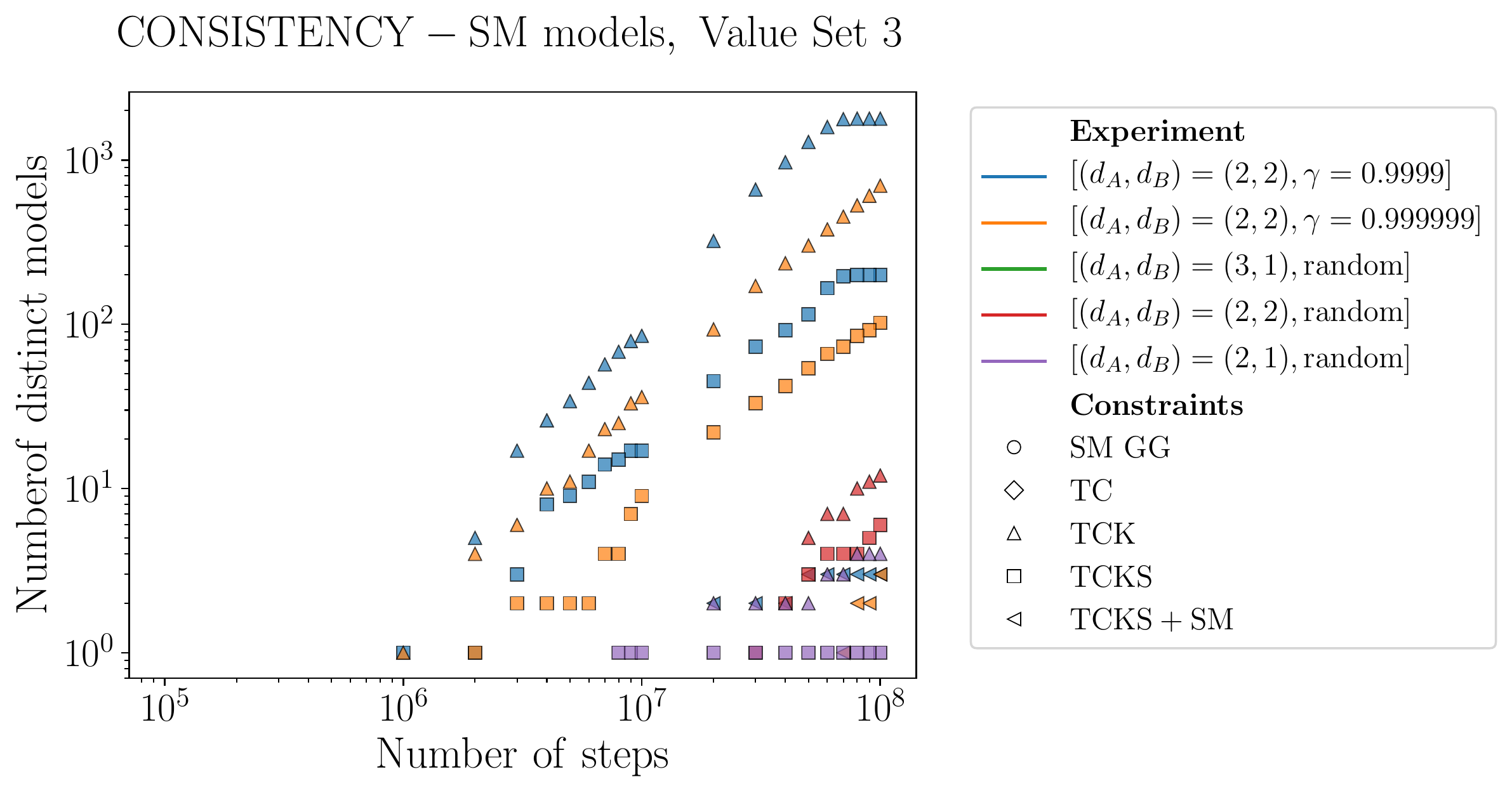}
\caption{Experimental results for reward function CONSISTENCY-SM. Results depend on choices of
value sets as in Table \ref{tab:valsconsec1}. Results for value sets $1$, $2$, and $3$ are
at top, middle, and bottom, respectively.}
\label{fig:stackvsrandomconsec1}
\end{figure}

\begin{figure}[t]
\centering
\includegraphics[width=0.8\textwidth]{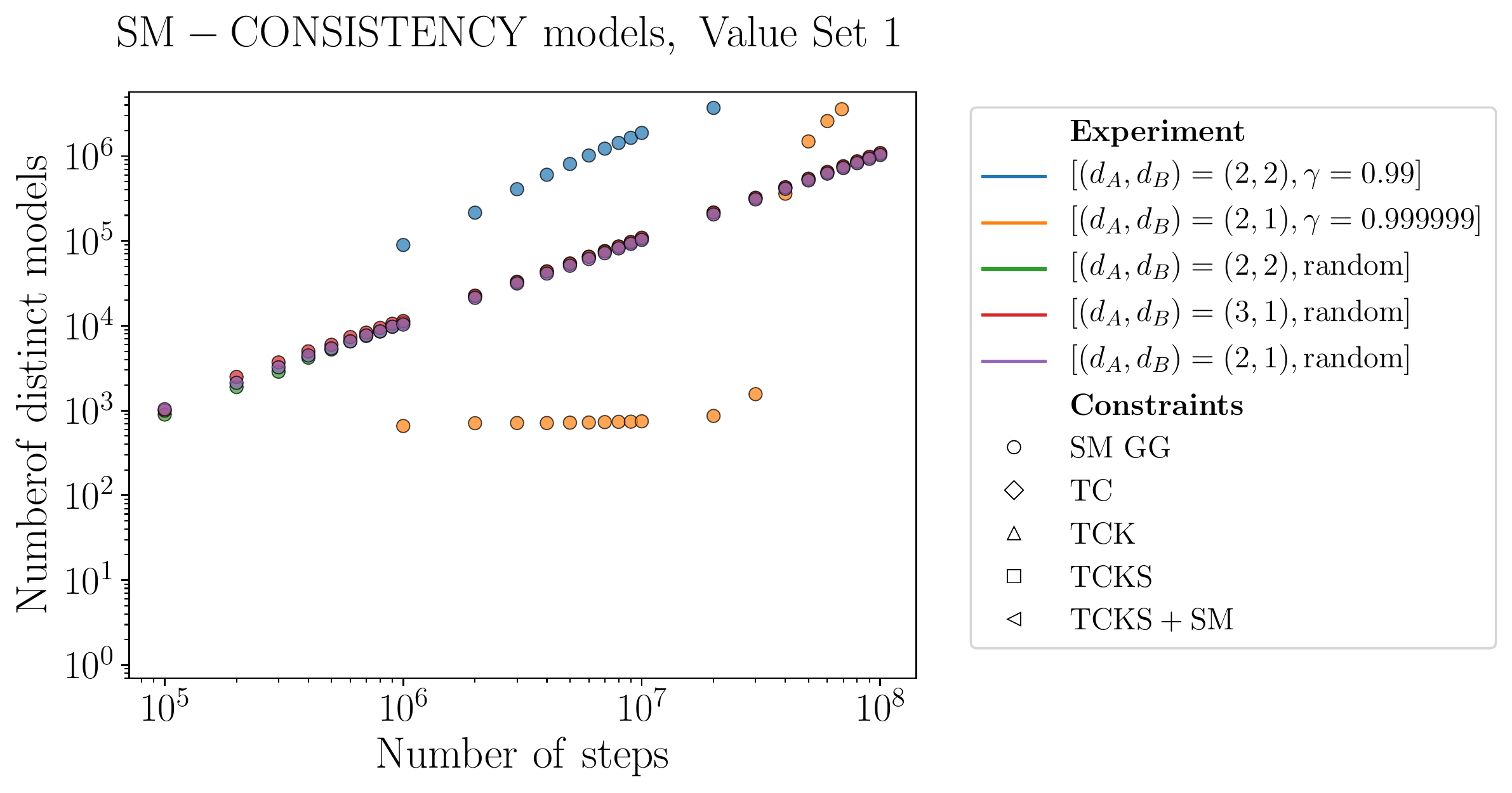} \\
\includegraphics[width=0.8\textwidth]{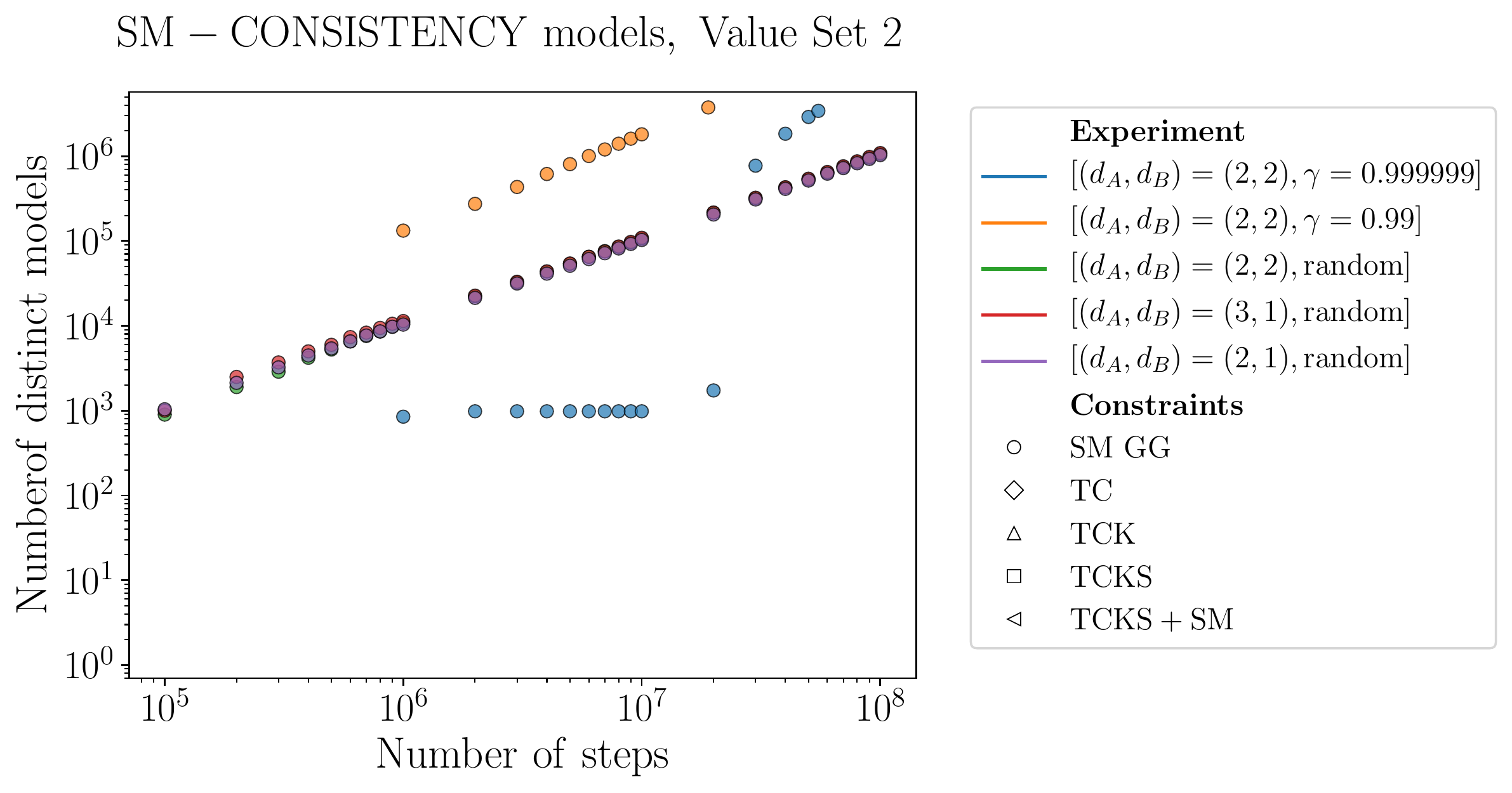} \\
\includegraphics[width=0.8\textwidth]{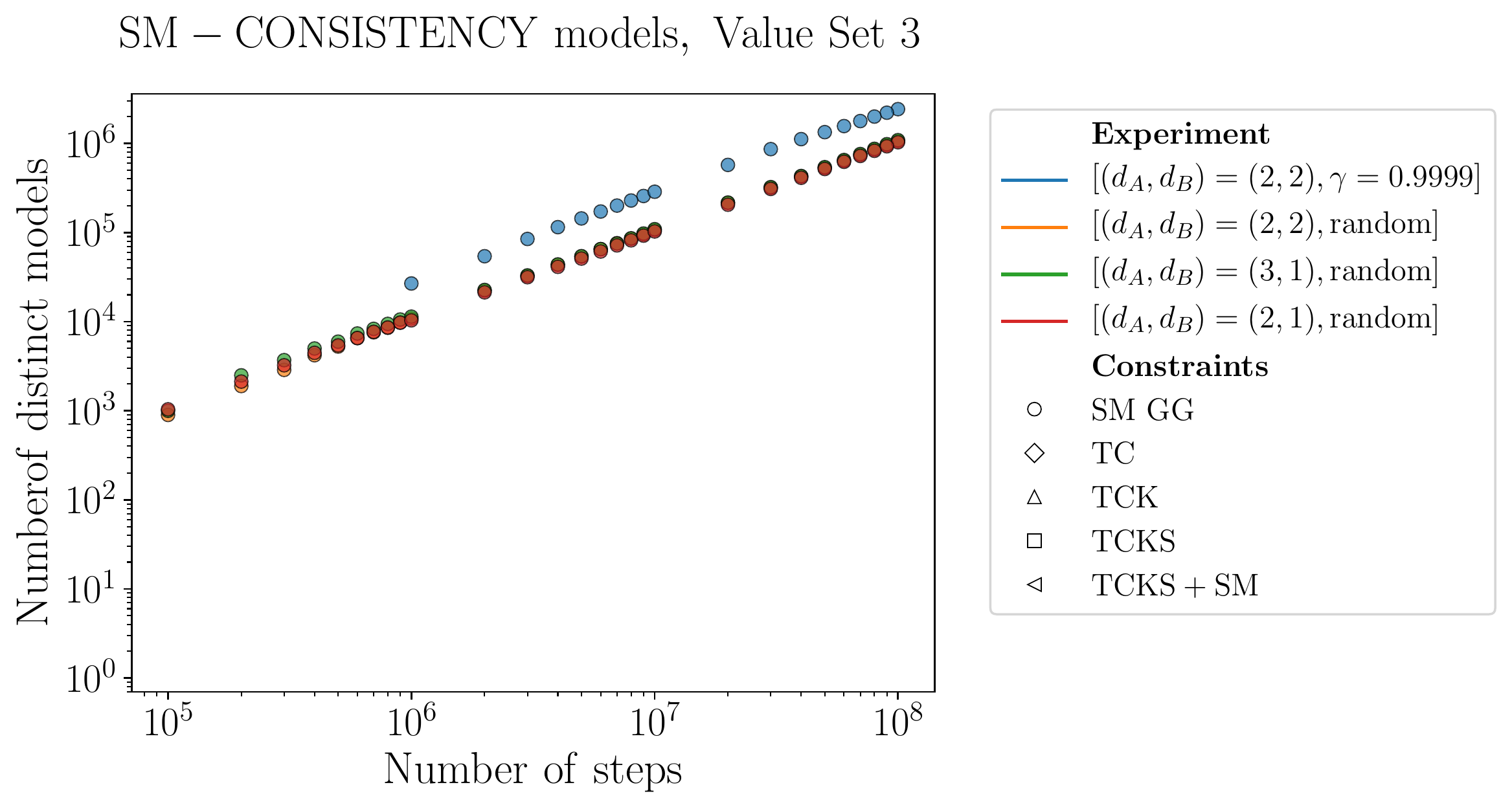}
\caption{Experimental results for reward function SM-CONSISTENCY. Results depend on choices of
value sets as in Table \ref{tab:valsconsec2}. Results for value sets $1$, $2$, and $3$ are
at top, middle, and bottom, respectively.}
\label{fig:stackvsrandomconsec2}
\end{figure}

\medskip

We now turn to a discussion of the results, as plotted in Figures \ref{fig:stackvsrandomsimult}-\ref{fig:stackvsrandomconsec2}.  In each of these figures, we monitor the progress of the agents each $10^6$ steps. In order to not overload the plots,  in the range $10^6$ to $10^7$ we plot each data point, whereas in the range $10^7$ to $10^8$ we average over $10$ data points, and in the interval $10^8$ to $10^9$ we average over $100$ points, and so on. This way, we get ten data points in each order of magnitude interval. The number of models satisfying various constraints are plotted. These include tadpole cancellation (TC), TC plus the K-theory constraint (TCK), TCK plus the supersymmetry conditions (TCKS)\footnote{TCKS models  are fully consistent supersymmetric string models.}, having the Standard Model gauge group (SM GG), and satisfying all consistency conditions while having the Standard Model gauge group (TCKS + SM).

Results for the stacking agent and random agent with reward function SIMULT are presented in Figure \ref{fig:stackvsrandomsimult}. For all value sets we present the results of the best-performing RL agents, and for value sets $1$ and $2$ we also present the results for an agent that demonstrates clear learning at late times, as demonstrated by sudden sharp increases. Results are comparable for all three value sets, but we discuss value set $1$ since results there are optimal by a small margin. 

Let us first compare the results for the best value of the discount factor $\gamma=.999999$. From the top plot in Figure \ref{fig:stackvsrandomsimult}, we see that the stacking agent has found $4\times 10^8$ models with SM GG after $2\times 10^7$ steps; by contrast, the random agents have factor of $\mathcal{O}(20)$ fewer models with SM GG after the same number of steps. The stacking agent has found $\mathcal{O}(200)$ models with TCK after $2\times 10^7$ steps, and the random agents have found a factor of $\mathcal{O}(100)$ fewer models after the same number of steps. By $2\times 10^7$ steps, the stacking agent has found $\mathcal{O}(50)$ fully consistent string models, i.e.\ those with TCKS, while the random agents find their first fully consistent model (a single one) at around $2.5\times 10^7$ steps. 

Let us next compare agents by their truncation parameters $(d_A,d_B)$. We note that for value sets $1$ and $2$, the $(d_A,d_B)=(3,1)$ agents learn to find Standard Model gauge groups much more slowly than the $(d_A,d_B)=(2,2)$ agents, but then increase their learning at late times so that the results are nearly comparable at the end of their runs; however, the difference in performance is also correlated with the different discount factors $\gamma$. We will discuss this more for the SM-CONSISTENCY experiments below. Note that the for larger and larger $(d_A,d_B)$, we truncate less and less the possible string configurations. On the other hand, the number of possible configurations increases exponentially, which requires much more runs for the agent to ``get acquainted'' with the landscape. However, already for $(d_A,d_B)=(2,2)$ there are $\mathcal{O}(10^{10})$ to $\mathcal{O}(10^{13})$ possible states, depending on the truncation. It is, however, not known whether the exact Standard Model is among them.

Results for the stacking agent and random agent with reward function CONSISTENCY-SM are presented in Figure \ref{fig:stackvsrandomconsec1}. Since consistency conditions (TCKS) are checked before particle physics conditions with this reward function, any model found with the Standard Model gauge group is necessarily consistent, which explains the presence of TCKS+SM models and the absence of models that have only the SM GG. The plots demonstrate the efficacy of the agents at solving various string consistency conditions. We see that there are many TCK models but no TC models, implying that all of the models that satisfy TC also satisfy the K-theory conditions. In general for the RL agents, there are about an order of magnitude more TCK models than TCKS models. Comparing the best-performing stacking agent to the best-performing random agent in all three plots, we find that the associated stacking agent finds a factor of $\mathcal{O}(200)$ more TCK models and a factor of $\mathcal{O}(50)$ TCKS models.

It is interesting to compare these CONSISTENCY-SM
results to the agent of Section \ref{sec:filler} that utilized
the STC reward function and learned the filler brane strategy.
After slightly under $10^7$ steps, which took $24$ hours since
SUSY was checked at every step, the agent found about $125$ models
that satisfied the SUSY and tadpole cancellation conditions.
In contrast, the best CONSISTENCY-SM agents (see Figure \ref{fig:stackvsrandomconsec1}), found over $200$ TCKS models in
$24$ hours, albeit utilizing $10^8$ steps. Since CONSISTENCY-SM
checks the tadpole conditions first, it \emph{cannot} be utilizing
the filler brane strategy, for reasons discussed above.
In addition, checking tadpole cancellation first saves time
from the costly evaluation of the SUSY conditions. We 
conclude that the CONSISTENCY-SM agents have learned a new
strategy for finding consistent string models that is about twice as efficient per unit time as 
the filler brane strategy.

Results for the stacking and random agent with reward function SM-CONSISTENCY are presented in Figure \ref{fig:stackvsrandomconsec2}. Here, consistency conditions would only be checked for models that have the Standard Model gauge group and three Standard Model families, possibly with exotics. Since no models that satisfy the latter constraints were found in these runs, the experiments never check for consistency, and therefore all output of these experiments is limited to features related to being SM-like. Accordingly, the plots appear to be much simpler, but this is an artifact of the ordering of when conditions are checked. We find that the best-performing stacking agent finds a factor of $\mathcal{O}(20)$ more models with SM GG than the best-performing random agent. This factor of improvement is consistent with the results of the SIMULT experiments. As with the SIMULT experiments, we again see that some of the RL agents learn to find the Standard Model gauge group very slowly at early times, but then learning picks up at late times. From the plot with value set $2$, we see that the key effect must be the different in the discount factor $\gamma$, since $(d_A,d_B)$ have the same values for the two discrepant RL plots. We therefore conclude that it is likely the large $\gamma$ factor ($\gamma = .999999$ vs $\gamma = .99$) that leads to sharp learning at late times in the SIMULT case. This could indicate a property of the landscape as perceived by the stacking agent: at late time, the agent has learned how to get to a good state, but getting there requires moving through states with a smaller reward. For larger $\gamma$, the agent takes future rewards more into consideration for its current policy, which means it temporarily accepts going through states with small rewards.

Summarizing these experiments, we find that the best RL agent often picks up factors of $\mathcal{O}(20)$, $\mathcal{O}(200)$, and $\mathcal{O}(50)$ relative to random agents in finding models with SM GG, TCK, and TCKS, respectively. The best-performing agents have $(d_A,d_B)=(2,2)$. We also find that smaller discount factors $\gamma$ leads to faster learning, but large $\gamma$ can lead to rapid learning at late times that may lead to optimal performance.

\subsection{Additional Stacking Agent Experiments}
\begin{figure}[t]
\centering
\includegraphics[width=.85\textwidth]{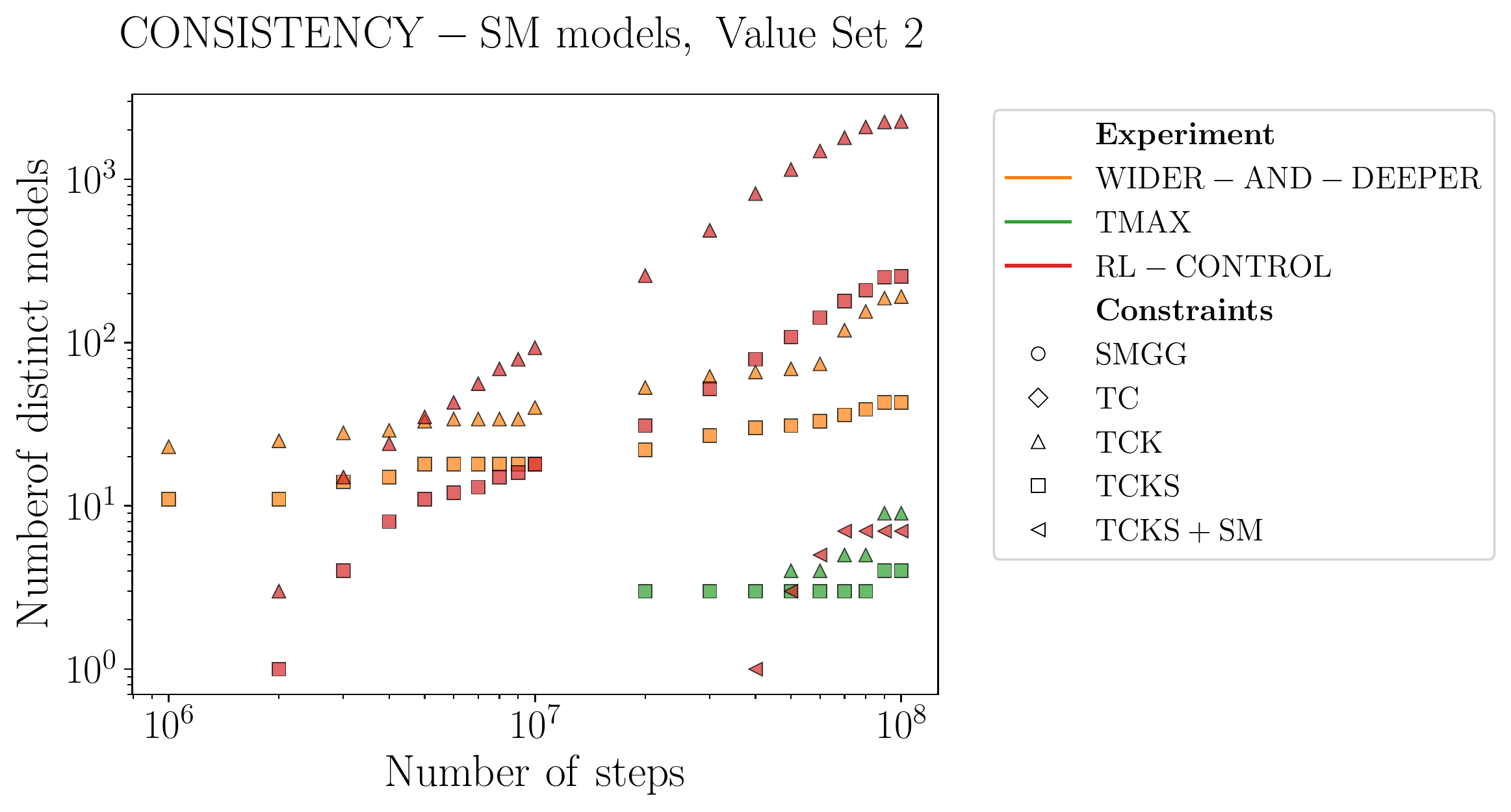}
\caption{Number of models of different types found for the additional stacking experiments WIDER-AND-DEEPER and TMAX.}
\label{fig:stackingext}
\end{figure}

Having performed systematic experiments to find promising hyperparameters for stacking agents, we would like to perform two additional experiments using the best-performing hyperparameters. Specifically, we will perform three new experiments. In each, $32$ A3C workers take $10^8$ total steps (or $24$ hours, whichever comes first) using $(d_A,d_B)=(2,2)$, $\gamma=0.9999$, and reward CONSISTENCY-SM with value assignment $2$. The two experiments differ in the following ways, named according to what is different about them:
\begin{itemize}
\item Experiment WIDER-AND-DEEPER: Throughout this paper, the policy and value function neural networks have $4$ hidden layers, three with $50$ nodes and one with $200$, or in some cases two hidden layers  with 200 nodes each. In WIDER-AND-DEEPER, they have $4$ layers with $2000$, $2000$, $2000$, and $200$ hidden nodes, respectively.
\item Experiment TMAX: instead of using the default value $t_{\text{max}}=5$, this experiment takes $t_{\text{max}}=20$. The parameter $t_{\text{max}}$ is the number of steps that are taken in between policy and value function updates, so
by increasing $t_{\text{max}}$ in this way the agent
sees four times as many states before updating its behavior.
\end{itemize}
Results of these experiments are presented in Figure \ref{fig:stackingext}, where results from WIDER-AND-DEEPER and TMAX are presented in green and orange, respectively. 

In order to see how well these new experiments
perform, we would like to compare each to 
a previous RL experiment that performed
well. We call the latter RL-CONTROL, which is the experiment with $(d_A,d_B)=(2,2)$, $\gamma=0.9999$, and reward CONSISTENCY-SM with value assignment $2$. The results of RL-CONTROL are  the orange data points in the middle
plot of Figure \ref{fig:stackvsrandomconsec1}. RL-CONTROL found $40$ TCK and $10$ TCK models after $6$ million steps; after $10^8$ steps, it found $\mathcal{O}(2000)$ TCK and $\mathcal{O}(250)$ TCKS models. For convenient comparison,
RL-CONTROL is plotted in Figure \ref{fig:stackingext}
in red.

Examining Figure \ref{fig:stackingext}, we see that WIDER-AND-DEEPER performs poorly compared to RL-CONTROL.
Specifically, after $10^8$ steps, WIDER-AND-DEEPER finds
a factor of $\mathcal{O}(10)$ fewer TCK and TCKS models. This is perhaps not surprising, since
wider and deeper networks are expected to have better
behavior at late times, but take longer to train.
Therefore, if training times are not long enough, wider
networks networks can actually have lower performance,
as seen here.
We see that TMAX performs poorly compared to both
WIDER-AND-DEEPER and RL-CONTROL. 
After $10^8$ steps, it has found only $8$ TCK models and $3$ TCKS models, which is orders of magnitude below CONTROL. A priori in a given environment it is not clear what value
of $t_\text{max}$ is optimal. Since immediate updates ($t_\text{max} = 1$) and end of episode
updates ($t_\text{max} = t_\text{end}$) are both
typically sub-optimal choices, in a given
environment the optimal value of $t_\text{max}$ is, a priori, unclear. From the
TMAX experiment, we see that in this environment  the optimal
value is likely below $t_\text{max} = 20$.

\subsection{Flipping and one-in-a-billion agents}

Let us discuss the results of the flipping agent as well as the two one-in-a-billion agents. They were run with 
good hyperparameters as determined by the stacking
agent, $(d_A,d_B)=(2,2)$, $\gamma=0.99$,
and reward structure CONSISTENCY-SM.
\begin{figure}[t]
\centering
\includegraphics[width=0.48\textwidth]{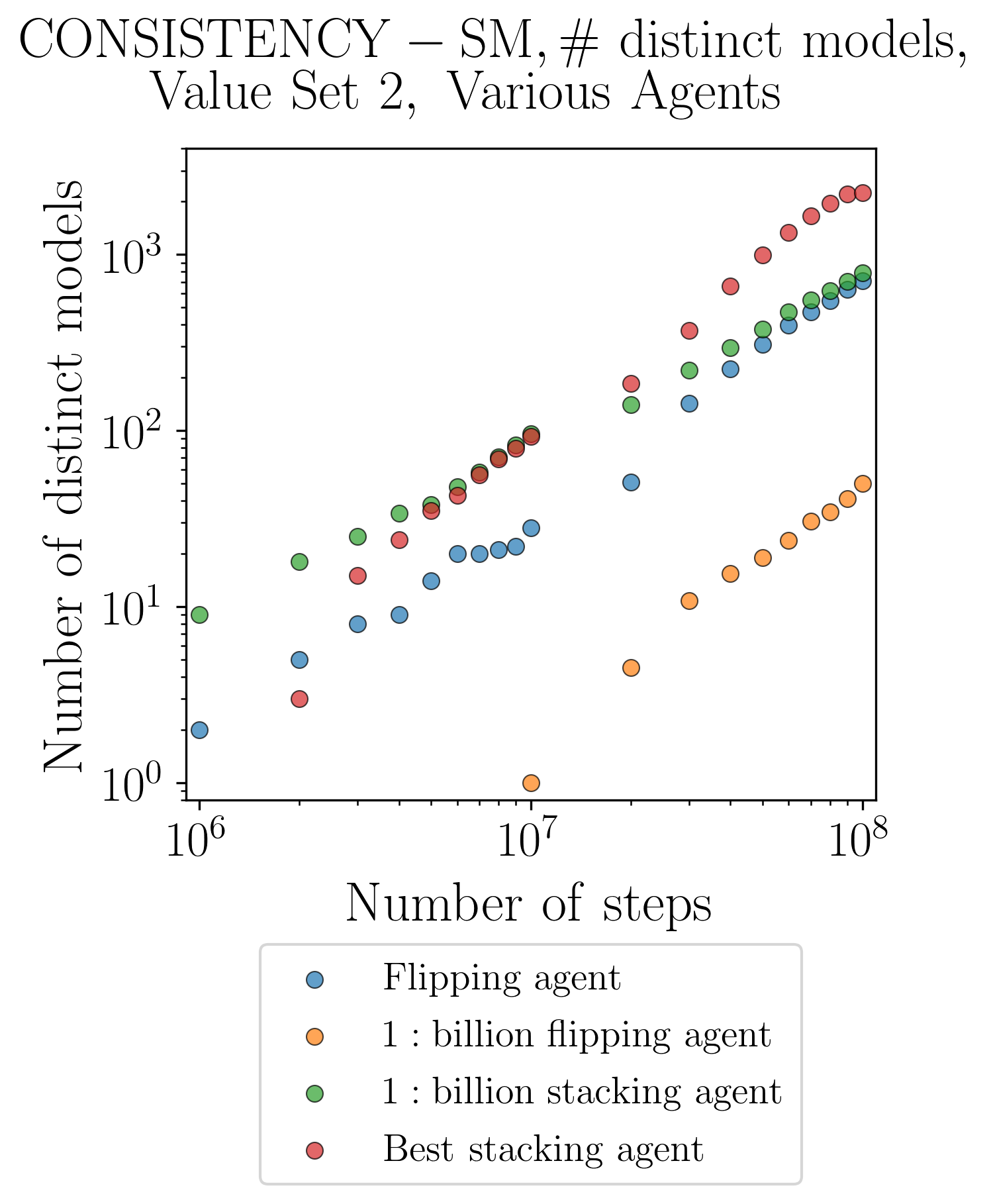}~~
\includegraphics[width=0.48\textwidth]{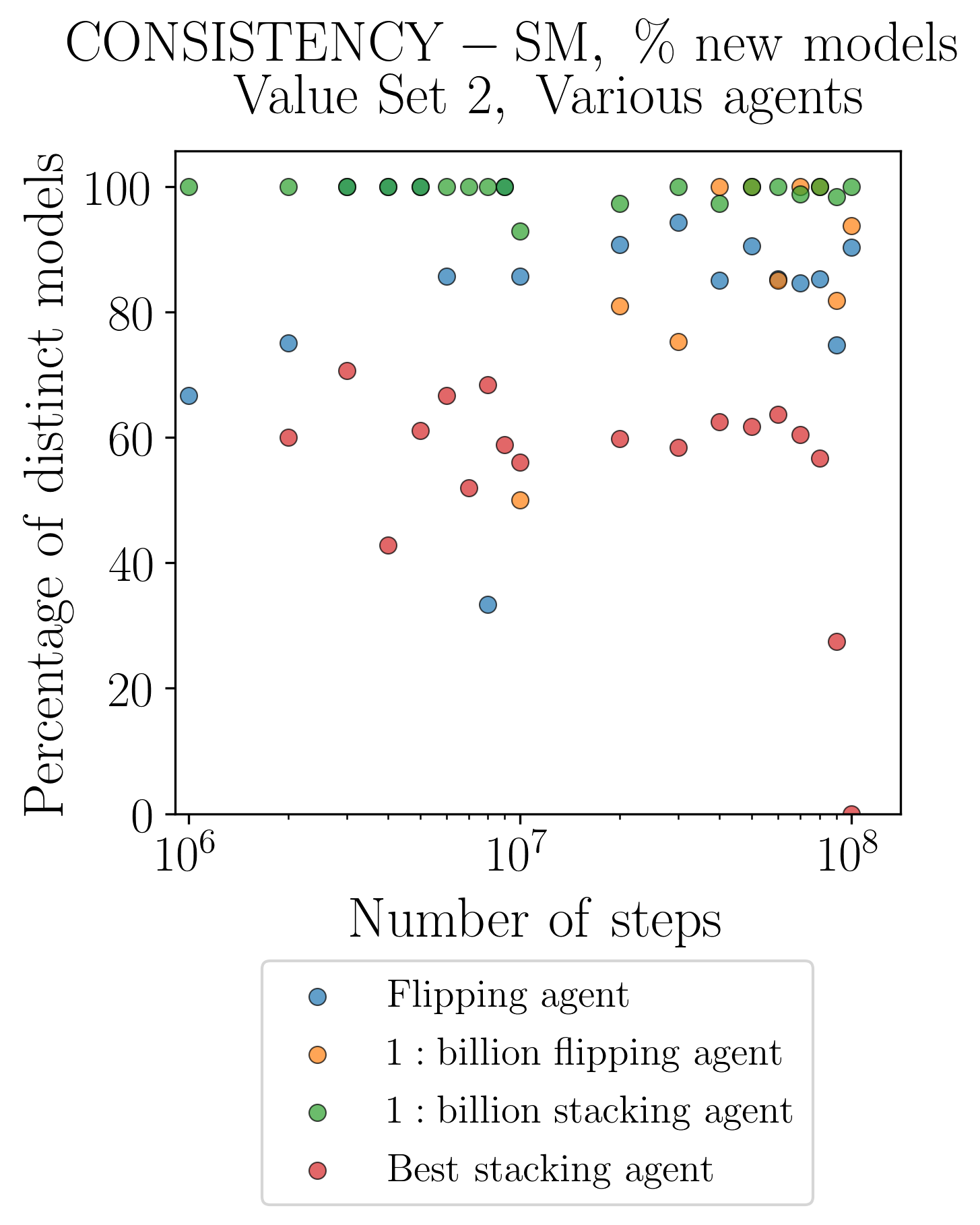} \\[2ex]
\includegraphics[width=0.98\textwidth]{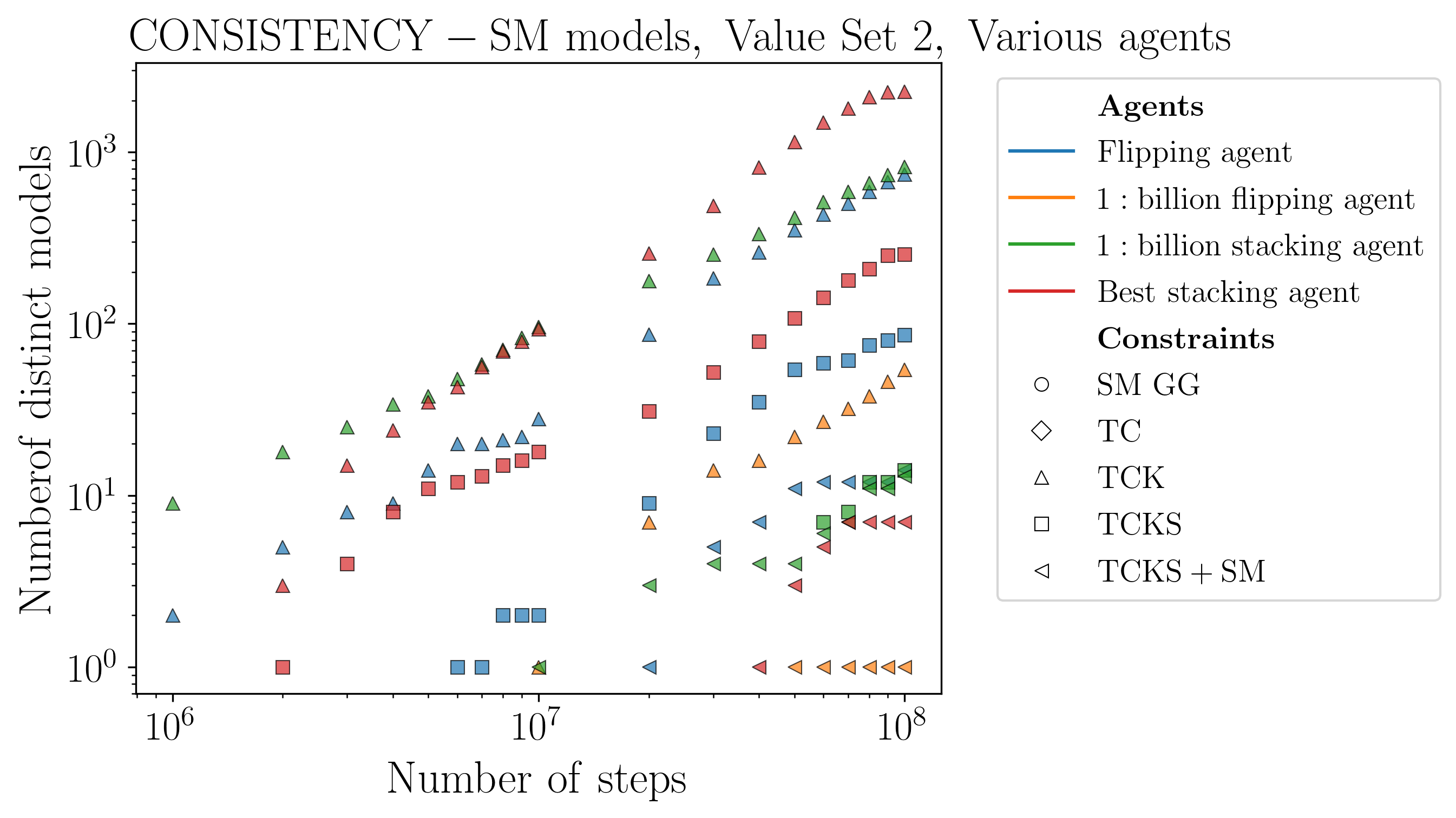}
\caption{Results for the flipping and the two one-in-a-billion agents, with the best stacking agent included for comparison. We show the overall number of models (top left), the rate of finding new models (top right), and the properties of the found models (bottom) for reward function CONSISTENCY-SM and hyperparameter $\gamma=0.99$.}
\label{fig:flippingAndOIABResults}
\end{figure}
The top left plot in Figure ~\ref{fig:flippingAndOIABResults} shows the overall number of steps on the $x$-axis and the total number of distinct models that satisfy at least one of the phenomenological or consistency constraints on the $y$-axis. This means, each plotted model has to satisfy at least TC or SM; however, due to the reward structure CONSISTENCY-SM used in the experiments, SM is only checked after all consistency constraints TCKS are checked, so the points correspond to tadpole-canceling models.

As we can see, the 1:billion stacking agent finds the most models of these type, roughly at a constant rate of $10^{-5}$; after $10^6$ steps, the agent has found $10$ different models and at $10^8$ steps the agent has found 1000 models. The flipping agent performs slightly weaker at the beginning, but catches up to the 1:billion stacking agent at around $10^8$ steps. Lastly, the 1:billion flipping agent performs worse by roughly a factor of 10. We again include the best stacking agent (called RL-CONTROL in the previous section) for comparison. We find that it outperforms the best agent in this section, i.e.\ the 1:billion stacking agent, by roughly a factor of 2 when it comes to satisfying TC. Since the stacking agent has more freedom in satisfying the tadpole as compared to the 1:billion stacking agent (the latter cannot introduce a hidden sector to satisfy the tadpole), this result is to be expected. However, we find that the 1:billion stacking agent outperforms the stacking agent once we impose all constraints TCKS+SM. The 1:billion stacking agent  is able to catch up to the other stacking agent since it has the last constraint already built in, while the stacking agent has not had too much time to learn to take the SM constraint into account.

It is somewhat surprising that the 1:billion stacking agent initially outperforms the flipping agent. This tells us that the agent can learn traversing the landscape by exchanging entire stacks more easily as compared to changing single winding numbers. However, after a while the agent becomes equally effective with both methods. This is probably due to the fact that flipping can lead to many more illegal moves than stacking. The fact that the 1:billion flipping agent performs worse than the unconstrained flipping agent is  to be expected, since the latter flipping agent has more degrees of freedom (i.e.\ the hidden sector stacks) to adjust and can hence cancel the tadpole more easily.

The top right plot of Figure~\ref{fig:flippingAndOIABResults} shows the rate at which the agents discover new models (as compared to reproducing a  model they have found previously). This can be used to monitor the exploration of the agents. All agents explore the landscape very well, with a new-model-rate of 80 to 100 percent for most of the training time. The rate of the flipping agents is lower than the one of the stacking agent. The reason for this is that the agents are reset every $10^5$ steps to their initial configuration. The flipping agents probably have to traverse the same models from their start configuration each time before they can branch out and reach new models. The stacking agent, in contrast can ``jump'' from the start configuration to any other winding configuration.

The bottom plot of Figure~\ref{fig:flippingAndOIABResults} shows how the three agents  learn to satisfy the different constraints over time (i.e.\ over number of steps). First we note that all agents never find TC models but always TCK. So, the K-theory constraint is automatically satisfied in all models. The TCK markers follow those of plot 1, which is due to the fact that the points in plot 1 correspond to the models that satisfy at least TC. At around $5\times10^6$, the flipping agent starts finding models that also satisfy the SUSY constraints. Most of the TCKS models which the 1:billion agents find automatically also satisfy the SM constraint, since by construction their winding numbers are such that the SM gauge group is already built in. After $10^8$ steps, both the flipping and the 1:billion stacking agent have found around 10 TCKS+SM models.

In summary, we find that the agent learns to traverse the landscape more easily by stacking. As an added bonus, the exploration rate of the stacking agent is higher than that of the flipping agent due to the structure of the landscape when traversed by flipping single winding numbers. The flipping agent learns quickly to satisfy SM, such that after $10^8$ steps, the agents find the same number of TCKS+SM models. We thus conclude that the agents can learn the properties needed for SM rather easily. The best model we could find,  however, was found by the 1:billion flipping agent, which satisfied all constraints and had 8 exotics. Here the number of exotics is the minimum number of exotics as computed for each of the four hypercharge embeddings studied in \cite{Gmeiner:2005vz}; as there, we do not impose a massless hypercharge for the sake of comparison. The best models of the flipping and 1:billion stacking agent had 14 and 18 exotics, respectively. So while the agents have not found the exact MSSM, they get rather close within 24h of running time.

\subsection{Comparison with earlier work}
It is instructive to compare our results with the results of~\cite{Gmeiner:2005vz}. Let us briefly recap their approach, which is different from ours. The authors fix the complex structure parameters to successively increasing values and find all winding numbers compatible with the SUSY constraints for this fixed complex structure. For these values, they then check which assignments satisfy the tadpole constraints. In this way they find $\mathcal{O}(10^8)$ models with TC and S, and $\mathcal{O}(2\times10^7)$ models with TCKS. However, the authors allow for untilted and tilted tori, quoting that 1.6 percent of the models live on tilted tori. In our analysis of the SM quantities, we have focused on tilted tori, since untilted tori cannot accommodate the Standard Model since they do not allow for an odd number of generations.

The authors ran their search on a cluster a decade ago for $4 \times 10^5$ CPU hours. We compare their result with the TCKS models as given in Figure~\ref{fig:stackvsrandomconsec1} for CONSISTENCY-SM reward structure, which correspond to 32 agents running for at most 24h (we did not keep track of the precise running time, but since the agent reaches $10^8$ steps, it has to be less than 24h). Since we run 32 agents on 8 hyperthreaded cores, we get at most 24 CPU hours (or 192 core hours). Making the conservative assumption that in the case of~\cite{Gmeiner:2005vz} all models on tilted tori correspond to exactly one tilted torus, the authors find around $\mathcal{O}(10^5)$ TCKS models in $\mathcal{O}(10^5)$ hours, while our agents finds $\mathcal{O}(10^2)$ models in $\mathcal{O}(10$) hours.  It is very difficult to account for the increase in speed of CPUs, RAM and storage, as well as for the fact that they searched over tilted and untilted tori, while our run was for tilted tori only but also included all Standard Model checks. This makes a more detailed comparison impossible. However, since the authors construct all winding number combinations within a given box (whose size is set by the complex structure parameters), we assume that their findings are comparable with our random search, which is outperformed by our agent by a factor $\mathcal{O}(10)$ to $\mathcal{O}(100)$, which seems plausible from the above comparison.

Based on their observations, the authors of~\cite{Gmeiner:2005vz} estimate that the chance of finding a Standard Model among a TCKS model is one in a billion; they construct a total of $10^8$ states, which leaves them with $\mathcal{O}(0.1)$ Standard Models in their ensemble. We want to use this landscape statistic for a very rough estimate of what to expect in our case. For $(d_A,d_B)=(2,2)$, there are an estimated $10^{13}$ states in our truncation according to~\eqref{eq:NStatesSymm}. We note that this number of states was computed for three untwisted tori; for two untwisted and one twisted, which we looked at, there are less permutation symmetries and the number will be somewhat larger. The random agent finds TCKS states at a rate of $10^{-8}$, so we expect roughly $10^5$ TCKS states in this truncation. In order to estimate how many Standard Models are among these, we make the assumption that the statistics of~\cite{Gmeiner:2005vz} carries over to our case. They find various suppression factors leading to the overall suppression of $10^{-9}$. The largest two suppression factors of $10^{-5}$ and $10^{-3}$ come from demanding three generations of quarks and leptons, respectively. As explained around~\eqref{eq:EvenChi}, this can only be achieved if at least one torus is tilted, which is the case in one percent of the examples of~\cite{Gmeiner:2005vz}. We thus estimate this suppression factors to be $10^{-3}$ and $10^{-1}$ for cases with tilted tori, respectively. This puts the overall likelihood of finding a Standard model amongst TCKS states on geometries where some tori are twisted at around $1:10^{5}$. Since we have $10^5$ TCKS states in the $(2,2)$ truncation, we expect this to contain $\mathcal{O}(1)$ Standard Models. For larger values of $(d_a,d_B)$ in the truncation, these numbers go up significantly, cf.~\eqref{eq:NStatesSymm}.

The authors of \cite{Gmeiner:2005vz} do not find models with three generations of quarks and leptons, even though they do not impose a massless  hypercharge. We impose a massless hypercharge and find models with a net number of three generations (i.e.\ number of generations minus number of anti-generations) of quarks, plus $\mathcal{O}(10)$ exotics. We do not monitor further the irreducible representations with respect to the Standard Model and the hidden sector of these exotics.

\section{Discussion and Summary}
\label{Sec:conc}

In this paper we have proposed deep reinforcement learning (RL) as a model-free way to explore the string landscape
with artificial intelligence (AI). 
In deep RL, an AI agent explores an environment in which
it may receive both positive and negative rewards, teaching
itself strategies that lead to improved behavior over
time. This is a mature field in computer science that has
led to state-of-the-art results in other fields, 
famously in playing Go \cite{silver2017mastering} and folding proteins, but also in physics, e.g.\ for quantum
control \cite{PhysRevX.8.031086} and quantum error
correction \cite{2018arXiv181007207S}.

Our general RL proposal for the string landscape 
consisted of three concrete ideas:
that RL is suitable for studying the string landscape since it may
be used in environments with exponentially large
numbers of states; that RL could discover
new strategies that could lead to superior results; and
that RL is model-free, in the sense that RL algorithms
can be applied to many different environments,
which allows string theorists to focus on the 
string environment rather than developing a new
algorithm. In our environment of choice for
this paper, which was an orientifold compactification
of type IIA string theory, we 
demonstrated that RL can make progress towards each
of these goals.

More broadly, understanding the particle physics and cosmology
implications of string theory requires grappling
with its large and computationally complex
landscape of vacua. Though
formal progress is certainly necessary, 
it is difficult to imagine obtaining a complete
understanding without
concrete and intelligent exploration. Our results
demonstrate for the first time that progress can be
achieved with artificial intelligence in
the context of reinforcement learning.
It is easy to imagine the use of reinforcement 
learning in other string theoretic contexts,
including outside of landscape studies, due to
the model-free nature of many of its algorithms.

\bigskip

\noindent Let us summarize the main results of our RL string landscape analysis.

Our A3C agents explored the environment of compactifications of type IIA superstring theory with intersecting D6-branes. This is a doubly constrained system: for the sake of consistency and stability we impose tadpole cancellation, K-theory, and supersymmetry constraints; for the sake of particle physics we attempt to find a model as close to the Standard Model of particle physics as possible. Together, these constraints give rise to a coupled system of Diophantine equations\footnote{Strictly speaking, the SUSY constraints in general need to be solved over $\mathbbm{R}$, but since we are not stabilizing moduli we can consider solutions in $\mathbbm{Q}$ (or equivalently $\mathbbm{Z}$) that lead to Diophantine equations.}, which are notoriously difficult to solve (see, e.g., \cite{Cvetic:2010ky} for a study of Diophantine undecidability in string theory), in a space of possible states that grows exponentially with input size. Nevertheless, our A3C agents perform very well. They significantly outperform random walkers, sometimes by several orders of magnitude, despite having learned only from their experience rather than being explicitly programmed.

Our agents explore a concrete set of type IIA compactifications on a fixed geometry known as a toroidal orbifold. We test and discuss different but equivalent ways of describing and traversing this landscape. Since we need to solve a multi-task reinforcement learning problem, we try different approaches that differ in the order in which the various physical and string consistency conditions are imposed and checked and discuss their influence on the agent's ability to learn strategies to solve these tasks. We also change the way in which the agent perceives the landscape, i.e.\ what it means to take a step. This heavily impacts which states are ``close'' to one another in the landscape, i.e.\ which string configurations can be reached from any
given configuration with a few number of steps. Consequently, this also impacts the strategy the agent has to learn for solving the various conditions.

We find that the best results are obtained using the CONSISTENCY-SM reward function in the stacking environment. The former denotes the order in which we impose the various constraints, which is checking successively for tadpole cancellation, K-theory, supersymmetry, Standard Model gauge group, and the Standard Model spectrum. The latter refers to how the agent traverses and  thus perceives the landscape. The agent takes steps by either adding or removing entire brane stacks, which means it chooses the homology class of cycles (which is characterized by the winding numbers around each of the 6 torus cycles, up to a certain truncation) on which to wrap the D6 branes. For each stack, the agent can furthermore increase or decrease the number of branes in the stack.

Compared to the best-performing random walkers, we find that this RL agents pick up factors of $\mathcal{O}(20)$, $\mathcal{O}(200)$, and $\mathcal{O}(50)$ in the number of models found with the Standard Model gauge group (without imposing consistency), tadpole cancellation and K-theory, and those two conditions plus supersymmetry, respectively. Unfortunately, we did not succeed in finding an exact realization of the Standard Model, as
in each case we found a number of exotics. 
Note that it is possible that type IIA compactifications on a $\mathbbm{Z}_2\times\mathbbm{Z}_2$ toroidal orbifold with one tilted torus does not admit a single Standard Model solution; the brute-force search of~\cite{Gmeiner:2005vz}, which ran for over $10^5$ CPU hours, also did not find a consistent Standard Model solution in this compactification.

We have demonstrated that  in the STC approach of solving the string consistency conditions, the RL agent can learn human-derived heuristic strategies, 
while in the TCKS approach, where no strategy of solving the Diophantine equations is known to humans, the agent derived
a new strategy that is about twice as efficient at finding fully consistent string models per unit time. Specifically, using the STC reward we demonstrated that the agent learns to find tadpole canceling supersymmetric models by disproportionately using so-called filler branes, which contribute to the tadpole cancellation conditions but not the SUSY conditions. It finds about $125$ such models in $24$ hours, taking $10^7$ steps.
A CONSISTENCY-SM agent finds over $200$ such models in $24$ hours, however, and the choice of reward function forbids the agent from utilizing the filler brane strategy. Instead, the agent found a strategy that is more efficient per unit time. 

It would be interesting to understand whether there is a simple interpretation of the efficient strategy. We leave this and other studies to future work, but anticipate exciting progress in a number of directions.

\vspace{1cm}
\noindent \textbf{Acknowledgments.} We thank Liam Fitzpatrick, Sergei Gukov, Gabriele Honecker, Sven Krippendorf, Cody Long, Dieter L\"ust, Liam McAllister, Hans Peter Nilles, Piotr Su{\l}kowski, Jiahua Tian, Patrick Vaudrevange, Lorenzo Vitale, Fernando Quevedo, and Yuan Xin for  useful discussions. J.H. is supported by NSF grant PHY-1620526. The work of F.R.\ was partly supported by the EPSRC network grant EP/N007158/1. F.R.\ further thanks Northeastern for kind hospitality and support during his visit. The work of B.N. is supported by NSF Grant PHY-1620575.

\clearpage
\appendix

\section{Value Sets for Reward Functions}

\begin{table}[h]
\centering
\begin{tabular}{|c|c|c|c|}
\hline
Property & Value $1$ & Value $2$ & Value $3$ \\ \hline
\tw{tadpoleDistanceMultiplier} & $1$ & $1$ & $1$\\ 
\tw{TC\_Reward} & $10^6$ & $10^6$ & $10^6$\\ 
\tw{TCK\_Reward} & $10^7$ & $10^8$ & $10^7$\\ 
\tw{TCKS\_Reward} & $10^8$ & $10^{10}$ & $10^8$\\
\tw{missingGroupFactorDistance} & $100$ & $100$ & $10^6$\\ 
\tw{missingParticleDistance} & $10$ & $10$ & $10$\\  
\tw{SMlike\_Reward} & $10^8$ & $10^{10}$ & $10^8$\\ 
\tw{SM\_Reward} & $10^9$ & $10^{12}$ & $10^9$\\ 
\hline
\end{tabular}
\caption{Value assignments for the SIMULT reward function.}
\label{tab:valssimult}
\end{table}

\begin{table}[ht]
\centering
\begin{tabular}{|c|c|c|c|}
\hline
Property & Value $1$ & Value $2$ & Value $3$ \\ \hline
\tw{tadpoleDistanceMultiplier} & $1$ & $1$ & $1$\\ 
\tw{TC\_Reward} & $10^7$ & $10^7$ & $10^7$\\ 
\tw{TCK\_Reward} & $10^8$ & $10^9$ & $10^8$\\ 
\tw{TCKS\_Reward} & $10^9$ & $10^{11}$ & $10^9$\\
\tw{missingGroupFactorDistance} & $10^4$ & $10^4$ & $10^6$\\ 
\tw{missingParticleDistance} & $10^4$ & $10^4$ & $10^6$\\  
\tw{SMlike\_Reward} & $10^{11}$ & $10^{13}$ & $10^{11}$\\ 
\tw{SM\_Reward} & $10^{13}$ & $10^{15}$ & $10^{13}$\\ 
\hline
\end{tabular}
\caption{Value assignments the CONSISTENCY-SM reward function.}
\label{tab:valsconsec1}
\end{table}

\begin{table}[ht]
\centering
\begin{tabular}{|c|c|c|c|}
\hline
Property & Value $1$ & Value $2$ & Value $3$ \\ \hline
\tw{tadpoleDistanceMultiplier} & $500$ & $500$ & $500$\\ 
\tw{TC\_Reward} & $10^{11}$ & $10^{13}$ & $10^{11}$\\ 
\tw{TCK\_Reward} & $10^{12}$ & $10^{15}$ & $10^{12}$\\ 
\tw{TCKS\_Reward} & $10^{13}$ & $10^{17}$ & $10^{13}$\\
\tw{missingGroupFactorDistance} & $50$ & $50$ & $10^6$\\ 
\tw{missingParticleDistance} & $1$ & $1$ & $10^6$\\  
\tw{SMlike\_Reward} & $10^{6}$ & $10^{6}$ & $10^{8}$\\ 
\tw{SM\_Reward} & $10^{9}$ & $10^{9}$ & $10^{9}$\\ 
\hline
\end{tabular}
\caption{Value assignments for the SM-CONSISTENCY reward function.}
\label{tab:valsconsec2}
\end{table}

\begin{table}[ht]
\centering
\begin{tabular}{|c|c|c|c|}
\hline
Property & Value $1$ & Value $2$ & Value $3$ \\ \hline
\tw{tadpoleDistanceMultiplier} & $1$ & $1$ & $1$\\ 
\tw{TC\_Reward} & $10^{7}$ & $10^{7}$ & $10^{7}$\\ 
\tw{TCK\_Reward} & $10^{8}$ & $10^{9}$ & $10^{8}$\\ 
\tw{TCKS\_Reward} & $10^{9}$ & $10^{11}$ & $10^{9}$\\
\hline
\end{tabular}
\caption{Value assignments for the CONSISTENCY reward function.}
\label{tab:valsconsist}
\end{table}

\begin{table}[ht]
\centering
\begin{tabular}{|c|c|c|c|}
\hline
Property & Value $1$ & Value $2$ & Value $3$ \\ \hline
\tw{missingGroupFactorDistance} & $50$ & $50$ & $10^6$\\ 
\tw{missingParticleDistance} & $1$ & $1$ & $10^6$\\  
\tw{SMlike\_Reward} & $10^{7}$ & $10^{9}$ & $10^{8}$\\ 
\tw{SM\_Reward} & $10^{9}$ & $10^{11}$ & $10^{9}$\\ 
\hline
\end{tabular}
\caption{Value assignments for the SM reward function.}
\label{tab:valssm}
\end{table}

\FloatBarrier
\providecommand{\href}[2]{#2}\begingroup\raggedright\endgroup

\end{document}